\documentclass[fleqn,usenatbib]{mnras}

\usepackage{newtxtext,newtxmath}

\usepackage[T1]{fontenc}

\DeclareRobustCommand{\VAN}[3]{#2}
\let\VANthebibliography\thebibliography
\def\thebibliography{\DeclareRobustCommand{\VAN}[3]{##3}\VANthebibliography}

\usepackage{graphicx}	
\usepackage{amsmath} 
 
\usepackage{amssymb}	

\usepackage{siunitx}
\usepackage{color,colortbl}
\usepackage{multirow}
\usepackage{multicol}
\usepackage{hyperref}

\DeclareSIUnit \parsec {pc}

\DeclareMathOperator\erf{erf}

\title[DEVILS: SED Fitting in D10-COSMOS]{Deep Extragalactic VIsible Legacy Survey (DEVILS): SED Fitting in the D10-COSMOS Field and the Evolution of the Stellar Mass Function and SFR-$M_\star$ relation}

\author[J. E. Thorne et al.]{
Jessica E. Thorne,$^{1}$\thanks{E-mail: jessica.thorne@icrar.org}
Aaron S. G. Robotham,$^{1,2}$
Luke J. M. Davies,$^{1}$
Sabine Bellstedt,$^{1}$
Simon P. Driver,$^{1}$\newauthor
Mat\'ias Bravo,$^{1}$
Malcolm N. Bremer,$^{3}$
Benne W. Holwerda,$^{4}$
Andrew M. Hopkins,$^{5}$
Claudia del P. Lagos,$^{1,2}$\newauthor
Steven Phillipps,$^{3}$
Malgorzata Siudek,$^{6,7}$
Edward N. Taylor,$^{8}$
Angus H. Wright$^{9}$
\\
$^{1}$ ICRAR, The University of Western Australia, 35 Stirling Highway, Crawley, WA 6009, Australia\\
$^{2}$ ARC Centre of Excellence for All Sky Astrophysics in 3 Dimensions (ASTRO 3D)\\
$^{3}$ Astrophysics Group, School of Physics, University of Bristol, Bristol BS8 1TL, UK\\
$^{4}$ Department of Physics and Astronomy, 102 Natural Science Building, University of Louisville, Louisville KY 40292, USA\\
$^{5}$ Australian Astronomical Optics, Macquarie University, 105 Delhi Rd, North Ryde, NSW 2113, Australia\\
$^{6}$ Institut de F\'{\i}sica d'Altes Energies (IFAE), The Barcelona Institute of Science and Technology, 08193 Bellaterra (Barcelona), Spain \\
$^{7}$ National Centre for Nuclear Research, ul. Hoza 69, 00-681 Warsaw, Poland\\
$^{8}$ Centre for Astrophysics and Supercomputing, Swinburne University of Technology, John Street, Hawthorn, VIC 3122, Australia\\
$^{9}$ Ruhr-University Bochum, Astronomical Institute, German Center for Cosmological Lensing, Universitätsstraße 150, 44780, Bochum, Germany\\
}

\date{Accepted XXX. Received YYY; in original form ZZZ}

\pubyear{2020}

\begin{document}
\label{firstpage}
\pagerange{\pageref{firstpage}--\pageref{lastpage}}
\maketitle

\begin{abstract}
We present catalogues of stellar masses, star formation rates, and ancillary stellar population parameters for galaxies spanning $0<z<9$ from the Deep Extragalactic VIsible Legacy Survey (DEVILS).
 DEVILS is a deep spectroscopic redshift survey with very high completeness, covering several premier deep fields including COSMOS (D10).
 Our stellar mass and star formation rate estimates are self-consistently derived using the spectral energy distribution (SED) modelling code ProSpect, using well-motivated parameterisations for dust attenuation, star formation histories, and metallicity evolution.
 We show how these improvements, and especially our physically motivated assumptions about metallicity evolution, have an appreciable systematic effect on the inferred stellar masses, at the level of $\sim$\,0.2 dex.
 To illustrate the scientific value of these data, we map the evolving galaxy stellar mass function (SMF) and the SFR-$M_\star$ relation for $0<z<4.25$.
 In agreement with past studies, we find that most of the evolution in the SMF is driven by the characteristic density parameter, with little evolution in the characteristic mass and low-mass slopes.
 Where the SFR-$M_\star$ relation is indistinguishable from a power-law at $z>2.6$, we see evidence of a bend in the relation at low redshifts ($z<0.45$).
 This suggests evolution in both the normalisation {\em and shape} of the SFR-$M_\star$ relation since cosmic noon.
 It is significant that we only clearly see this bend when combining our new DEVILS measurements with consistently derived values for lower redshift galaxies from the Galaxy And Mass Assembly (GAMA) survey: this shows the power of having consistent treatment for galaxies at all redshifts.
\end{abstract}

\begin{keywords}
galaxies: general -- 
galaxies: evolution -- 
galaxies: star formation --  
galaxies: stellar content
\end{keywords}





\section{Introduction}\label{sec:Intro}

\begin{figure*}
    \centering
    \includegraphics[width = \linewidth]{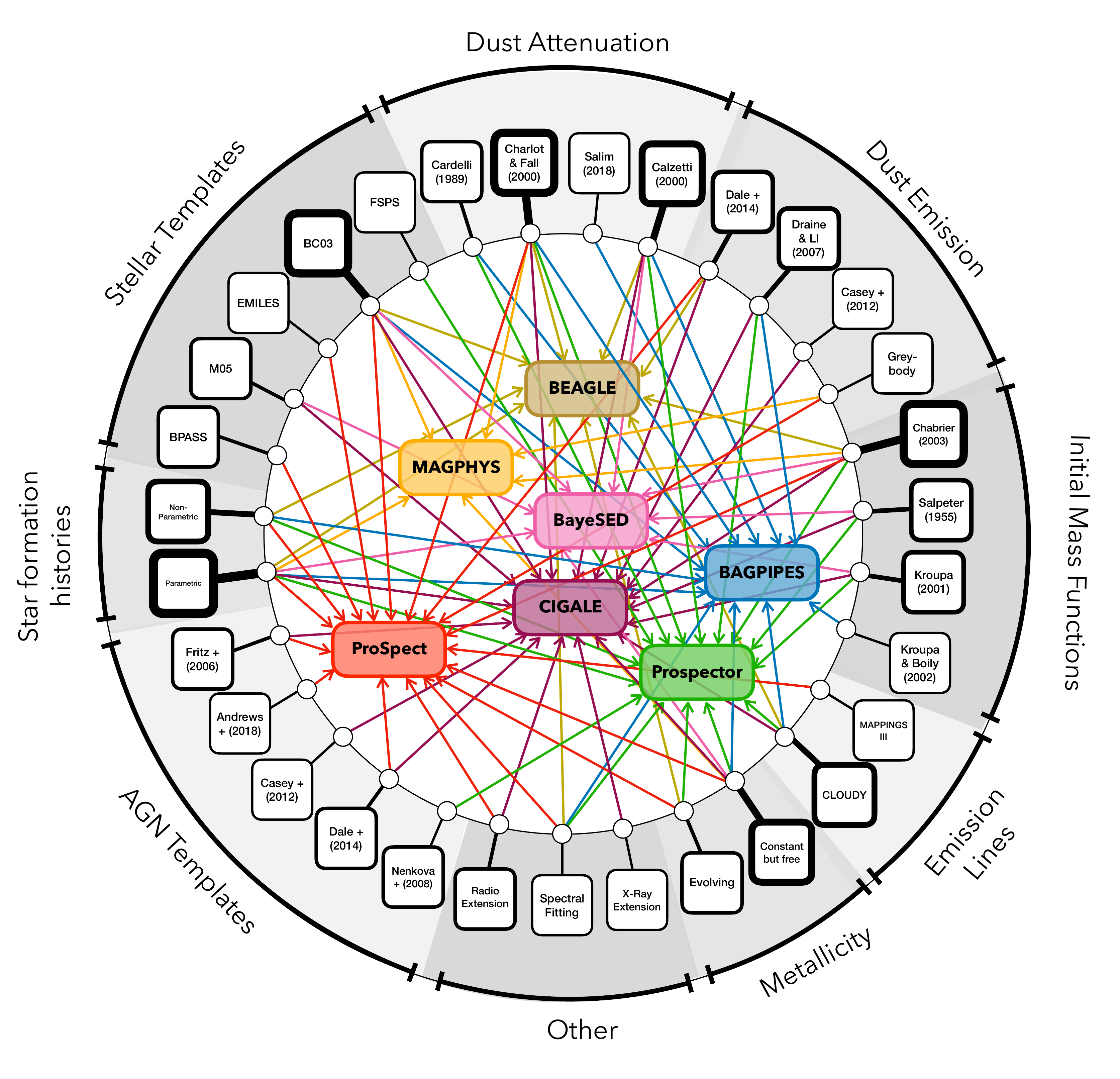}
    \caption{Schematic depicting some of the most popular FUV-FIR broadband SED fitting codes in the literature and the input models they employ, including stellar templates (models used to describe the emission from stars), the initial mass functions (used to describe the mass distribution of stars that form from a single birth cloud), dust attenuation and emission models, AGN models, and the parameterisations of the star formation and metallicity histories. 
    The thickness of the borders of each model correlates directly to the number of SED fitting codes that employ that particular model. 
    We present the tabular form in Table~\ref{tab:SED_Diagram}.
    The red represents \textsc{ProSpect} \citep{RobothamProSpectgeneratingspectral2020}, dark gold is \textsc{beagle} \citep{ChevallardModellinginterpretingspectral2016}, blue is \textsc{bagpipes} \citep{CarnallInferringstarformation2018}, purple is \textsc{cigale} \citep{CIGALE,BoquienCIGALEpythonCode2019,YangxcigalefittingAGN2020}, green is \textsc{Prospector} \citep{Prospector, JohnsonStellarPopulationInference2020}, orange represents \textsc{magphys} \citep{MAGPHYS}, and pink is \textsc{BayeSED} \citep{HanDecodingSpectralEnergy2012,HanBayeSEDGeneralApproach2014,HanComprehensiveBayesianDiscrimination2019}. 
    Additional stellar template references: BC03 refers to the templates from \citet{BC03}, EMILES are the templates from \citet{VazdekisUVextendedEMILESstellar2016}, M05 is \citet{MarastonEvolutionarypopulationsynthesis2005} and FSPS is \citet{ConroyPropagationUncertaintiesStellar2009}.
    Other references: The MAPPINGS-III photoionization tables are from \citet{LevesqueTheoreticalModelingStarForming2010} and CLOUDY is described in \citet{FerlandCLOUDY90Numerical1998,Ferland2013ReleaseCloudy2013}.}
    \label{fig:SED_Diagram}
\end{figure*}

Galaxies emit radiation over the full electromagnetic spectrum, from gamma-rays to radio, due to contributions from stellar populations, dust, active galactic nuclei (AGN) etc (see \citealt{WalcherFittingintegratedSpectral2011,ConroyModelingPanchromaticSpectral2013}). 
The different processes occurring within galaxies each dominate and contribute at different wavelengths, leaving their imprint on the galaxy spectrum. 
This distribution of energy emitted as a function of wavelength is called a spectral energy distribution (SED) and can be the primary source of information about properties of spatially unresolved galaxies (e.g. \citealt{MAGPHYS}).
Because different emission mechanisms can be dominant in different wavelength regimes, certain wavelengths have been used as proxies for various astrophysical quantities of interest.
For example, the near-ultraviolet (NUV) is dominated by light from short-lived OB stars and can be used to trace the star formation rate (SFR) of galaxies averaged over $\sim100$\,Myr timescales \citep{KennicuttStarFormationMilky2012}. 
This is made problematic by dust, which preferentially absorbs blue wavelengths, and re-radiates absorbed energy into the infrared (see \citealt{DraineInterstellarDustGrains2003}).
Because of this, the combination of UV and total infrared emission can also be used to trace star formation over longer temporal baselines ($\sim300$\,Myrs), but this measurement can be biased by dust heating from numerous low mass stars. 
There is considerable interplay between components that have a significant impact on the overall galaxy SED. 
By obtaining a simultaneous description of the whole picture, we hope to get an internally consistent and more accurate understanding of each of the constituent processes.
Over the past decade, considerable effort has been devoted to extracting information from galaxy SEDs by simultaneously modelling stellar, dust and AGN components, exploiting information from the far-ultraviolet (FUV) to the far-infrared (FIR) (see reviews by \citealt{WalcherFittingintegratedSpectral2011, ConroyModelingPanchromaticSpectral2013}).

The FUV-FIR SED of a galaxy can be broken down into its contributions from stellar emission, dust and other processes to extract information about the star formation history (SFH), stellar mass, dust mass, and dust properties (i.e. \citealt{MAGPHYS,TomczakGalaxyStellarMass2014,GAMA_MAGPHYS,BellstedtGalaxyMassAssembly2020b}).
Modelling the stellar component of a galaxy is done by combining various stellar templates to describe the age and metallicity of stellar populations (i.e. \citealt{BC03,MarastonEvolutionarypopulationsynthesis2005,ConroyPropagationUncertaintiesStellar2009,EldridgeSpectralpopulationsynthesis2009,EldridgePopulationspectralsynthesis2019}), initial mass functions (IMFs) to describe the mass distributions of stars as they form (i.e. \citealt{SalpeterLuminosityFunctionStellar1955,Kroupavariationinitialmass2001,Kroupamassfunctionstar2002,ChabrierGalacticStellarSubstellar2003}) and a parameterisation of the SFH of the galaxy. 
In order to properly model the FUV-FIR SED of galaxies, the contribution of dust must also be considered to correctly model both the attenuation of stellar emission in the FUV-optical but also the re-emission into the FIR. 
This is often done by assuming a model for dust attenuation (i.e. \citealt{Cardellirelationshipinfraredoptical1989,CalzettiDustContentOpacity2000,CharlotSimpleModelAbsorption2000,SalimDustAttenuationCurves2018}), for dust emission (i.e. \citealt{DraineDustMassesPAH2007,CaseyFarinfraredspectralenergy2012,DaleTwoParameterModelInfrared2014} or a grey-body spectrum) and assuming energy balance between the attenuated stellar light and the re-emission in the FIR (as per \textsc{magphys}; \citealt{MAGPHYS}).
While dust luminosities, and therefore masses, can be obtained through FUV-FIR SED fitting, these are often poorly constrained due to the lack of deep FIR imaging. 

Despite the effort invested over the past several decades there are still a number of outstanding problems in astronomy that limit the extraction of information from galaxy SEDs. These include the evolution, or otherwise, of the IMF \citep{Kroupavariationinitialmass2001}, the full and accurate mapping of stellar isochrones \citep{BertelliTheoreticalisochronesmodels1994,GirardiEvolutionarytracksisochrones2000}, 
the accurate production of stellar atmospheres over a suitably dense grid of temperatures and metallicities \citep{1992IAUS..149..225K,PicklesStellarSpectralFlux1998,LeBorgneSTELIBlibrarystellar2003,IvanovMUSElibrarystellar2019}, the proper treatment of stellar binary evolution \citep{EldridgeSpectralpopulationsynthesis2009}, and the treatment of dust for a broad range of galaxy types \citep{CharlotSimpleModelAbsorption2000,TrayfordFadegreysystematic2020}.
Despite these limitations, the past decade has seen the development of many codes designed to infer galaxy properties from broad-band galaxy SEDs, including \textsc{magphys} \citep{MAGPHYS}, \textsc{cigale} \citep{CIGALE,BoquienCIGALEpythonCode2019,YangxcigalefittingAGN2020}, \textsc{bagpipes} \citep{CarnallInferringstarformation2018}, \textsc{Prospector} \citep{Prospector,JohnsonStellarPopulationInference2020}, \textsc{Beagle} \citep{ChevallardModellinginterpretingspectral2016}, \textsc{BayeSED} \citep{HanDecodingSpectralEnergy2012,HanBayeSEDGeneralApproach2014,HanComprehensiveBayesianDiscrimination2019} and \textsc{ProSpect} \citep{RobothamProSpectgeneratingspectral2020}. Figure~\ref{fig:SED_Diagram} shows a schematic of some of the popular FUV-FIR broadband SED fitting codes and the input models and templates as described above. 
Figure~\ref{fig:SPS} shows a similar schematic describing the isochrones, atmospheres and IMFs that are combined to generate each of the stellar templates in Figure~\ref{fig:SED_Diagram}. 

In order to correctly model the amount of stellar mass with the right age distribution, it is necessary to model the SFH of each galaxy. 
Despite the significant progress in SED modelling over the last 10 years, there is still ongoing debate as how to best model the SFH of galaxies. 
This is largely split into two approaches - using a parametric or non-parametric SFH (see \citealt{LejaHowMeasureGalaxy2019,CarnallHowMeasureGalaxy2019}). 
This naming convention is misleading as both parametric and non-parametric SFHs are modelled by parameters. 
Parametric SFHs are those that assume a functional form, e.g. exponentially declining, delayed-exponentially declining, double power laws, exponentially increasing \citep{LeeBiasesUncertaintiesPhysical2009,LeeEstimationStarFormation2010, MarastonStarformationrates2010,PforrRecoveringgalaxystellar2012,SmethurstGalaxyZooevidence2015}. 
Non-parametric SFHs are those in which the star formation rate (SFR) as a function of time is described using piecewise constant functions (i.e., a step function; see \citealt{CidFernandesSemiempiricalanalysisSloan2005, Prospector, LejaHowMeasureGalaxy2019}).
Non-parametric SFHs allow for more flexibility in the shape of the SFH but typically require more free parameters and can more easily produce highly unphysical solutions.
Non-parametric SFHs are also limited by the difficulty of distinguishing the ages of old stars and require the use of much larger time bins for the earlier part of the SFH. 

While some comparisons have highlighted that parametric implementations are not as successful at recovering properties as non-parametric implementations \citep{CarnallHowMeasureGalaxy2019,LejaHowMeasureGalaxy2019,LowerHowWellCan2020}, such comparisons are highly dependent on the type of parametric SFH assumed.
A well selected parametric SFH can produce comparable results to a non-parametric SFH. 
\cite{RobothamProSpectgeneratingspectral2020} investigated the differences between parametric SFH forms within \textsc{ProSpect} in comparison to the \textsc{Shark} semi-analytic model of galaxy formation \citep{LagosSharkintroducingopen2018}. 
They found that it is not possible, using the implemented SFHs, to capture the fine details in the simulated star formation or metallicity history, but that the general smoothed form is readily recoverable with a well selected parametric SFH, as would be the case with non-parametric step functions. 

A second important ingredient in generating an SED is the assumed metallicity of the gas from which the stars form.
Measuring gas or stellar metallicity from photometric data alone is difficult due to the age-metallicity-dust degeneracy (see \citealt{WortheyComprehensivestellarpopulation1994, PapovichStellarPopulationsEvolution2001}): a galaxy can appear red either because it does not form stars anymore, because it has a high metallicity, or because it is strongly attenuated. 
Because of this difficulty, many SED fitting codes take the simple approach of fixing the assumed metallicity to the solar value or assuming a constant metallicity over the whole cosmic time, but allowing this constant metallicity value to be a modelled as a free parameter \citep{MAGPHYS,BoquienCIGALEpythonCode2019,LejaHowMeasureGalaxy2019}. 
However, it is known that the mean metallicity of galaxies increases with age as galaxies undergo chemical enrichment (\citealt{PeiCosmicChemicalEvolution1995,SomervilleSemianalyticmodellinggalaxy1999,NagamineStarFormationHistory2001}, see \citealt{Maiolinoremetallicacosmic2019} for a recent review). 
The metallicity, $Z$, affects the SED in two distinct ways. Firstly, as new stars form from the newly enriched interstellar medium, they begin their lives with slightly higher metal content, which results in lower effective temperatures, including a cooler main sequence and giant branch. 
Secondly, at fixed effective temperature ($T_\text{eff}$) an increase in metallicity results in strong spectroscopic absorption features and generally redder colours \citep{ConroyModelingPanchromaticSpectral2013}.
Both of these effects contribute to the overall reddening of an SED with increasing metallicity.
\cite{WortheyComprehensivestellarpopulation1994} studied the degeneracy between metallicity and age and introduced his ``3/2 rule'', whereby an increase/decrease in the population age by a factor of three is almost perfectly degenerate with an increase/decrease in metallicity by a factor of two. 

The danger in assuming that the metal content in galaxies is constant is that this assumption will directly affect other parameters of interest that are derived from the SFH, such as the stellar mass and SFR. 
Recent work by \cite{BellstedtGalaxyMassAssembly2020b} showed that poor implementations of metallicity in SED fitting can have a large impact on the shape of the derived cosmic star formation history (CSFH) as predicted by \cite{WortheyComprehensivestellarpopulation1994}, but that making simple assumptions about chemical enrichment in galaxies can provide much more reasonable solutions.

In this work, we apply \textsc{ProSpect} \citep{RobothamProSpectgeneratingspectral2020} in a parametric mode to multiwavelength photometry from the Deep Extragalactic VIsible Legacy Survey (DEVILS, \citealt{DaviesDeepExtragalacticVIsible2018}) in order to measure stellar and dust masses, and SFRs for galaxies in the D10-COSMOS early science field. 
The stellar and dust mass estimates and SFRs derived in this work form a crucial part of the DEVILS value-added data-set and thus a primary goal of this paper is to provide a foundational understanding for users of these catalogues. 
We also use these new measurements to provide our best effort stellar mass functions and SFR-$M_\star$ relations in the redshift range $0 < z < 9$.

The structure of this paper is as follows. 
After describing the DEVILS survey and related data sets in Section~\ref{sec:Data}, we describe the SED fitting method utilised in this work  and compare our stellar masses and SFRs to previous measurements in Section~\ref{sec:ProSpect}. 
Using these new values, we derive the stellar mass function and its evolution in Section~\ref{sec:SMF} and the galaxy star formation main sequence in Section~\ref{sec:SFMS}. 
We summarise our results in Section~\ref{sec:conclusion} and provide a description of the data availability in Section~\ref{sec:DataAvailability}.
Throughout this work we use a \cite{ChabrierGalacticStellarSubstellar2003} IMF and all magnitudes are quoted in the AB system. 
We adopt the \cite{PlanckCollaborationPlanck2015results2016} cosmology with $H_0 = 67.8 \, \si{\kilo \meter \per \second \per \mega \parsec}$, $\Omega_{M} = 0.308$ and $\Omega_\Lambda = 0.692$. 

\section{Data} \label{sec:Data}
This paper presents the first catalogue of stellar mass, dust mass, SFR, and metallicity estimates for the D10 field of DEVILS \citep{DaviesDeepExtragalacticVIsible2018}. 
DEVILS is an on-going optical spectroscopic redshift survey using the Anglo-Australian Telescope specifically designed to have high spectroscopic completeness over a large redshift range ($z < 1$) in three well-studied extragalactic fields: COSMOS/D10 (1.5 deg$^2$),  ECDFS/D02 (3 deg$^2$), and XMM-LSS/D03 (1.5 deg$^2$) covering a total of 6 deg$^2$. 
DEVILS will build a sample of $\sim 60,000$ galaxies down to $Y_{\text{mag}} < 21.2$ to a high completeness ($> 85\%$), allowing for robust parameterization of group and pair environments in the distant universe. For a full description of the survey science goals, survey design, target selection and spectroscopic observations see \cite{DaviesDeepExtragalacticVIsible2018}. In this paper we focus on the D10-COSMOS region which is prioritised for early science. At the time of this analysis, 3,394 redshifts have been collected in D10 from DEVILS observations, which, when combined with other surveys results in a total of 13,787 spectroscopic redshifts in D10.

\subsection{Photometry Catalogue}

We use the new DEVILS D10 FUV-FIR photometry derived using \textsc{ProFound} \citep{RobothamProFoundSourceExtraction2018}, and described in detail in Davies et al. (submitted).
However, briefly, this photometry uses the newest imaging data sets (including Subaru-HSC and UltraVISTA-DR4) and includes measurements in the \textit{FUV NUV ugrizYJHK$_{s}$ IRAC1 IRAC2 IRAC3 IRAC4 MIPS24 MIPS70 P100 P160 S250 S350 S500} bands. 
Table~\ref{tab:photometry} lists the bands used in this work, and the corresponding facility, survey, central wavelength, and nominal depth of each band. 
The imaging in every band covers the entire 1.5 deg$^2$ of the D10 field, except for UltraVISTA Ks and MIPS24 imaging which have a small region of the D10 field missing. 
Narrow- and intermediate-band filters are not included in the DEVILS photometry catalogue to ensure consistency over the three fields, as D02 and D03 do not have imaging in these filters.

The new photometry catalogue was derived using a similar process to that employed to create the DEVILS input catalogues as outlined in \cite{DaviesDeepExtragalacticVIsible2018}, and to
derive new photometry for GAMA as described in \citet{BellstedtGalaxyMassAssembly2020a}.
For the new photometric catalogue, \textsc{ProFound} is applied on an RMS weighted stack of the $YJH$ bands for initial source detection, and definition of the initial segmentation map. 
While the \textsc{ProFound} derived segmentation map successfully identifies individual sources, it can fragment bright galaxies into multiple segments (a known problem with most automated source finding algorithms, e.g. Source Extractor). 
Bright fragmented galaxies were regrouped manually using the \texttt{profoundSegimFix} function within \textsc{ProFound}. 
In addition to segment regrouping, some highly clustered sources were merged into a single \textsc{ProFound} segment and thus required \textit{ungrouping}.
This ungrouping was also performed manually using \texttt{profoundSegimFix} where new segments were drawn onto an image. 
The new grouped and ungrouped segments were folded into the segmentation map to define the source locations which were used for the remainder of the photometry pipeline. 

After the initial segmentation map was defined and manual fixing applied, these segments were used to measure photometry in the UV-MIR bands (GALEX-FUV to \textit{Spitzer} IRAC 4). 
This initial process was applied only to the UV-MIR where the pixel-scale and seeing are comparable and source blending/ confusion between bands is low.
Extinction corrections were applied using the Planck E(B-V) map, and object classification was performed using the new photometry to derive star, artefact and mask flags. 
Photometry for the FIR bands was measured using the \texttt{FitMagPSF} mode in \textsc{ProFound} on a selection of optically-detected objects and additional objects detected in \textit{MIPS24} in order to obtain fluxes in the \textit{MIPS24 - S500} bands which are semi- to unresolved.
IRAC imaging can have significant source blending which often requires deconfusion techniques. 
Photometry for the IRAC 3 and 4 channels was measured using both the default segment mode and the PSF mode in \textsc{ProFound}. 
It was found that the default segment mode produced better (tighter) colours and agreed more closely with previous measurements from \cite{LaigleCOSMOS2015CATALOGEXPLORING2016} and \cite{AndrewsG10COSMOS382017}.
This was deemed to be the preferred solution and was used for the final photometry catalogue.

We find that the new photometry is consistent in colour-analysis to previous approaches (i.e. \citealt{LaigleCOSMOS2015CATALOGEXPLORING2016}) using fixed-size apertures (which are specifically tuned to derive colours), but produces superior total source photometry, which is essential for the derivation of stellar masses, SFRs and SFHs, as done in this work. 
The photometry catalogue will be released as part of the DEVILS data release 1 (DR1) in the DEVILS\_PhotomCat data management unit (DMU).

\begin{table*}
    \centering
    \caption{Overview of multi-wavelength data used for the DEVILS D10 photometry as described in detail in Davies et al. (submitted).}
    \label{tab:photometry}
    \begin{tabular}{l l l l l l}
    \hline
        Facility	&	Survey	&	Band	&	Central Wavelength ($\mu$m)	&	Nominal Depth ($5\sigma$ AB)	&	Reference 	\\
        \hline
GALEX	&	GALEX-DIS	&	FUV	&	0.154	&	26	&	\citet{ZamojskiDeepGALEXImaging2007}	\\
	&		&	NUV	&	0.231	&	35.6	&		\\
CFHT	&	CFHT-COSMOS	&	u	&	0.379	&	>25.4	&	\citet{CapakFirstReleaseCOSMOS2007}	\\
Subaru	&	HSC-SSP (DUD)	&	g	&	0.474	&	27.3	&	\citet{AiharaSeconddatarelease2019}	\\
	&		&	r	&	0.622	&	26.9	&		\\
	&		&	i	&	0.776	&	26.7	&		\\
	&		&	z	&	0.893	&	26.3	&		\\
VISTA	&	UltraVISTA	&	Y	&	1.02	&	>24.7	&	\citet{McCrackenUltraVISTAnewultradeep2012}	\\
	&		&	J	&	1.26	&	>24.5	&		\\
	&		&	H	&	1.65	&	>24.1	&		\\
	&		&	Ks	&	2.16	&	>24.5	&		\\
Spitzer	&	SPLASH	&	S36/IRAC 1	&	3.53	&	24.9	&	\citet{LaigleCOSMOS2015CATALOGEXPLORING2016}	\\
	&		&	S45/IRAC 2	&	4.47	&	24.9	&		\\
	&	S-COSMOS	&	S58/IRAC 3	&	5.68	&	22.4	&	\citet{SandersSCOSMOSSpitzerLegacy2007}	\\
	&		&	S80/IRAC 4	&	7.75	&	22.3	&		\\
	&		&	MIPS24	&	23.5	&	19.3	&		\\
	&		&	MIPS70	&	70.4	&	14.2	&		\\
\textit{Herschel}	&	PEP	&	P100	&	98.9	&	14.1	&	\citet{LutzPACSEvolutionaryProbe2011}	\\
	&	PEP	&	P160	&	156	&	13.3	&		\\
	&	HerMES	&	S250	&	250	&	14.1	&	\citet{OliverHerschelMultitieredExtragalactic2012}	\\
	&	HerMES	&	S350	&	350	&	14.4	&		\\
	&	HerMES	&	S500	&	504	&	13.9	&		\\
	\hline
    \end{tabular}
\end{table*}

\subsection{Redshift Measurements}
DEVILS is currently $50\% $ complete and is due to finish spectroscopic observations in 2021. 
In our spectroscopic observing program, we prioritised the D10 field to obtain full spectroscopic completeness ($>85\%$) prior to completing the other fields. 
In this work we combine the current DEVILS redshifts with other spectroscopic, grism and photometric redshifts to recover stellar masses and SFRs for as many objects in the D10 field as possible. The list of redshift sources and references is included in Table \ref{tab:redshifts}.
We combine two approaches to match redshift measurements to sources in our photometry catalogue. 
Spectroscopic redshifts are matched using the segmentation maps from \textsc{ProFound} and are allocated if a redshift lies within an object's segment. 
If two sources are in the same segment the source with the largest pixel flux divided by distance-to-segment centre is taken. 
Photometric and grism catalogues are matched using the RA and Dec from the redshift source catalogue and the RAcen and Deccen (the flux weighted centre of each segment) from \textsc{ProFound} using a 2 arcsec nearest neighbour match using \texttt{coordmatch} (from \textsc{Celestial}, \citealt{RobothamCelestialCommonastronomical2016}).

In many cases there are multiple redshift measurements for a single segment from multiple spectroscopic redshifts from different programs, or numerous photometric redshifts. In order to select \texttt{zBest}, we rank the redshift sources as per Table \ref{tab:redshifts} and adopt the highest ranked redshift, except in the case where we only have to choose between a grism or photometric redshift. In those cases we adopt the grism redshift if:
\begin{equation}
    \frac{ |z_\text{photo} - z_\text{grism}|}{1+z_\text{photo}} \le 0.05,
\end{equation}
otherwise we adopt the photometric redshift. 
This is due to the fact that there is a large catastrophic failure rate in the grism redshifts but we want to capitalise on the precision of the grism measurements when they are in agreement with the photometric redshift.  

Although DEVILS related science will be predominantly focused on $z<1$ we include all redshifts where available to ensure we have comprehensive stellar mass and star formation rate estimates for the field.
We use include photometric redshifts from the existing catalogues regardless of $\chi^2$ value for completeness.
This compilation of redshifts will be made publicly available as part of DEVILS DR1 in the DEVILS\_D10MasterRedshiftCat DMU. 

We select sources to derive stellar properties if they are not classed as stars (\textit{starflag} column of DEVILS\_PhotomCat), artefacts (\textit{artefactflag}) and are not masked (\textit{mask}).
We also exclude any sources that have negative or zero redshifts. 
This results in 494,084 galaxies of which 24,099 have spectroscopic redshifts, 7,307 have grism redshifts, and the remaining 462,678 have photometric redshifts (see Appendix \ref{app:redshifts}).

\section{SED Modelling with \textsc{ProSpect}} \label{sec:ProSpect}

\begin{table*}
    \centering
        \caption{The units,  fitting regime (whether fitting in logarithmic or linear space), ranges, priors, and references for the free parameters used in this work. If there is no explicitly stated prior then we are assuming a uniform prior over the allowed parameter ranges. }
        \label{tab:ProSpectParameters}
    \begin{tabular}{l l l l l c}
        \hline
         Parameter & Units & Type & Range & Prior & Reference \\
         \hline
         \texttt{mSFR} & $M_\odot \text{yr}^{-1}$ & Log & [-3,4] & & Section \ref{sec:SFHs} \\ 
         \texttt{mpeak} & Gyr & Linear &[-2, 13.38] &  &  Section \ref{sec:SFHs} \\
          \texttt{mperiod} & Gyr & Log & [$\log_{10}(0.3)$, 2]& $100 \erf (\texttt{mperiod}+2) - 100$&  Section \ref{sec:SFHs} \\
         \texttt{mskew} & - & Linear & [-0.5,1] & & Section \ref{sec:SFHs} \\
          \texttt{Zfinal} & & Log & [-4, -1.3] & &  Section \ref{sec:ZHs} \\
          \texttt{tau\_birth} & & Log& [-2.5, 1.5 ]& $\exp{(-\frac{1}{2} (\frac{\tau_\text{birth} - 0.2}{0.5})^2 ) }$ &  Section \ref{sec:DustPriors}\\
         \texttt{tau\_screen} & & Log & [-5, 1] &  $-20\erf(\tau_\text{screen}-2) $ &  Section \ref{sec:DustPriors} \\
         \texttt{alpha\_birth} & & Linear & [0,4] &$\exp{(-\frac{1}{2} (\frac{\alpha_\text{birth} + 2}{1})^2 ) }$ & Section \ref{sec:DustPriors}\\
         \texttt{alpha\_screen} & & Linear & [0,4] & $\exp{(-\frac{1}{2} (\frac{\alpha_\text{screen} + 2}{1})^2 ) }$&  Section \ref{sec:DustPriors} \\
         \hline
    \end{tabular}
\end{table*}

To extract stellar mass and SFR estimates for our galaxies we use \textsc{ProSpect} \citep{RobothamProSpectgeneratingspectral2020}, a new state-of-the-art SED modelling program. 
\textsc{ProSpect} uses the \citealt{BC03} (BC03) stellar libraries and the \cite{ChabrierGalacticStellarSubstellar2003} IMF to model the stellar components. 
To model dust attenuation in galaxies, \textsc{ProSpect} uses the \cite{CharlotSimpleModelAbsorption2000} model which consists of a two component description of the interstellar medium, a diffuse dust component that attenuates emission for all stars, and a birth cloud which just attenuates emission from stars less than 10 Myr old. 
\textsc{ProSpect} utilises the \cite{DaleTwoParameterModelInfrared2014} templates to model the re-radiation of photons absorbed by dust into the infrared. 
Whilst \textsc{ProSpect} can model an AGN component through a number of templates \citep{FritzRevisitinginfraredspectra2006,DaleTwoParameterModelInfrared2014,AndrewsModellingcosmicspectral2018} we do not fit for the presence of AGN in this work. 
\cite{BellstedtGalaxyMassAssembly2020b} note that powerful AGN are expected to dominate the mid-IR portion of the SED where photometric uncertainties and modelling floors provide little constraining power to the fit. 
As such, AGN emission will result in larger mid-IR residuals without having a large impact on the derived stellar properties of the galaxy.

Many other SED modelling codes use the same underlying templates (e.g. \textsc{magphys}; \citealt{MAGPHYS}, etc.) but the benefit of \textsc{ProSpect} lies in the fact that \textsc{ProSpect} is extremely flexible in how it can process star formation histories and because it incorporates evolving metallicities. We describe the key assumptions and models used by \textsc{ProSpect} in the next sections (Sections~\ref{sec:SFHs}~-~\ref{sec:DustPriors}) and present all the free parameters, the allowed ranges and imposed priors in Table~\ref{tab:ProSpectParameters}. 


\subsection{Star Formation Histories} \label{sec:SFHs}
In order to obtain estimates for the stellar mass and SFR of galaxies, a parameterisation of the SFH needs to be adopted. 
Following the analysis by \cite{RobothamProSpectgeneratingspectral2020} and the implementation by \cite{BellstedtGalaxyMassAssembly2020b}, we use the \texttt{massfunc\_snorm\_trunc} function to model star formation histories with \textsc{ProSpect}. 
This models star formation histories using a skewed-Normal distribution with a truncation at the beginning of the Universe to force galaxies to have a SFR = 0 at the beginning of the Universe. 
As shown in figure 10 of \cite{RobothamProSpectgeneratingspectral2020}, the \texttt{massfunc\_snorm\_trunc} parameterisation can reproduce a diverse range of SFHs without the bias seen in SFHs when an exponentially declining SFH is adopted (which is popular in the literature). 
The \texttt{massfunc\_snorm\_trunc} parameterisation was deemed to be the best option, based on the fact that it can appropriately model the smoothed form of a diverse range of simulated SFHs (see \citealt{RobothamProSpectgeneratingspectral2020}) and that the inferred average SFHs across a large population are consistent with measurements of the CSFH (see \citealt{BellstedtGalaxyMassAssembly2020b}).

The parameterisation of the \texttt{massfunc\_snorm\_trunc} is explained thoroughly in equations 1-5 (section 3.1.1) of \cite{BellstedtGalaxyMassAssembly2020b} and a variety of possible SFHs are shown in figure 10 of \cite{RobothamProSpectgeneratingspectral2020}.
Briefly, the \texttt{massfunc\_snorm\_trunc} parameterisation is a skewed Normal distribution modelled by four free parameters:
\begin{itemize}
    \item \texttt{mSFR} - the peak SFR of the SFH,
    \item \texttt{mpeak} - the age of the SFH peak,
    \item \texttt{mperiod} - the width of the Normal distribution,
    \item \texttt{mskew} - the skewness of the Normal distribution.
\end{itemize}

This parameterisation achieves a smooth truncation between the peak of the SFH and the beginning of the Universe. For this work, we use a fixed value of \texttt{mtrunc} = 2\,Gyr and \texttt{magemax} = 13.38\,Gyr. 
The \texttt{magemax} parameter has been selected to fix the start of star formation to the epoch at which the highest-$z$ galaxies are known to exist (z = 11, \citealt{OeschREMARKABLYLUMINOUSGALAXY2016}), corresponding to a lookback time of 13.38\,Gyr. 
We allow the \texttt{mpeak} parameter to take negative values, allowing the SFH to peak up to 2\,Gyr after the observation point of our galaxies which allows for rising SFRs at the time of observation. 
Allowing the star formation to peak at negative values introduces more degeneracies in the parameter space. 
The lower limit of the \texttt{mperiod} parameter was selected due to the sampling of the BC03 templates.

The \texttt{massfunc\_snorm\_trunc} parameterisation is inherently unimodal, and will achieve the best results for galaxies that have experienced a single epoch of star formation. 
For galaxies that may experience multiple distinct periods of star formation, this parameterisation will not be entirely accurate but as described above, \cite{RobothamProSpectgeneratingspectral2020} found that this parameterisation of the SFH is able to recover the SFH of a population of simulated galaxies accurately (see figures 28, 29 and C1 of \citealt{RobothamProSpectgeneratingspectral2020}). 
While this will be reflected in the uncertainties derived for individual galaxies, we do not expect this assumption to have a significant impact at a statistical level for a large population study.

\subsection{Modelling Metallicity} \label{sec:ZHs}

\begin{figure}
    \centering
    \includegraphics[width = \linewidth]{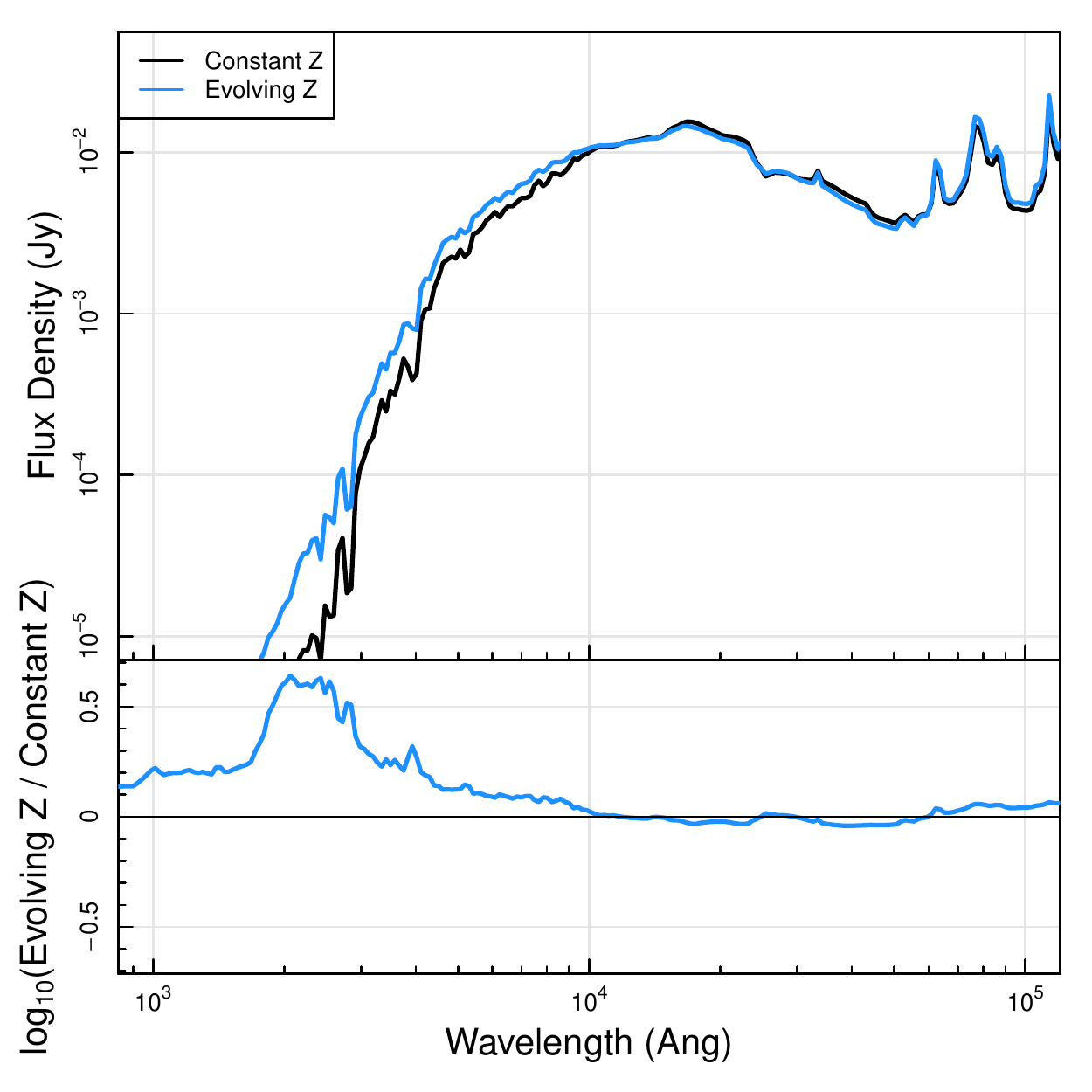}
    \caption{Variation in the NUV - NIR portion of a galaxy SED when changing from assumed constant (black, $Z = 0.02$) metallicity to a linearly evolving metallicity with final metallicity $Z = 0.02$ (blue). 
    The SEDs are generated for a galaxy with an old stellar population and a formed stellar mass of $10^{10} M_\odot$ and star formation that peaked 8 Gyrs ago. We also show the ratio of the two SEDs in the lower panel.}
    \label{fig:metallicityinfluence}
\end{figure}

As described in Section~\ref{sec:Intro} the SEDs of galaxies have typically been fit assuming constant metallicity. 
In Figure \ref{fig:metallicityinfluence} we show the effect of changing from a constant metallicity to an evolving metallicity with the same final metallicity on the shape of the NUV-NIR SED. 
The SED shown in Figure~\ref{fig:metallicityinfluence} was generated for a galaxy with a very old stellar population, which will highlight the largest differences between SEDs with evolving and constant metallicity histories. 
Depending on the SFH of the galaxy, the SED produced by assuming a constant or evolving metallicity can vary by a factor of a few in the UV-optical especially for very old, red galaxies.
This difference is measurable in broad-band photometry and will have a direct impact on the extracted SFH and resulting stellar mass and SFR. 
The effect of different metallicity assumptions on the overall CSFH is explored by \cite{BellstedtGalaxyMassAssembly2020b} using GAMA data. 
\cite{BellstedtGalaxyMassAssembly2020b} show that assuming constant metallicity can be catastrophic on the overall shape of the CSFH, but that making simple, yet well informed, assumptions about the evolution of metals within galaxies can provide much more reasonable solutions when compared with the empirical data at high redshift.
\cite{BellstedtGalaxymassassembly2021} also showed that the resulting metallicity measurements from \textsc{ProSpect} produced a mass-metallicity relation consistent with previous measurements.
It is for these reasons that we deliberately do not assume a constant metallicity and implement an evolving metallicity in which metal enrichment follows 1:1 the stellar mass build-up, so when e.g. half of a galaxy's stellar mass has been assembled half of its chemical enrichment will also have occurred. 
This is similar to the closed-box model of metallicity growth, but the linear model allows for a reasonable amount of inflow which is nearer to the actual Universe and is not modelled when assuming a closed-box. 
Analysis using the semi-analytic model \textsc{Shark} \citep{LagosSharkintroducingopen2018} suggests this is a reasonable approximation to make in practice  (see \citealt{RobothamProSpectgeneratingspectral2020}). 
This mapping of metallicity to mass build up naturally introduces low initial metallicity for the earliest phases of star formation, as expected, and higher metallicity for the later phases of star formation. 
Unless there is extreme gas inflow of extremely metal poor gas, it is hard to drastically break this type of metal evolution for realistic galaxy formation \citep{NomotoNucleosynthesisStarsChemical2013}. 
This model of metallicity evolution is implemented in \textsc{ProSpect} through the \texttt{Zfunc\_massmap\_lin} function which linearly maps the stellar mass build-up onto the metal build-up. 
We use a fixed initial metallicity value of $0.0001$ (as this is the lowest metallicity template in BC03), and allow the final metallicity to be a free parameter, which we fit for within the range of metallicities in the BC03 templates ($0.0001$ to $0.05$). 
A variety of metallicity histories generated using \texttt{Zfunc\_massmap\_lin} are shown in figure 12 of \cite{RobothamProSpectgeneratingspectral2020} but differ slightly from our implementation due to the varying initial metallicty which we do not fit for.

\subsection{Modelling Dust} \label{sec:DustPriors} 
\begin{figure}
    \centering
    \includegraphics[width = \linewidth]{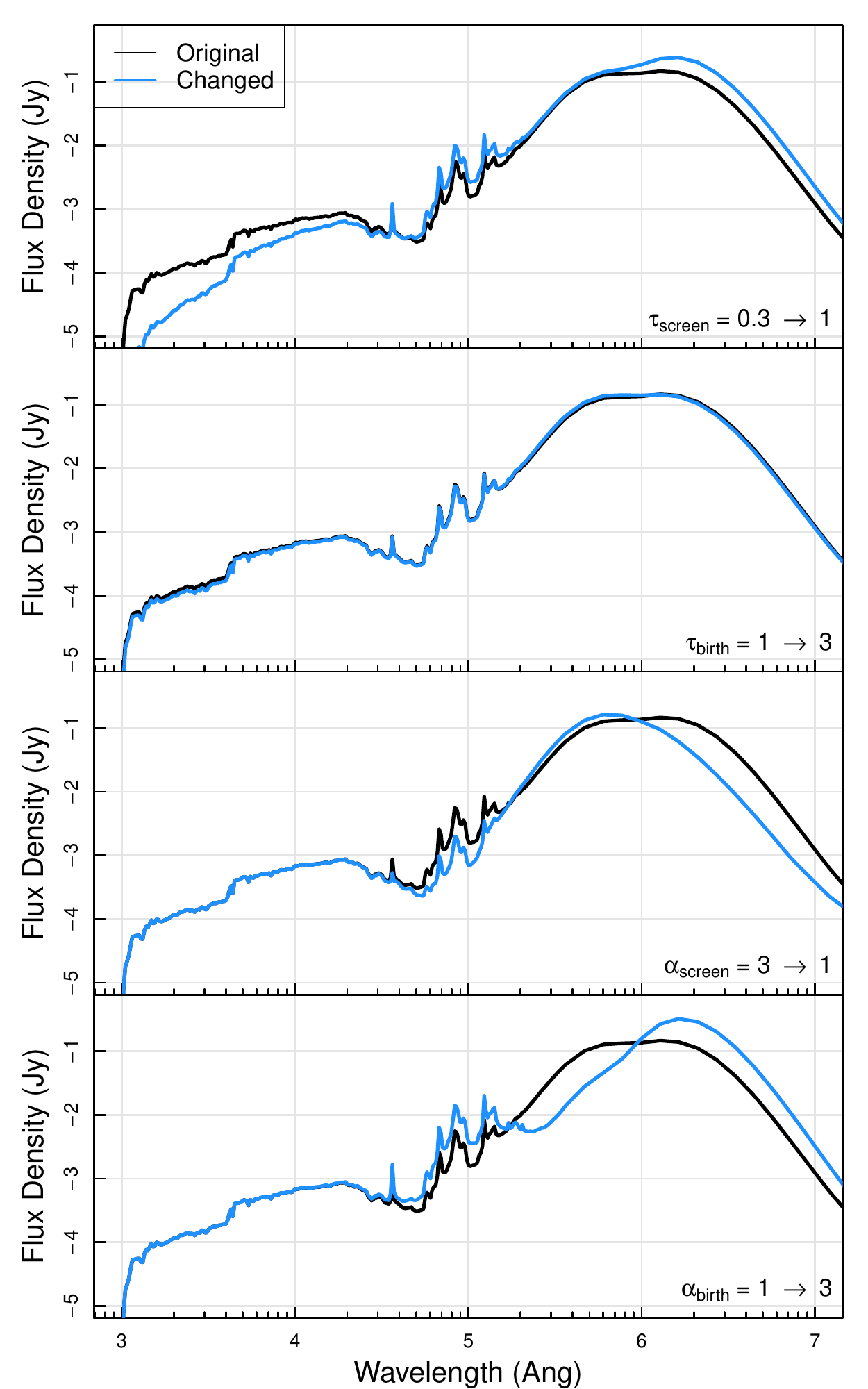}
    \caption{Here we show the effect of changing the four dust parameters ($\tau_{\text{screen}}$, $\tau_{\text{birth}}$, $\alpha_\text{screen}$, $\alpha_\text{birth}$ in the four panels from top to bottom respectively). The black represents the original model galaxy with a stellar mass formed (not remaining) of $10^{10} M_\odot$ with very recent star formation and typical dust parameters. The blue line shows the effect of changing each of the dust parameters in turn with the change in value displayed the bottom right corner of each panel. Essentially, the $\tau$ parameters change the amount of dust (high $\tau$ values represent more higher dust column densities) and the $\alpha$ values represent the temperature of the dust.}
    \label{fig:dustparameterinfluence}
\end{figure}

To model dust in \textsc{ProSpect} we use the \cite{CharlotSimpleModelAbsorption2000} two phase model for attenuation where the flux observed at a given
wavelength ($\lambda$) is modified by the attenuation factor $A$:
\begin{equation}
    A(\lambda) = e^{-\tau (\lambda/\lambda_0)^\nu}
\end{equation}
where $\lambda_0$ is the pivot wavelength ($5500$\,{\AA} by default), $\tau$ is the effective optical depth of attenuation, and $\nu$ is the modifying power which we set to 0.7.  
The $\tau$ parameter effectively changes the column density of dust, with larger $\tau$ values representing larger dust column densities. 
In practice the $\nu$ and $\tau$ parameters are quite degenerate when fitting to observational data.
We use the \cite{DaleTwoParameterModelInfrared2014} templates for the re-emission of energy in the far-infrared using an energy balance approach. These templates have a free parameter $\alpha$ that specifies the power law of the radiation field heating the dust, where lower values of $\alpha$ correspond to hotter dust. 

To model the effect of dust in our galaxies we include four free parameters, two that control the reddening of the dust ($\tau_{\text{birth}}$  and $\tau_{\text{screen}}$) and two that model the radiation field heating the dust ($\alpha_{\text{birth}}$  and $\alpha_{\text{screen}}$). 
The `birth' parameters affect the dust in `birth clouds' and only affect emission from stars that are less than 10\,Myr old, whilst the `screen' parameters represent the dust in the diffuse interstellar medium and affect emission from all stars. 
Figure~\ref{fig:dustparameterinfluence} shows the differences made to the overall galaxy SED by changing the values of each of the dust parameters given a mock galaxy with stellar mass of $10^{10} M_\odot$ and a SFH that is peaking at observation.
As shown in the upper two panels of Figure \ref{fig:dustparameterinfluence}, the value of each of the $\tau$ parameters influences the amount of UV-NIR light attenuated and re-emitted into the FIR. 
$\tau_\text{screen}$ has a larger impact on the overall shape of the distribution as it affects the emission from all stars in the galaxy, whereas, in this case, $\tau_\text{birth}$ has a much smaller effect due to the smaller number of stars it impacts. 
The $\alpha$ parameters essentially represent the temperature distribution of the dust, with smaller values of alpha representing higher temperatures, and affect the location of the FIR peak. 
The effect of each of the parameters on the overall SED can be explored in more detail using the interactive \textsc{ProSpect} tool\footnote{\url{prospect.icrar.org}}.

To guide our fits towards physical dust parameters, and to assist in the convergence of our stellar parameters, we impose priors on our four dust parameters. 
In setting the priors we assume that the column density of dust in the ISM is lower than in birth clouds. We present all the free parameters, the allowed ranges and imposed priors in Table~\ref{tab:ProSpectParameters}. 

\subsection{MCMC Set-up}
We implement \textsc{ProSpect} in a Bayesian manner using MCMC fitting in order to ensure we have well-fitted SEDs with the ability to extract realistic uncertainties instead of using a simple $\chi ^2$ minimisation. 
Simple $\chi ^2$ fitting routines cannot properly estimate the uncertainties when there are substantial degeneracies, as is the case when fitting broadband photometry of galaxies. 

We fit our galaxies in a two-stage process using the \textsc{Highlander} R package\footnote{\url{https://github.com/asgr/highlander}}.
\textsc{Highlander} alternates between genetic optimisation using the \texttt{cmaeshpc}\footnote{\url{https://github.com/asgr/cmaeshpc}} package and an MCMC chain using the \texttt{LaplacesDemon}\footnote{\url{https://cran.r-project.org/web/packages/LaplacesDemon/index.html}} package.
By alternating between the two different phases, \textsc{Highlander} is able to more efficiently sample the posterior parameter space, especially in scenarios that are highly multi-modal such as SED fitting, whilst still retaining the ability to extract uncertainties for each of the galaxy properties. 
Within \texttt{LaplacesDemon} we utilise the CHARM\footnote{Component-wise hit and run metropolis} algorithm using a student-t likelihood. 
We fit with 500 steps for the genetic optimisation and 100 steps of MCMC and then repeat this process for a second time but instead fit for 200 steps in the final MCMC phase. 
In this configuration, \textsc{ProSpect} takes approximately 2 minutes to fit each galaxy using a modern CPU. 

Table \ref{tab:ProSpectParameters} presents each of our free parameters, whether it is fit in logarithmic or linear space, the allowed values, the imposed priors and the related section. 
To determine the initial values for each galaxy we take the midpoint of allowed values for each parameter.

To account for offsets between facilities and instruments, and to allow for small fluctuations in the zero-point offset, we adopt a 10\% error floor in all bands. 
\cite{RobothamProSpectgeneratingspectral2020} showed that there is little difference within \textsc{ProSpect} in the convergence of fits when moving from introduced errors of 0.01 mag to 0.1 mag errors, but there is noticeable difference when moving to 0.5 mag errors. 
We also remove bands that fall within the polycyclic aromatic hydrocarbon (PAH) dust features between rest-frame 5-15\,$\mu$m as we find large residuals at these wavelengths when compared to the \cite{DaleTwoParameterModelInfrared2014} templates. 
This results in fitting without \textit{IRAC3} and \textit{IRAC4} at the lowest redshifts ($z < 0.6$), without \textit{MIPS24} for $0.6 < z < 3$ and \textit{MIPS70} for our highest redshift objects ($z > 3$).
Bands within the PAH features were also dropped for the \textsc{ProSpect} analysis of GAMA by \cite{BellstedtGalaxyMassAssembly2020b}.

\begin{figure*}
    \centering
    \includegraphics[width = \linewidth]{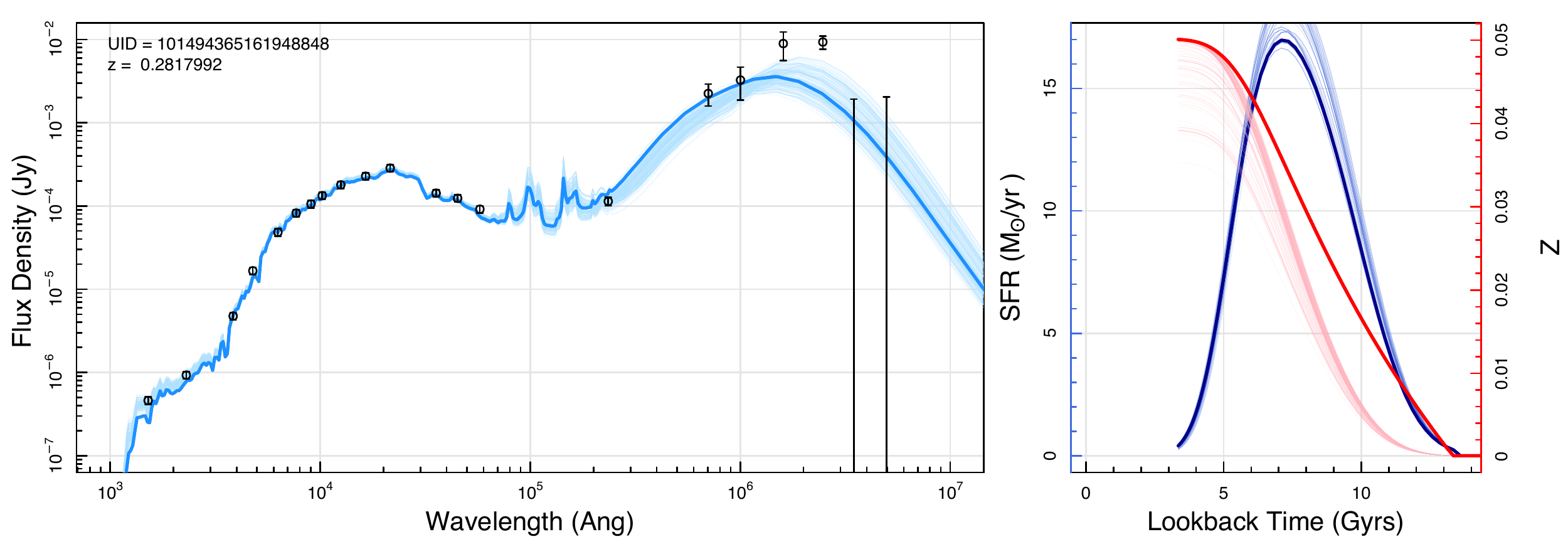}
    \includegraphics[width = \linewidth]{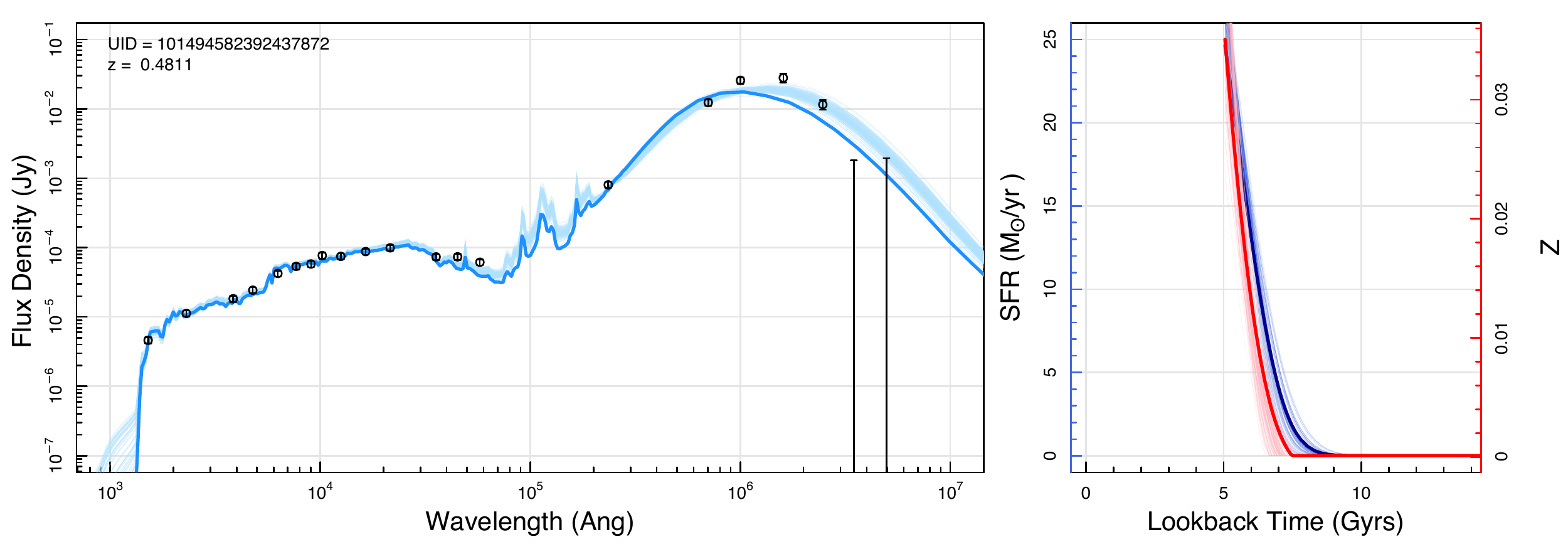}
    \caption{Here we present two example outputs from ProSpect for galaxies 101494365161948848 and 101494582392437872. The left panel for both galaxies shows the input photometry (black circles and error bars), the best fit SED in blue and the SED generated from each step of the final MCMC chain in pale blue. The right panel shows the best fit SFH in blue and the posterior sampling in pale blue with the scale given by the left axis. The metallicity history for each galaxy is also presented in the right panel as the red line for the best fit solution and as the pink lines for the sampling of the posterior. The scale for the metallicity history is shown on the right axis.}
    \label{fig:ProSpect_Examples}
\end{figure*}

\subsection{ProSpect Outputs}

Figure \ref{fig:ProSpect_Examples} shows examples of the \textsc{ProSpect} outputs for two galaxies, 101494365161948848 and 101494582392437872 (respectively an early- and late-type), both of which have spectroscopic redshift measurements. 
These galaxies were selected to show an example of a currently star forming galaxy, and a galaxy that peaked in star formation early and has subsequently quenched.
We show the input photometry and resulting SED in the left panel and the extracted star formation and metallicity histories in the right panel for each galaxy. 
The early-type galaxy (top-panel) has a SFH for which star formation peaked $> 11$\,Gyr ago and stopped forming stars in the $\sim1$\,Gyr before observation. 
The late-type galaxy (bottom-panel) has a rapidly rising SFH at the time of observation and only began forming stars $\sim3$\,Gyr before observation. 
For the late-type galaxy (101494365161948848), the very rapid star formation influences the resulting metallicity evolution, where the metallicity continues to increase until the present time. 
For the quenched early-type galaxy (101494582392437872), the maximum metallicity was reached at the time of quenching, with the metallicity remaining constant from then on. 
We do note that a number of the steps in the posterior rise more rapidly and finish with \texttt{Zfinal} (the final metallicity) reaching the upper limit of the metallicities covered by the \cite{BC03} templates ($5 \times10^{-2}$). 
Approximately 15 per cent of the galaxies in this sample have a best-fit final metallicity of 
$5 \times10^{-2}$.

The fits to our SEDs are good in general, especially in the optical-NIR regime for objects with spectroscopic redshifts, but there is variation in the extracted star formation and metallicity histories. 
Despite this, the current SFR and overall stellar mass estimates are well constrained for all galaxies. 
In addition to the examples shown in Figure \ref{fig:ProSpect_Examples}, individual inspections were made of several hundred other sources drawn randomly from the sample to verify the fits. In the vast majority of cases ($> 99$ per cent), the \textsc{ProSpect} outputs appear appropriate and the attenuated data accurately describe the measured flux values. In the minority of cases that do not look appropriate, this can be attributed to incorrect photometric redshift measurements or objects whose segment does not appropriately capture all the flux, or whose segment also includes other sources. 

\begin{figure}
    \centering
    \includegraphics[width=\linewidth]{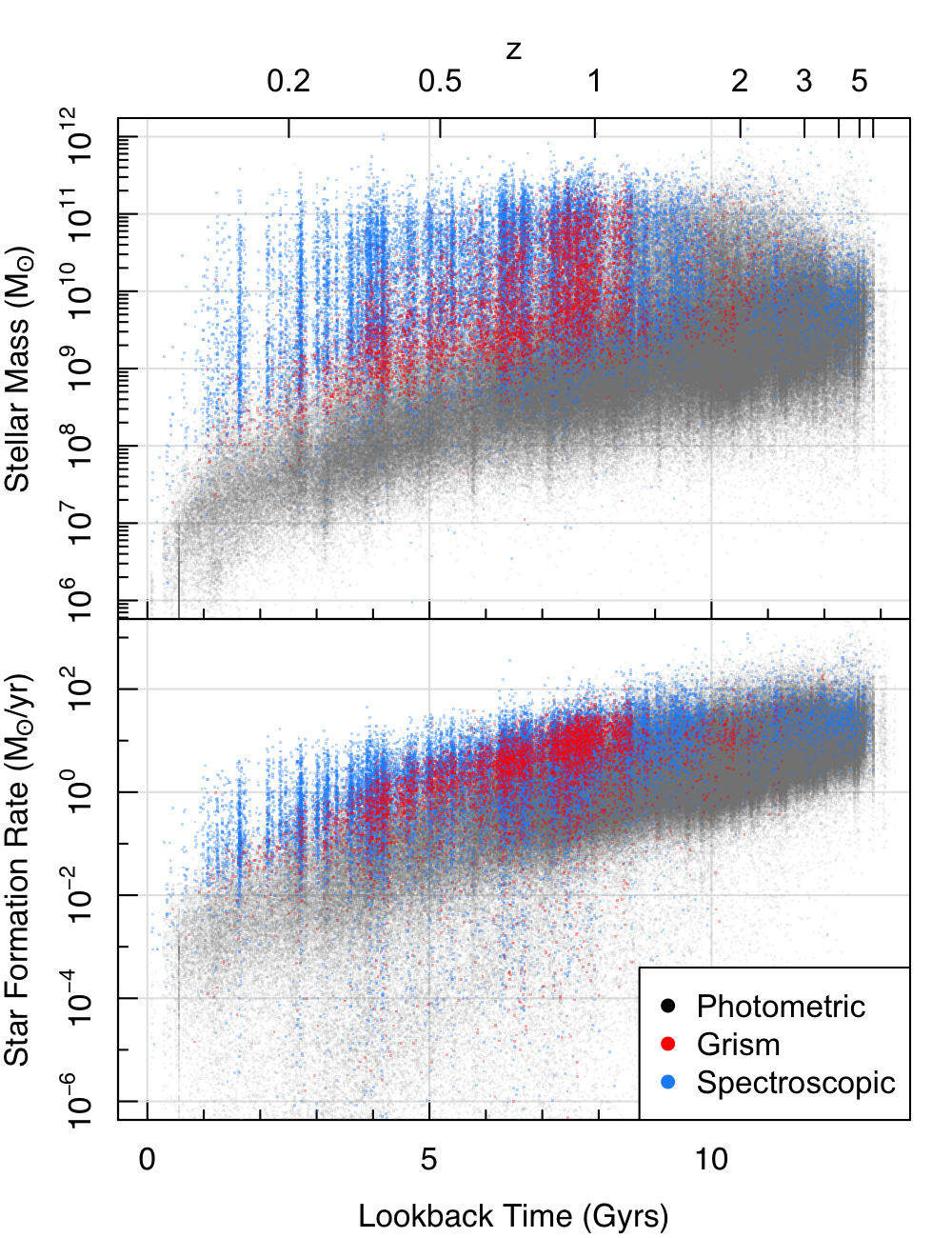}
    \caption{The range of stellar masses (\textit{upper panel}) and SFRs (\textit{lower}) as a function of lookback time coloured by redshift type. We show the corresponding redshifts on the top axis. }
    \label{fig:SM_DM_SFR_byLBT}
\end{figure}

\begin{figure}
    \centering
    \includegraphics[width=\linewidth]{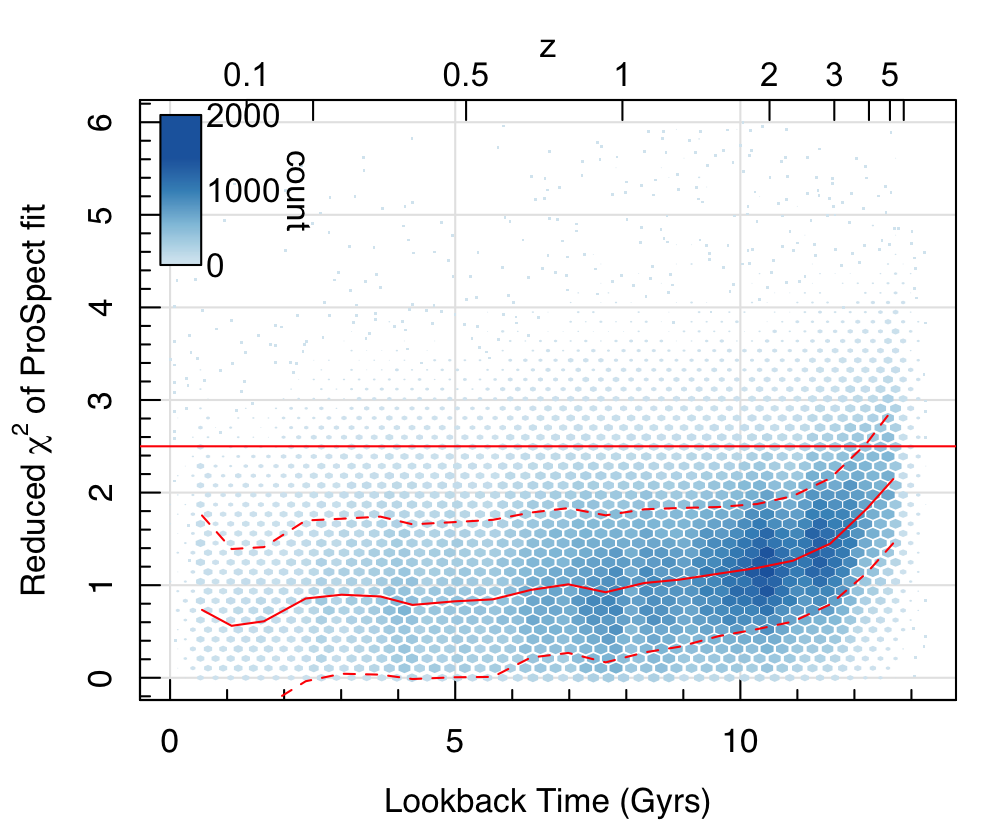}
    \caption{The range of $\chi^2$ values as a function of lookback time. We show the running median as the solid line and the 16th and 84th percentiles as the dashed lines.
    The horizontal red line shows the $\chi^2 = 2.5$ cut described in Appendix~\ref{App:chi2}.}
    \label{fig:chi2_redshift}
\end{figure}

Through this implementation of \textsc{ProSpect} we do not only measure stellar masses and SFRs for each of our galaxies, but also make estimates of the SFH, final metallicity, dust mass, and dust luminosity.
Figure \ref{fig:SM_DM_SFR_byLBT} shows the resulting distribution of stellar masses and SFRs as a function of lookback time. 
The stellar mass and SFR estimates are the most robust measurements made using \textsc{ProSpect} and will be the focus of this paper.  
We do, however, include the metallicity and dust estimates in the DEVILS\_D10ProSpectCat DMU, but caution that these have large model-dependent uncertainties that are not truly reflected in the uncertainties from the fitting process. 
The metallicity measurements and star formation histories will be the focus of future work (Thorne et al. in prep). 
We also show the reduced $\chi^2$ values of the \textsc{ProSpect} fits as a function of lookback time in Figure~\ref{fig:chi2_redshift} with the median shown as the solid red line and the 16th and 84th percentiles shown as the dashed lines.
The median $\chi^2$ values represent a good fit for the vast majority of the sample but deteriorate for $z>4$ as the 84th percentile reaches the $\chi^2$ cut described in Appendix~\ref{App:chi2}.
Objects with $z>4.25$ are included in the D10\_ProSpectCat DMU for completeness but should only be used with caution.

\subsection{Cross-Checking Measurements with Existing Samples}

\begin{figure*}
    \centering
    \includegraphics[width = \linewidth]{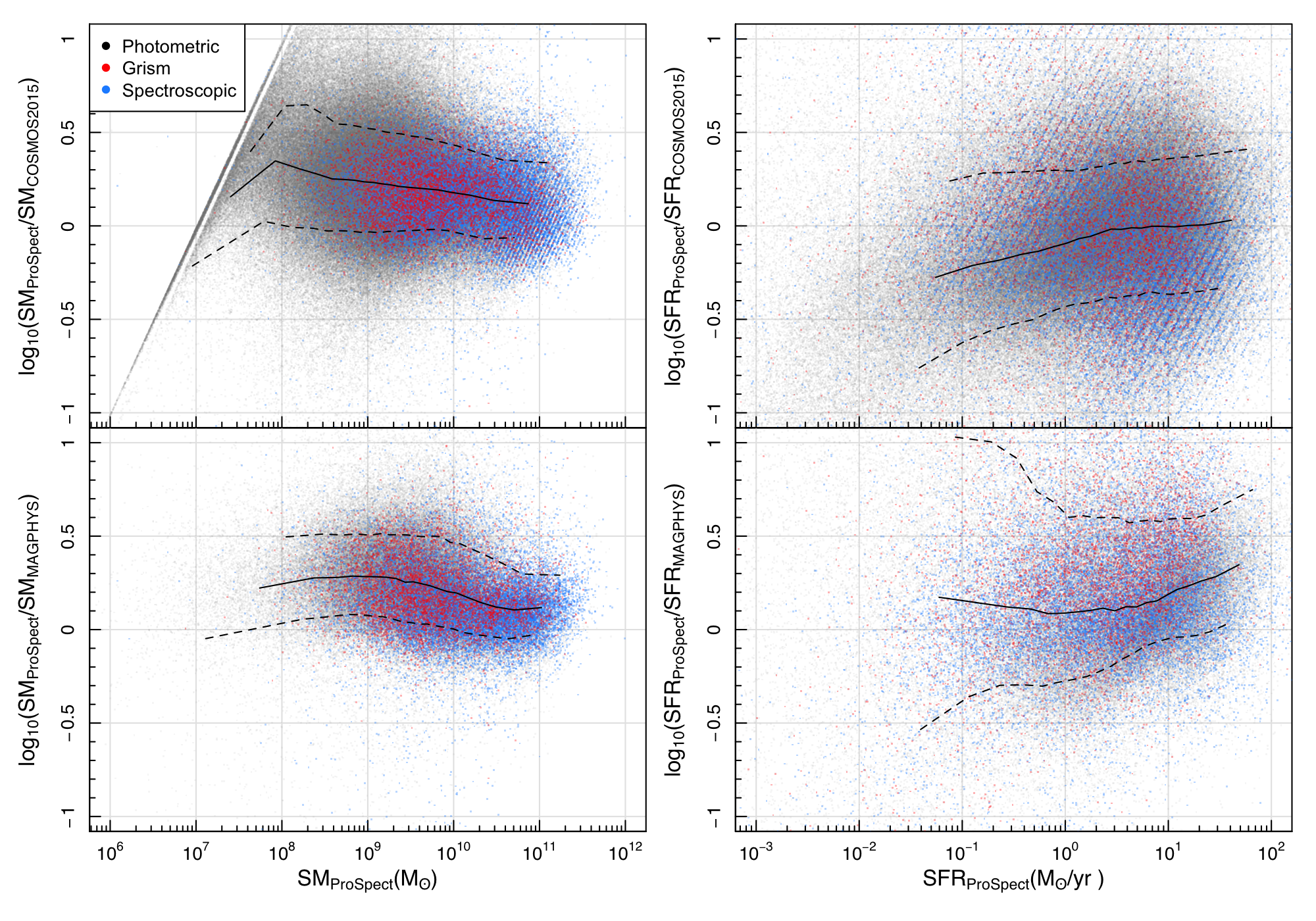}
    \caption{Comparisons between our derived stellar mass and SFR estimates to those derived by \citet{LaigleCOSMOS2015CATALOGEXPLORING2016} (\textit{upper panels}) and those derived by \citet{GAMA_MAGPHYS} (\textit{lower panels}) coloured by redshift type.  We show the running median as the solid black line, and the 16th and 84th percentiles as the dashed black lines.}
    \label{fig:SMandSFRComp}
\end{figure*}

In the D10 field there are previously published stellar masses and SFRs from \cite{LaigleCOSMOS2015CATALOGEXPLORING2016} (COSMOS2015) and \cite{GAMA_MAGPHYS}. 
The \cite{GAMA_MAGPHYS} catalogue uses the \textsc{magphys} SED fitting code \citep{MAGPHYS} to derive stellar masses and SFRs, whilst the COSMOS2015 stellar masses and SFRs are derived using LEPHARE \citep{ArnoutsLePHAREPhotometricAnalysis2011}.
The COSMOS2015 measurements use a library of synthetic spectra generated from BC03, assuming a \cite{ChabrierGalacticStellarSubstellar2003} IMF and combine the exponentially declining SFH and delayed SFH ($\text{SFR} \propto \tau ^{-2} t e^{-t/\tau}$, where $\tau$ is the timescale of the decline). 
The SED fitting method employed for the COSMOS2015 catalogue considers two metallicities, solar and half solar, and includes two attenuation curves: the starburst curve of \cite{CalzettiDustContentOpacity2000} and a curve with a slope of $\lambda^{0.9}$ (appendix A of \citealt{ArnoutsEncodinginfraredexcess2013}). 
Emission lines are added following \cite{IlbertCosmosPhotometricRedshifts2009}. 
The SFRs from COSMOS2015 have large associated errors due to the lack of infra-red data which will affect the influence of dust on the SFRs. 
The COSMOS2015 photometry catalogue is a compilation of different measurement techniques - \textsc{Source Extractor} for the optical - NIR, PSF fitting from \cite{CapakFirstReleaseCOSMOS2007} for the ultraviolet, PSF fitting as per IRACLEAN \citep{HsiehTaiwanECDFSNearInfrared2012} for the Spitzer-IRAC bands and position matched with FIR photometry from \cite{LeFlochDeepSpitzer242009,LutzPACSEvolutionaryProbe2011} and \cite{OliverHerschelMultitieredExtragalactic2012}. 
\cite{GAMA_MAGPHYS} use LAMBDAR \citep{WrightGalaxyMassAssembly2016} derived photometry for their measurements. 
This photometry catalogue has known problems which drove the development of \textsc{ProFound}  \citep{RobothamProFoundSourceExtraction2018}, and as such photometry used in this work is superior. So not only are the  stellar mass and SFR derivation techniques different, so too are the photometric data on which those estimates are based.

We present comparisons between our \textsc{ProSpect} derived stellar masses and SFRs to those from COSMOS2015 and \cite{GAMA_MAGPHYS} in Figure \ref{fig:SMandSFRComp}.
Our \textsc{ProSpect}-derived measurements of stellar mass are higher than previous measurements by approximately 0.21\,dex. 
This is a direct result of our physically motivated implementation of metallicity (see Section \ref{sec:ZHs}).
By allowing the metallicity of the galaxy to grow with the stellar mass, an older, and therefore redder, stellar population is obtained \citep{RobothamProSpectgeneratingspectral2020}. 
This was also found in GAMA by \cite{BellstedtGalaxyMassAssembly2020b} who found an offset of 0.18\,dex when comparing to \textsc{magphys}-derived stellar masses.
Older populations have a higher mass-to-light ratio resulting in a higher stellar mass measurement overall \citep{RobothamProSpectgeneratingspectral2020}. 
The $1\sigma$ scatter from the running median is 0.25\,dex in both cases and the sharp boundary in the stellar mass comparisons to COSMOS2015 is due to a lower limit of $M_\star = 10^7 M_\odot$ in their fitting. 
We find that this mass offset is constant across the D10 field and is not impacted by the differing depth of the UltraVISTA photometry. 
The mass offset is greatest for objects at low redshift ($\sim$0.3\,dex) due to larger allowed variations in the stellar age, and due to more extended emission being recovered for these objects by \textsc{ProFound}.

We find that our derived SFRs are lower than those from COSMOS2015 by 0.05\,dex and higher than those derived by \cite{GAMA_MAGPHYS} using \textsc{magphys} by 0.12\,dex on average. 
The $1\sigma$ scatter on the SFRs is higher than the scatter on the stellar masses (0.3\,dex and 0.4\,dex for COSMOS2015 and \textsc{magphys} respectively). 
We do note that there is structure in the comparison to the COSMOS2015 measurements in the form of the diagonal striping in both the stellar mass and SFR estimates. 
We believe this is due to the discrete stellar and metallicity templates implemented by COSMOS2015 to estimate the SFRs. 

Although these comparisons provide insight into the different stellar masses and SFRs obtained via different SED fitting techniques, we do wish to reiterate that there are also significant improvements in the underlying photometry catalogue used in this work from the photometry catalogues used by COSMOS2015 and \cite{GAMA_MAGPHYS}. 
For comparisons of \textsc{ProSpect}'s performance on the same galaxies but with different photometry catalogues see section 5 of \cite{RobothamProSpectgeneratingspectral2020}. 

\section{Stellar Mass Functions}\label{sec:SMF}
\begin{figure}
    \centering
    \includegraphics[width = \linewidth]{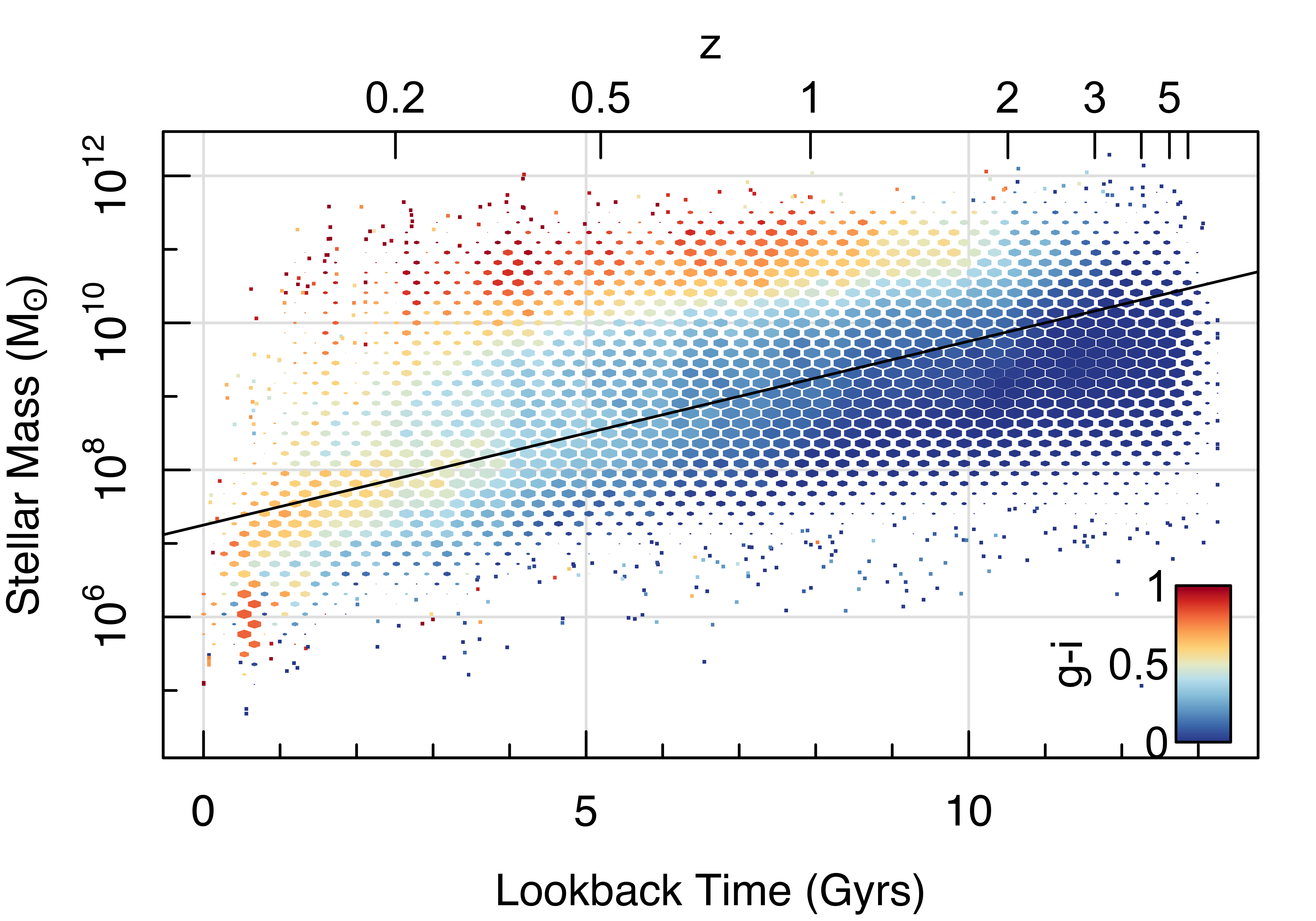}
    \caption{The stellar mass selection limit as a function of lookback time in Gyrs. The size of the hexagons is proportional to the number of galaxies in each bin and each bin is coloured by the median rest-frame $g-i$ colour. We show our stellar mass completeness cut as the solid black line which is given by Equation~\ref{eq:MassComplete}}
    \label{fig:colors}
\end{figure}

The galaxy stellar mass function \citep{BellOpticalNearInfraredProperties2003,Baldrygalaxystellarmass2008,BaldryGalaxyMassAssembly2012}, is a fundamental tool for studying the evolution of galaxies over cosmic time. 
Its integral returns the density of baryonic mass currently bound in stars while the shape of the distribution relates to the evolutionary pathways of galaxy growth and assembly. 
The redshift zero stellar mass function is a key calibration for most galaxy formation models that are carefully tuned to best reproduce the latest measurements (e.g. \citealt{CrainEAGLEsimulationsgalaxy2015,SchayeEAGLEprojectsimulating2015,Laceyunifiedmultiwavelengthmodel2016, LagosSharkintroducingopen2018}; Proctor et al. in prep). 
To compare the galaxy stellar mass distribution measurements obtained from \textsc{ProSpect} to other measurements, we derive the stellar mass function in 15 redshift bins roughly evenly distributed in lookback time.
We use all galaxies in these redshift bins regardless of the photometric redshift $\chi^2$ or the likelihood of the  \textsc{ProSpect} fit, but show the impact of making cuts in both in Appendix~\ref{App:chi2}.

In this work, as in \cite{GAMA_MAGPHYS,WrightGAMAG10COSMOS3DHST2018}, we use only volume-complete samples of the full data set at each redshift interval, thus significantly reducing the possible number of systematic biases that may affect our analysis. 
Mass completeness limits in each of the redshift bins have been calculated using the unattenuated \textit{g-i} rest-frame colour as calculated from the \textsc{ProSpect} fits.
We show the distribution of stellar mass as a function of lookback time coloured by the median \textit{g-i} colour in Figure~\ref{fig:colors}. 
Using this approach, per-bin completeness limits used in this work were estimated by making a linear cut in Figure~\ref{fig:colors} to limit to mass and redshift bins that were complete (the reddest region in the top left corner).
We began with a very conservative cut removing everything below $M_\star = 10^9 M_\odot$ at $z=0$ and compared the differences to the recovered stellar mass functions if we lowered the $z=0$ intercept in 0.25\,dex increments.
We find no difference in our stellar mass functions when we decrease our $z=0$ intercept to $M_\star = 10^{7.25} M_\odot$.
Where therefore use this relation to make our completeness cut:
\begin{equation} \label{eq:MassComplete}
    \log_{10} (M_\star / M_\odot) = \frac{1}{4} t_\text{lb} + 7.25,
\end{equation}
where $t_\text{lb}$ is the lookback time in Gyrs.
In each redshift bin we truncate to stellar masses above this cut. 
This approach is less rigorous than other methods of mass completeness estimate \citep{MarchesiniEvolutionStellarMass2009, MuzzinEvolutionStellarMass2013,TomczakGalaxyStellarMass2014}, however is unlikely to bias our analysis over the mass ranges we explore.

\begin{figure*}
    \centering
    \includegraphics[width = \linewidth]{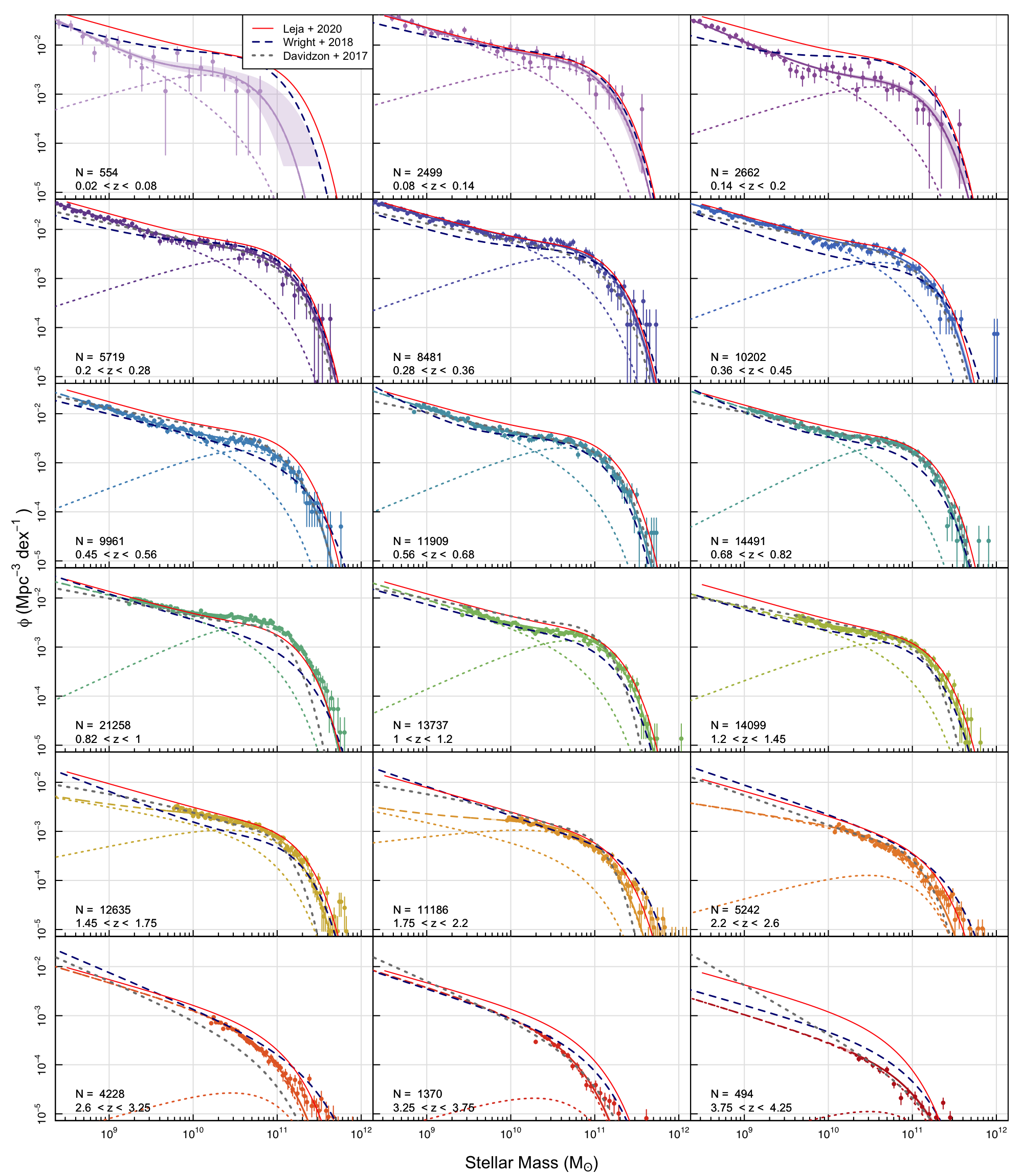}
    \caption{The two-component Schechter function fits to our stellar mass measurements. Each panel shows a redshift bin (limits are annotated) with the fitted data (points), the best fit from \texttt{dftools} (solid lines), the extrapolation to low masses (dashed line), and the 1-$\sigma$ range from the bootlegs (shaded region). 
    We also show the two individual Schechter components as the dotted lines in the same colour as the best fit. 
    We also show the Schechter fits that include data from the COSMOS field from \citet{LejaNewCensusUniverse2020} in red, \citet{WrightGAMAG10COSMOS3DHST2018} in dashed blue and \citet{DavidzonCOSMOS2015galaxystellar2017} in dotted grey. We show the number of galaxies above the mass completeness cut in each redshift bin and the redshift range of each bin in the lower left corner of each panel.}
    \label{fig:SMF}
\end{figure*}

\begin{figure*}
    \centering
    \includegraphics[width = \linewidth]{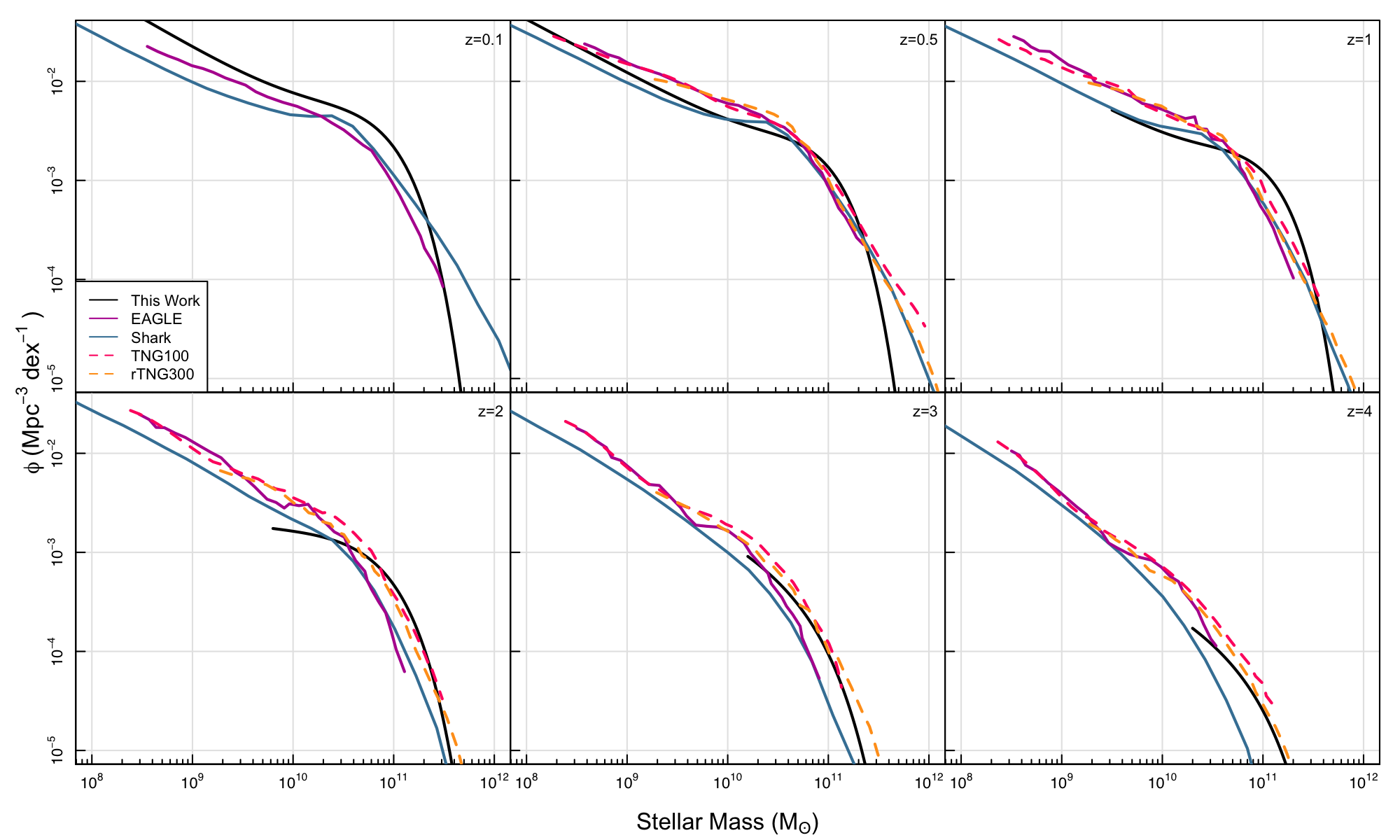}
    \caption{The stellar mass function derived in this work compared to theoretical stellar mass functions from simulations at six redshifts. We show the best fit two-component Schechter function from Figure~\ref{fig:SMF} in black with comparisons to the stellar mass function from the \textsc{Shark} semi-analytic model \citep{LagosSharkintroducingopen2018} in blue, \textsc{eagle} results from \citet{FurlongEvolutiongalaxystellar2015} in purple, and results from the TNG100 and TNG300 IllustrisTNG simulations from \citet{PillepichFirstresultsIllustrisTNG2018} in dashed red and orange.}
    \label{fig:SMFplusSims}
\end{figure*}

\cite{WrightGAMAG10COSMOS3DHST2018} motivate using a two component \cite{Schechteranalyticexpressionluminosity1976} function to model the stellar mass function even out to high redshifts based on the biases induced on the break mass ($M^*$) parameter by a single component Schechter function. 
They found that fits using a single component Schechter function move to significantly higher values of $M^*$ at early times, causing the regression to behave somewhat poorly.
Because of this, we also elect to fit a double Schechter function to our data set in all redshift bins as parameterised as follows:
\begin{equation}\label{eq:DoubleSchechter}
   \phi (\mathcal{M}) =  \ln(10)  e^{-\mu} \left( \phi^*_1 \mu^{\alpha_1 + 1} + \phi^*_2 \mu^{\alpha_2 + 1}\right),
\end{equation}
where $\mu = 10^{\mathcal{M}}/10^{M^*}$, $\mathcal{M} = \log_{10}(M_\star/M_\odot)$ and $\phi$ is the number density as a function of stellar mass.  

\begin{figure}
    \centering
    \includegraphics[width = \linewidth]{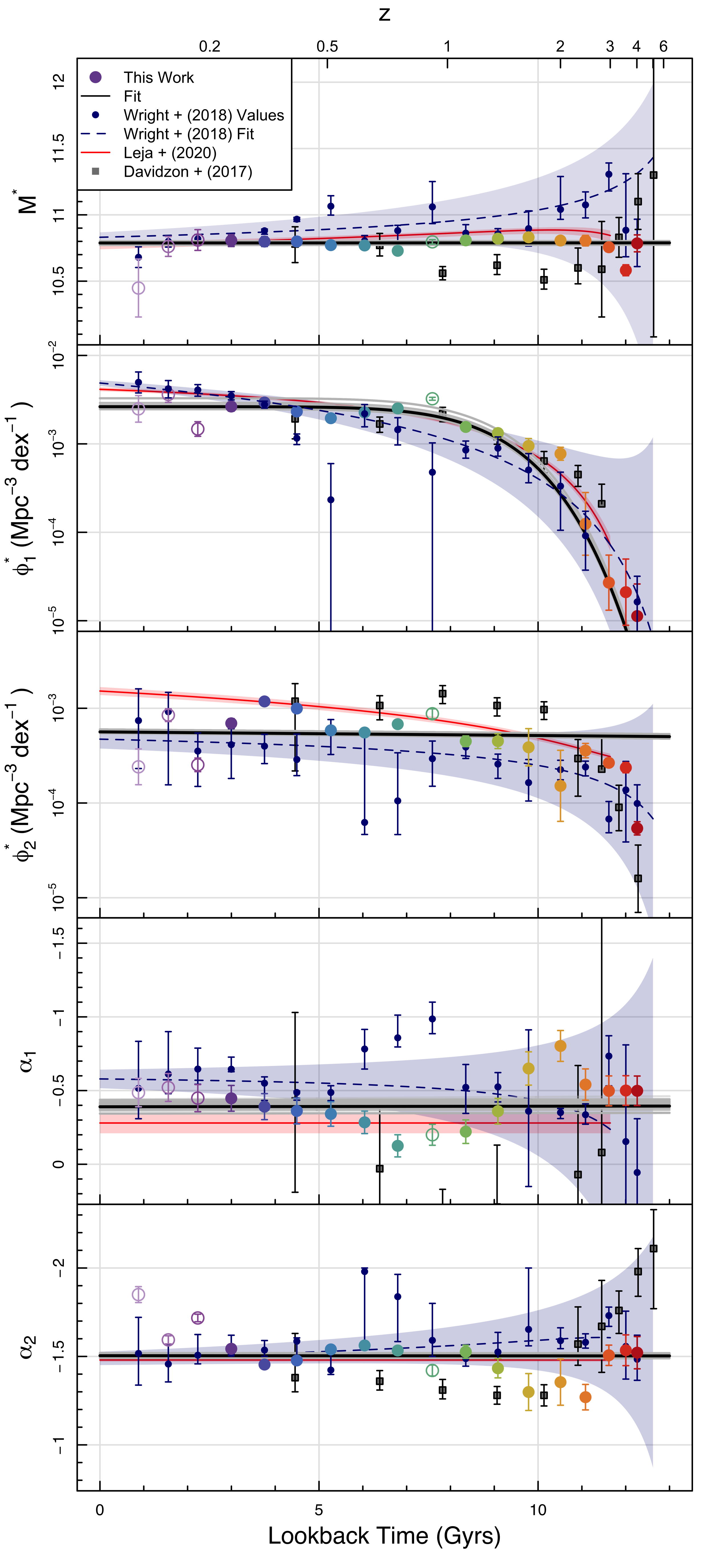}
    \caption{Evolution of the Schechter function parameters as a function of lookback time. 
    Open circles represent redshift bins that were determined to be over or under dense as described in the text.
    The black line shows the fits to each parameter and the grey lines show samples from the posterior. 
    We show the values (navy blue points), fits (navy blue line) and uncertainty range (pale blue shaded region) from \citet{WrightGAMAG10COSMOS3DHST2018} and the evolution of the continuity model and its uncertainties from \citet{LejaNewCensusUniverse2020} as the red lines and red shaded region respectively.
    We also show the results from \citet{DavidzonCOSMOS2015galaxystellar2017} using the COSMOS2015 stellar masses as the dark grey squares.
    All data points from this work are provided in Appendix~\ref{App:Vals}.
    }
    \label{fig:SMF_Evolving}
\end{figure}

\begin{figure*}
    \centering
    \includegraphics[width = \linewidth]{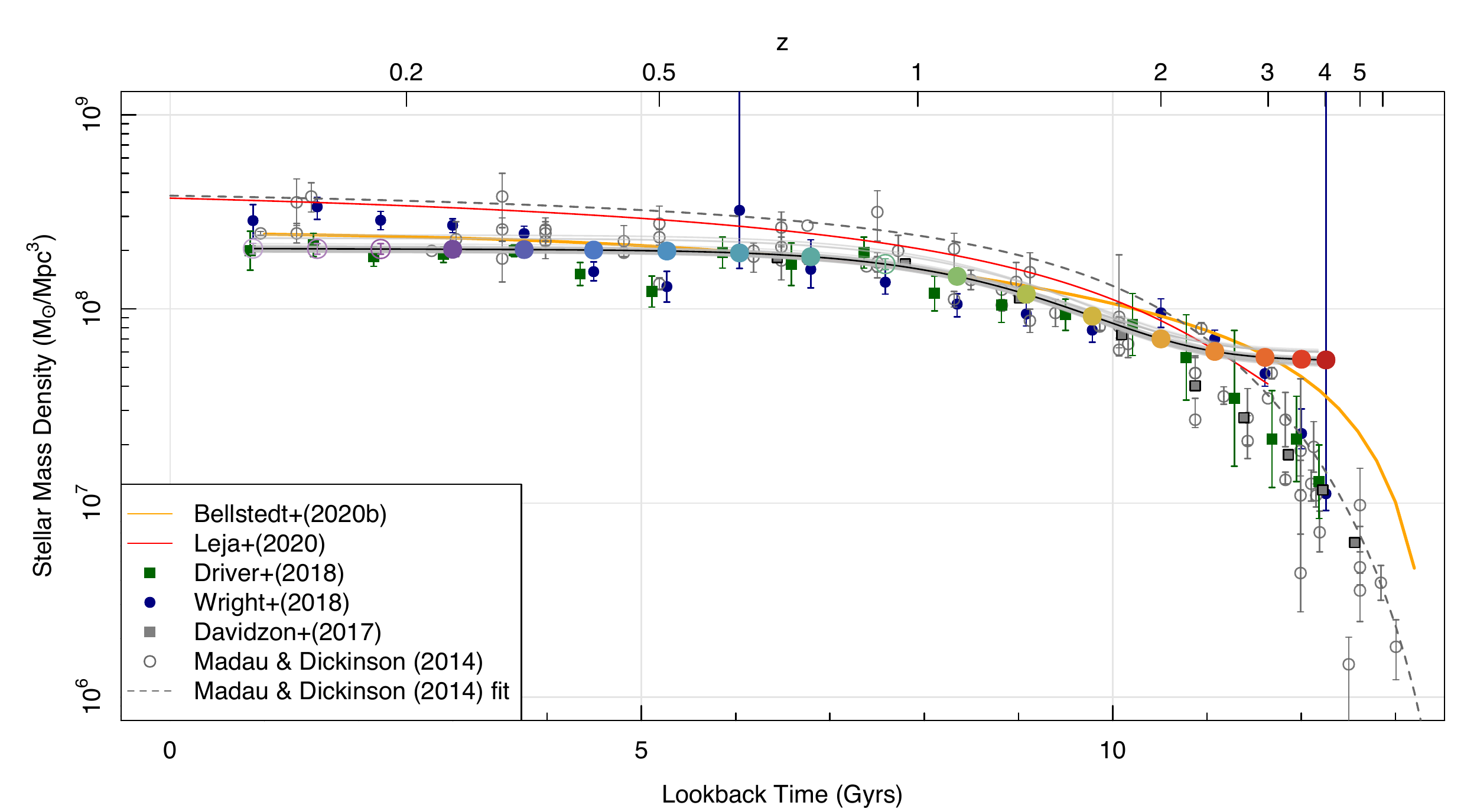}
    \caption{The evolution of stellar mass density over cosmic time from the analytic integral of the regressed double component Schechter parameters. 
    The black line shows the continuous evolution of the stellar mass density while the coloured points show the stellar mass density in each of our redshift bins. 
    The grey lines and uncertainties on each point are calculated using variations of the analytic integral using samples from the posterior of each parameter. 
    The open points represent redshift bins that were determined to be over or under dense compared to a smooth evolution of galaxies with $\log_{10} (M_\star / M_\odot) = M^* \pm 0.3$ and are not considered in the fitting of each Schechter parameter.
    We show comparisons to measurements from \citet{MadauCosmicStarFormationHistory2014,GAMA_MAGPHYS,WrightGAMAG10COSMOS3DHST2018,DavidzonCOSMOS2015galaxystellar2017} as the open grey circles, dark green squares, navy blue circles and dark grey squares respectively. 
    We also show the stellar mass density evolution derived from the SFHs of $\sim 7,000$ low redshift ($z < 0.06$) GAMA galaxies by \citet{BellstedtGalaxyMassAssembly2020b} in orange. }
    \label{fig:SMD}
\end{figure*}

We use the \texttt{dftools} R package \citep{ObreschkowEddingtondemoninferring2018} to fit Equation~\ref{eq:DoubleSchechter} in each of the redshift bins. \texttt{dftools} allows for a general modified maximum likelihood method for inferring generative distribution functions from uncertain and biased data. 
The benefits to using \texttt{dftools} are that it is free of binning and natively accounts for small number statistics, non-detections and simultaneously deals with observational uncertainties (Eddington bias). 
See \citet{ObreschkowEddingtondemoninferring2018} for a complete description of \texttt{dftools}.

When fitting the stellar mass function with \texttt{dftools} we combine the stellar mass uncertainties from \textsc{ProSpect} with the redshift uncertainties for photometric redshift sources to better account for the uncertainty in the stellar mass measurements.
We include broad Gaussian priors ($\sigma = 0.5$) on the $M^*$ and $\phi^*$ parameters based on the predicted values from \cite{WrightGAMAG10COSMOS3DHST2018} and tighter Gaussian priors ($\sigma = 0.1$) on the $\alpha$ parameters at $\alpha_1 = -0.5$ and $\alpha_2 = -1.5$ based on the values derived by \cite{WrightGAMAG10COSMOS3DHST2018} and \cite{LejaNewCensusUniverse2020}. 
This allows us to constrain the slopes of the two components especially in our higher redshift bins ($z > 1.75$) where we are not fitting galaxies below $M_\star = 10^{10} M_\odot$.
We bootstrap the fits 100 times in each redshift bin to produce more accurate covariances which we show as the shaded region in each panel.

Figure~\ref{fig:SMF} shows the observed number density, binned in stellar mass, with the uncertainties indicated as the per-bin error bars. 
Each panel shows the best fit double component Schechter as the solid line and the two individual Schechter components as a dotted line in the same colour. 
We show the extrapolation of the fit at the low mass end as a dashed line of the same colour. 
All individual fit parameters are presented in Appendix~\ref{App:Vals}.
We also show comparisons to the stellar mass functions measured by \cite{DavidzonCOSMOS2015galaxystellar2017,WrightGAMAG10COSMOS3DHST2018} and \cite{LejaNewCensusUniverse2020} as the dotted grey, dashed black and solid red lines respectively. 
\cite{DavidzonCOSMOS2015galaxystellar2017} use the COSMOS2015 catalogue \citep{LaigleCOSMOS2015CATALOGEXPLORING2016} to measure the stellar mass function out to $z\sim6$. 
The measurement from \cite{WrightGAMAG10COSMOS3DHST2018} utilised a combined data set consisting of GAMA, G10-COSMOS \citep{DaviesGalaxyMassAssembly2015} and 3D-HST \citep{Skelton3DHSTWFC3selectedPhotometric2014}.
 They use GAMA to supplement measurements at the low redshift end and use solely 3D-HST for $z>1.75$. 
 Note that these stellar mass functions are measured using the masses derived by \cite{GAMA_MAGPHYS} which are found to be 0.21\,dex lower than the stellar mass estimates derived in this work (Figure~\ref{fig:SMandSFRComp}).
 Recently \cite{LejaNewCensusUniverse2020} remeasured the stellar mass function using SED fits to the COSMOS2015 \citep{LaigleCOSMOS2015CATALOGEXPLORING2016} and 3D-HST \citep{Skelton3DHSTWFC3selectedPhotometric2014} photometry for $\sim 100,000$ galaxies between $0.2 < z < 3$. 
 \cite{LejaNewCensusUniverse2020} fit the stellar mass function over this redshift range using a two-component Schechter function but assuming a `continuity model' which directly fits the evolution of the stellar mass function and ensures a smooth evolution of the parameters. 
 They assume no evolution of the $\alpha$ parameters but allow the $M^*$ and $\phi^*$ parameters to vary in three `anchor' redshifts and assume a quadratic evolution between these redshifts.

We do not correct for large scale structure in any redshift bin which leads to the underestimation of the stellar mass function in the lowest redshift bins ($z < 0.20$) when compared to \cite{WrightGAMAG10COSMOS3DHST2018}. 
This is because \cite{WrightGAMAG10COSMOS3DHST2018} supplement their low redshift measurements with the much larger area of the GAMA survey which is far less prone to impacts from large scale structure.
The overestimation of the stellar mass function compared to \cite{WrightGAMAG10COSMOS3DHST2018} between $ 0.36 < z < 1$ is due to the large clusters known in the COSMOS field \citep{BellagambaOptimalfilteringoptical2011}, especially between $0.82 < z < 1$ where there are a large number of known clusters \citep{FinoguenovXMMNewtonWideFieldSurvey2007}. 
\cite{WrightGAMAG10COSMOS3DHST2018} recover lower stellar mass functions across these redshifts due to constraints from GAMA and 3D-HST, whereas we agree closely with the results from \cite{DavidzonCOSMOS2015galaxystellar2017} as these were also calculated using data purely from the COSMOS field. 
We also measure lower stellar mass functions at the highest redshifts when compared to both \cite{WrightGAMAG10COSMOS3DHST2018} and \cite{LejaNewCensusUniverse2020} most likely due to the small samples and small mass range over which we can constrain the stellar mass function. 
In the highest redshift bins ($z > 2.6$) we find minimal contribution from the second Schechter component suggesting that a single Schechter component would be sufficient.

Figure~\ref{fig:SMFplusSims} shows a comparison of our derived stellar mass functions to those from simulations. 
We show comparisons to the stellar mass functions obtained from the semi-analytic model \textsc{Shark} \citep{LagosSharkintroducingopen2018}, and from the \textsc{eagle} \citep{SchayeEAGLEprojectsimulating2015} and IllustrisTNG \citep{PillepichFirstresultsIllustrisTNG2018} hydrodynamical simulations. 
At all redshifts there is reasonable agreement between the theoretical stellar mass functions and ours, with the largest discrepancies at high stellar masses.

\subsection{Evolution of the stellar mass function}
 Figure~\ref{fig:SMF_Evolving} shows the evolution of the Schechter function parameters, $M^*$, $\phi^*$ and $\alpha$.
We show comparisons to the double component Schechter fits from \cite{DavidzonCOSMOS2015galaxystellar2017}, \cite{WrightGAMAG10COSMOS3DHST2018} and \cite{LejaNewCensusUniverse2020}. 
 We also show the quadratic fit in redshift to the evolution of each parameter from \cite{WrightGAMAG10COSMOS3DHST2018}. 
 
As we have not corrected our individual stellar mass functions for the effects of large scale structure, we find that we recover quite different stellar mass functions to previous work in some redshift bins.  
We isolate redshift bins that are over- or under-dense by fitting a third order smooth spline through the evolution of the density of galaxies with $\log_{10} (M_\star/M_\odot) = M^* \pm 0.3$  and selecting bins that differ from this spline by more than 50 per cent. 
These points are shown as open circles in Figure~\ref{fig:SMF_Evolving} and are excluded from the fits to the evolution of each parameter. 

The $M^*$ parameter shows little to no evolution over the redshift range examined and is in very close agreement with the measurements from \cite{WrightGAMAG10COSMOS3DHST2018} and \cite{LejaNewCensusUniverse2020}.
This is despite the known differences between the stellar masses derived in this work and those used by \cite{WrightGAMAG10COSMOS3DHST2018} (shown in Figure~\ref{fig:SMandSFRComp}) and \cite{LejaNewCensusUniverse2020}.
We believe this is due to over estimations of the `fluxscale' factor used to correct the aperture derived stellar masses for missing flux as implemented by \cite{WrightGalaxyMassAssembly2017,WrightGAMAG10COSMOS3DHST2018}.

The value of $\phi^*_1$ shows the strongest evolution of any of our fitted parameters with a steep increase over the first $\sim4$\,Gyr of the Universe, and flattening since.
We find no evolution of the $\alpha$ parameters as per \cite{WrightGAMAG10COSMOS3DHST2018} and \cite{LejaNewCensusUniverse2020}. 
We find that each of the double component Schechter parameters derived in this work are in agreement with the previous measurements from \cite{WrightGAMAG10COSMOS3DHST2018} and \cite{LejaNewCensusUniverse2020} and differ the most in bins where we find over- or under-densities. 
We fit the evolution of each of the parameters with a linear fit except for $\phi^*_1$ which we fit with a seventh-order polynomial in lookback time as this best recovers the sharp downturn at high lookback time. 
All SMF fit results and fits to the evolution of each of the parameters are provided in Appendix~\ref{App:Vals}. 

The stellar mass density shown in Figure~\ref{fig:SMD} is derived using the analytical integration of the regressed Schechter parameters over all masses. 
By using the regressed values of each of the parameters we assume smooth evolution over cosmic time and are therefore not subject to differences caused by large scale structure. 
We compare our results to those from \cite{WrightGAMAG10COSMOS3DHST2018}, \cite{GAMA_MAGPHYS}, \cite{DavidzonCOSMOS2015galaxystellar2017}, and the compilation from \cite{MadauCosmicStarFormationHistory2014}. 
We also show comparisons to the inferred stellar mass density evolution from \cite{BellstedtGalaxyMassAssembly2020b} in orange.
This evolution was measured from a sample of $\sim7,000$ low redshift galaxies from the GAMA survey using the \textsc{ProSpect} derived SFHs to trace the entire cosmic SFR and stellar mass density evolution. 
Our fits show a similar evolution and reasonable agreement with previous work at all redshifts, despite the known 0.2\,dex offset between the stellar masses estimates. 
We expect to be most consistent with the results from \cite{BellstedtGalaxyMassAssembly2020b} and \cite{LejaNewCensusUniverse2020} as they are both known to recover higher stellar masses than previous work by $0.1-0.3$\,dex. 
Over most of cosmic time we find very close agreement with \cite{BellstedtGalaxyMassAssembly2020b} and differ only at the highest lookback times, where they find a higher stellar mass density than previous measurements. 
This is unsurprising, as the constraint from SED fitting at this epoch is relatively hard, and hence the \cite{BellstedtGalaxyMassAssembly2020b} values are most uncertain at this epoch.
At high lookback times we agree more closely with the results from \cite{GAMA_MAGPHYS} and \cite{MadauCosmicStarFormationHistory2014}. 
We do, however, recover a lower stellar mass density over all cosmic time compared to \cite{LejaNewCensusUniverse2020}. 
This is due to the fact that we recover lower fitted values of $\phi_1^*$ at low lookback times, and of $\phi_2^*$ at lookback times $>4$\,Gyrs. 
This results in a slightly lower stellar mass density across all of cosmic time. 
Despite the higher estimates of stellar mass derived in this work, we find no offset in any of the stellar mass function parameters or resulting stellar mass density from previous measurements. 

\section{SFR-$M\star$ Relation}\label{sec:SFMS}

\begin{figure*}
    \centering
    \includegraphics[width=\linewidth]{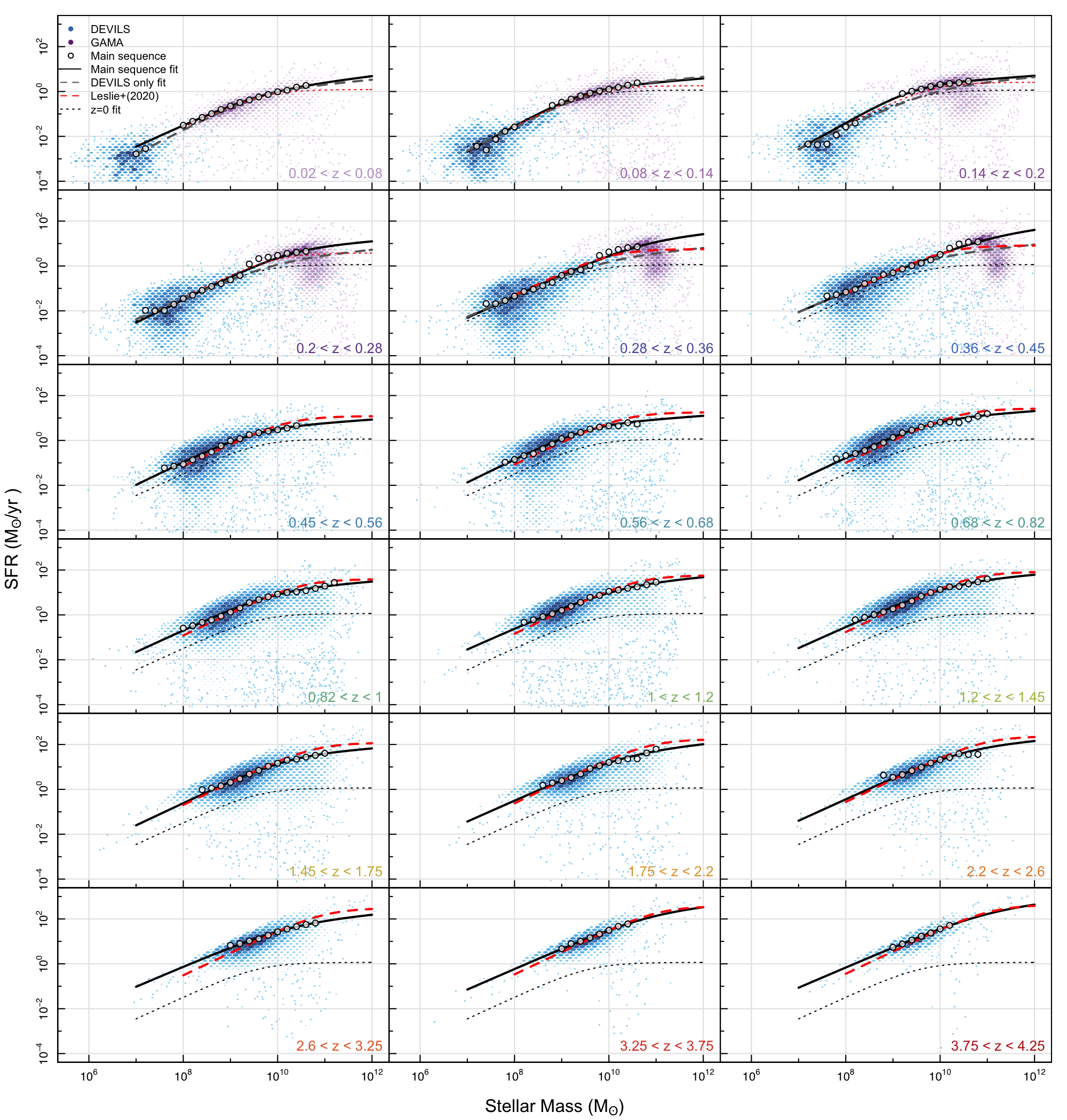}
    \caption{The evolution of the SFR-$M_\star$ plane as a function of redshift for the combined sample of all star-forming and passive DEVILS (blue) and GAMA (purple) galaxies. 
    We show the density of galaxies as a 2D-histogram where the darkness and size of the hexagons corresponds to the number of objects in each bin. 
    We show the medians of the upper Gaussian (i.e. the star forming population) from the mixture model as the white circles and the fit of Equation~\ref{eq:LeeSFMS} to these points in solid black. 
    For $z<0.45$ we show the fit using only DEVILS data as the dashed grey line and we show the lowest redshift ($z < 0.08$) result in each panel as the dotted black line to highlight the evolution in the normalisation.
    We also show the main sequence from \citet{LeslieVLACOSMOSGHzLarge2020} at each redshift as the dashed (dotted when extrapolated for $z<0.3$) red line but shifted 0.2\,dex higher in stellar mass to account for the known offset between \textsc{ProSpect} and COSMOS2015.}
    \label{fig:SFMS_z}
\end{figure*}

\begin{figure}
    \centering
    \includegraphics[width = \linewidth]{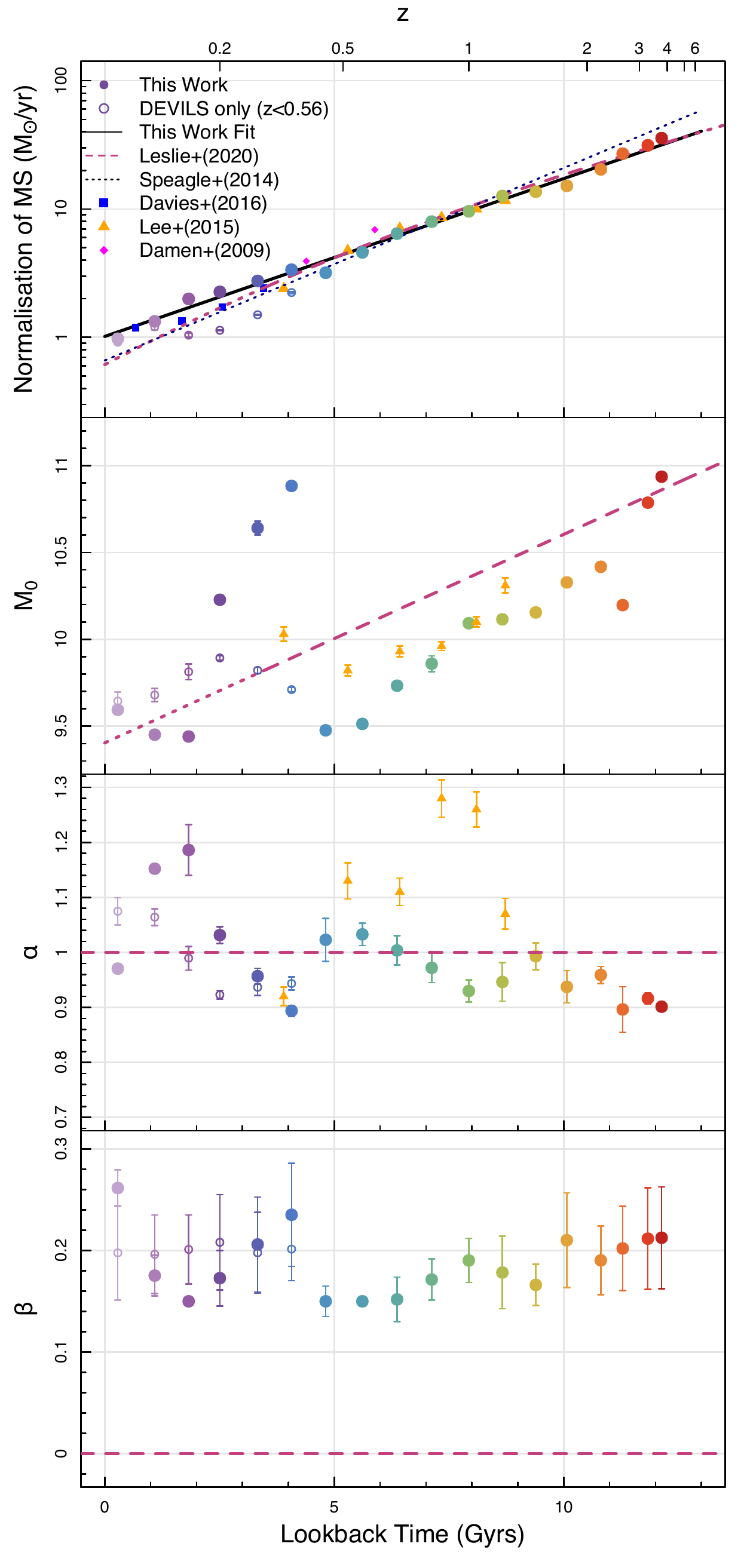}
    \caption{
    First: Evolution of the normalisation of the SFR-$M_\star$ relation at $\log_{10}(M_\star/M_\odot) = 10.0$. We show the measurements made using only DEVILS at $z<0.45$ as the open circles in each panel. 
    We show comparisons to the \citet{DaviesGAMAHATLASmetaanalysis2016} GAMA results at low redshifts (blue), and \citet{LeeTurnoverGalaxyMain2015} (orange) and \citet{LeslieVLACOSMOSGHzLarge2020} (dashed purple line, dotted when extrapolating) at comparable redshifts to this work also measured in the COSMOS field. 
    We shift the \citet{LeslieVLACOSMOSGHzLarge2020} and \citet{LeeTurnoverGalaxyMain2015} stellar mass values by 0.2\,dex before calculating the normalisation to account for the known offset in stellar mass between those derived in this work and COSMOS2015. 
    We also show comparisons to the measurements obtained by \citet{DamenEvolutionSpecificStar2009} in magenta and the fit using the compilation of data in \citet{SpeagleHighlyConsistentFramework2014} as the dotted black line.
    Second: Evolution of the turn-over mass ($\mathcal{M}_0$) in comparison to the \citet{LeeTurnoverGalaxyMain2015} (orange) and \citet{LeslieVLACOSMOSGHzLarge2020} measurements.
    Third and Forth: the same but for the power-law slope of the low ($\alpha$) and high ($\beta$) mass ends of the SFR-$M_\star$ relation respectively.
    }
    \label{fig:EvoSFMS}
\end{figure}

The SFR-$M_\star$ relation (or the star forming galaxy main sequence, e.g. \citealt{Brinchmannphysicalpropertiesstarforming2004,NoeskeStarFormationAEGIS2007,SalimUVStarFormation2007,WhitakerStarFormationMass2012, Lara-LopezGalaxyMassAssembly2013, LeeTurnoverGalaxyMain2015,DaviesGAMAHATLASmetaanalysis2016,DaviesStarFormingMainSequence2019}) is a key diagnostic of both the distribution and evolution of star formation in the Universe. This relation shows the tight correlation between stellar mass and star formation in actively star forming galaxies and is known to evolve in normalisation out to high redshift (e.g. \citealt{DaddiMultiwavelengthStudyMassive2007, Elbazreversalstarformationdensity2007, NoeskeStarFormationAEGIS2007, LeeTurnoverGalaxyMain2015, LeslieVLACOSMOSGHzLarge2020}). There is no established consensus in the literature on the proper form of the main sequence; whether it is linear across all redshifts 
(e.g. \citealt{WuytsGalaxyStructureMode2011, SpeagleHighlyConsistentFramework2014, PearsonMainsequencestar2018}), has a flattening or turn-over at stellar masses $\log_{10}(M_\star/M_\odot) > 10.5$ (e.g. \citealt{WhitakerConstrainingLowmassSlope2014,LeeTurnoverGalaxyMain2015,SchreiberHerschelviewdominant2015,LeslieVLACOSMOSGHzLarge2020}) or if any flattening evolves with time. This discrepancy seems to be driven by selection effects. Furthermore, the normalisation of the main sequence relation depends on the SFR tracer and calibrations used (e.g. \citealt{SpeagleHighlyConsistentFramework2014,DaviesGAMAHATLASmetaanalysis2016, DaviesStarFormingMainSequence2019}) 

As a final demonstration of our measurements, we present an analysis of the star-formation main sequence relation in the SFR-$M_\star$ plane in Figure~\ref{fig:SFMS_z}.
Whilst we need to be mass complete to measure the galaxy stellar mass function, we only need to be complete in the blue star forming population to measure the main sequence. 
Because of this we use all objects in our catalogue out to $z=4.25$ as this is where the median $\chi^2$ values deteriorate (see Figure~\ref{fig:chi2_redshift}).
To ensure that we have enough high-mass galaxies at low redshift to constrain a potential turn-over, we supplement the DEVILS stellar mass and SFR estimates with \textsc{ProSpect} fits of the subset of publicly available GAMA galaxies as presented in \cite{BellstedtGalaxyMassAssembly2020b}. 
The stellar mass and SFR estimates for GAMA were derived in much the same way as the DEVILS estimates derived in this work, with small changes to account for the much smaller redshift range and differing filter set in the photometry catalogue. We supplement the DEVILS measurements with GAMA for redshifts below $z\le0.45$.

Many methods have been employed in the literature to extract the star forming population including optical colour cuts \citep{TaylorGalaxyMassAssembly2015,DaviesGAMAHATLASmetaanalysis2016}, specific SFR selections \citep{GuoStarFormationMain2015}, or morphological selections \citep{DaviesStarFormingMainSequence2019}.  
For this work we elect to split the star forming and passive populations by first making a cut in specific SFR (sSFR) at $\text{sSFR} = 1\times10^{-13}\,\text{yr}^{-1}$ to remove objects with very low star formation rates that are possible due to the SFH parameterisation. 
We then fit a mixture model of two Gaussians implemented using the \texttt{MixTools} package \citep{Benagliamixtoolspackageanalyzing2009} to split the star forming galaxies from the rest of the quenched galaxies. 
We fit the mixture model in stellar mass bins of width 0.2 dex. 

To fit the main sequence we adapt the parameterisation from \cite{LeeTurnoverGalaxyMain2015} which has been shown to hold out to $z\approx 4$ \citep{TomczakSFRMRelationEmpirical2016}. 
Equation 2 from \cite{LeeTurnoverGalaxyMain2015} and the adaptation from \cite{LeslieVLACOSMOSGHzLarge2020} assume a constant SFR at high mass (i.e. a slope of zero), which we do not find evidence for in any redshift bin. 
We adapt equation 2 from \cite{LeeTurnoverGalaxyMain2015} to add an additional slope to freely model the SFR at high stellar masses:
\begin{equation}\label{eq:LeeSFMS}
    \log_{10} (\text{SFR}) = \mathcal{S}_0 - \log_{10}\left[ \left(\frac{10^\mathcal{M}}{10^{\mathcal{M}_0}} \right) ^{-\alpha}  + \left(\frac{10^\mathcal{M}}{10^{\mathcal{M}_0}} \right) ^{-\beta} \right],
\end{equation}
where $\mathcal{M} = \log_{10}(M_\star/M_\odot)$ and the SFR is measured in $M_\odot \text{yr}^{-1}$. 
This parameterisation allows us to quantify the interesting characteristics of the relation between stellar mass and SFR: $\alpha$ and $\beta$, the power-law slope at low and high stellar masses respectively, $\mathcal{M}_0$, the turnover mass (in $\log_{10}(M_\star/M_\odot)$), and $\mathcal{S}_0$, the maximum value of $\log_{10}(\text{SFR})$ that the function approaches at high stellar mass. 

We provide Normal priors of $\mu = 1$, $\sigma = 0.05$ and $\mu = 0.2$, $\sigma = 0.05$ on the low and high mass slopes respectively. 
The position of the prior on the low mass slope was selected based on the measurements from \cite{LeeTurnoverGalaxyMain2015} and the assumption of \cite{LeslieVLACOSMOSGHzLarge2020}, whilst the prior on the high mass slope was selected by independently fitting the GAMA data at low redshift with a purely linear relationship. 
We also implement a broad Normal prior on the turn-over mass, $M_0$, using the second line of equation 6 from \cite{LeslieVLACOSMOSGHzLarge2020} to determine the turn over mass at the median redshift of each bin. 
We use this turn over mass as the mean for the prior and assume a standard deviation of $\sigma = 0.3$.
When fitting our model to the main sequence values, we fit to only mass bins that have more than 300 galaxies. 
We fit our model to the data using \textsc{Highlander}, assuming a student-t likelihood as it is more robust to outliers due to the heavier tails. 
We fit Equation \ref{eq:LeeSFMS} to each redshift bin independently.
In the low redshift bins where we are supplementing with GAMA measurements, we also fit to only the DEVILS data for comparison. 
In these cases, the recovered turn-over mass and high-mass slope values are driven by the priors as DEVILS provides no constraint on these parameters on its own.
All fit values to the SFR-$M_\star$ main sequence are provided in Appendix~\ref{App:Vals}.

Figure~\ref{fig:SFMS_z} shows the distribution of all DEVILS and GAMA sources (i.e. star forming and passive) in the SFR-$M_\star$ plane as the hexagonal 2D-histogram and the best fit star formation main sequence as obtained in this work in comparison to the recent measurements as derived by \cite{LeslieVLACOSMOSGHzLarge2020}. 
The measurements from \cite{LeslieVLACOSMOSGHzLarge2020} are also obtained in the COSMOS field but instead use the COSMOS2015 \citep{LaigleCOSMOS2015CATALOGEXPLORING2016} stellar mass estimates, which are known to be smaller than those measured by \textsc{ProSpect} by approximately 0.2 dex. 
\cite{LeslieVLACOSMOSGHzLarge2020} use SFRs derived from 3\,GHz radio continuum imaging for their fits. 
We find very good agreement with the results from \cite{LeslieVLACOSMOSGHzLarge2020} in all redshift bins but vary most at the high mass end. 
We do see a discontinuity between the DEVILS and GAMA measurements at intermediate masses between $0.14 < z < 0.45$, where the GAMA star formation rates are higher than predicted from DEVILS on its own. 

Our fits show clear variation in normalisation across the redshift range with the normalisation increasing to higher redshifts. 
To compare this evolution to previous results, we take the normalisation at $M_\star = 10^{10} M_\odot$, the high- and low-mass slopes, and the turn-over mass and compare to the results from \cite{DamenEvolutionSpecificStar2009,LeeEstimationStarFormation2010,DaviesGAMAHATLASmetaanalysis2016,LeslieVLACOSMOSGHzLarge2020} in Figure~\ref{fig:EvoSFMS}.
At low redshift ($z < 0.45$) we show the values derived from fitting only DEVILS as the open circles, and the values including GAMA as the filled circles. 
We select $M_\star = 10^{10} M_\odot$ as our normalization point, as it is well sampled in all redshift bins and is above the incompleteness limits in almost all redshift bins. 
Note that in some of the low redshift bins $M_\text{star} = 10^{10} M_\odot$ is above the turn-over mass, but we find that this does not impact our results. 
We also overplot measurements from the sSFR evolution from \cite{DamenEvolutionSpecificStar2009}, the low redshift measurements using GAMA from \cite{DaviesGAMAHATLASmetaanalysis2016} and other measurements from the COSMOS field from \cite{LeeTurnoverGalaxyMain2015} and \cite{LeslieVLACOSMOSGHzLarge2020}. 
We correct for the known 0.2\,dex offset between the stellar masses used in this work and those used by \cite{LeeTurnoverGalaxyMain2015} and \cite{LeslieVLACOSMOSGHzLarge2020} by measuring the normalisation from their work at $10^{9.8}M_\odot$. 
We also show the fit for the evolution of the main sequence from \cite{SpeagleHighlyConsistentFramework2014}, who use a detailed compilation of 25 different samples to evaluate the main sequence out to $z\sim6$. 
Our results are consistent in normalisation with previous work, but we do see some small differences in the low redshift bins.
When we include the measurements from GAMA we recover higher normalisations than previous work, but recover lower normalisations when we consider DEVILS on its own. 

Whilst the normalisation of the relation is easily compared with other works, the low-mass slope and turn-over mass are harder to compare due to differences in the parameterisation in each work. 
We compare our measurements of the low-mass slope and turn-over mass to the values from \cite{LeeTurnoverGalaxyMain2015} and \cite{LeslieVLACOSMOSGHzLarge2020} as their parameterisations only differ from ours in the treatment of the two slope parameters. 
\cite{LeslieVLACOSMOSGHzLarge2020} use a fixed low-mass slope with $\alpha = 1$ which we find is similar to our measurements at low redshift, but our low-mass slope decreases slightly with increasing redshift despite the prior centered at $\mu = 1$. 
We also find that the low-mass slope measurements from \cite{LeeTurnoverGalaxyMain2015} are steeper than our measurements in all redshift bins. 
\cite{LeeTurnoverGalaxyMain2015} and \cite{LeslieVLACOSMOSGHzLarge2020} both assume a fixed high-mass slope of zero but we include the variation in our values as a function of lookback time in Figure~\ref{fig:EvoSFMS}.

The parameterisation used by \cite{LeslieVLACOSMOSGHzLarge2020} assumes a linear evolution of turn-over mass with lookback time, and whilst we find that our turn-over masses do increase with lookback time, we recover lower turn-over masses across most of the redshift range used by \cite{LeslieVLACOSMOSGHzLarge2020}. 
Our recovered turn-over mass values in the highest redshift bins ($z > 3.25$) agree very closely with \cite{LeslieVLACOSMOSGHzLarge2020}, but these values are driven by the imposed priors as there are no data at these masses to constrain a turn-over and show no evidence of a turn-over in Figure~\ref{fig:SFMS_z}.
We expect that this is due to evolution of the main sequence where bending only occurs at low redshift. 
This is expected to happen as massive galaxies start to undergo quenching at $z<1.5$ more systematically than at higher redshift \citep{KatsianisEvolvingMassdependentssSFRM2019} and could be due to the growth of bulge components that contribute to the stellar mass but not to the SFR. 
This naturally leads to a bending in the main sequence at high masses. 
The main sequence using DEVILS will be explored further in Thorne et al. (in prep.).

\section{Conclusions}\label{sec:conclusion}
We have applied the \textsc{ProSpect} SED-fitting code to 494,000 galaxies between $0<z<9$ in the D10-COSMOS field of the DEVILS survey. 
Through the use of a parametric SFH and an evolving metallicity tied to the growth of stellar mass we have recovered stellar and dust mass estimates, SFRs, star formation and metallicity histories and the current gas phase metallicity for each galaxy.
In this work we focus on the stellar mass and SFR estimates, but include the dust mass and metallicity estimates in the D10\_ProSpectCat DMU. 
Discussion of metallicities will be deferred to Thorne et al. (in prep) and we stress that the dust masses obtained in this work are heavily dependent on the assumed model and dust temperature which are ill-constrained for a large number of our galaxies due to lack of FIR data.  

The results are summarised as follow:

\begin{itemize}
    \item In this work we obtain stellar masses, SFRs and dust masses for 494,000 objects between $0 < z < 9$ which will be made publicly available in future DEVILS data releases in the D10\_ProSpectCat DMU. 
    \item  We show comparisons of the stellar masses and SFRs obtained in this work to previous measurements from \cite{LaigleCOSMOS2015CATALOGEXPLORING2016} and \cite{GAMA_MAGPHYS}. Using \textsc{ProSpect} we obtain stellar masses that are 0.2 dex higher than previous measurements due to our physically motivated treatment of metallicity (Section~\ref{sec:ProSpect}).
    \item We use our new stellar mass measurements to measure the stellar mass function for $0.02 < z < 4.25$ (Section~\ref{sec:SMF}). 
    We find good agreement with previous measurements from \cite{DavidzonCOSMOS2015galaxystellar2017,WrightGAMAG10COSMOS3DHST2018,LejaNewCensusUniverse2020} and find no evidence of evolution in the break mass $M^*$ or the two $\alpha$ slope parameters (Figure~\ref{fig:SMF_Evolving}).
    We also find good agreement with previous measurements of the evolution of stellar mass density (Figure~\ref{fig:SMD}).
    \item We compare our stellar mass and SFR estimates to previous measurements using the SFR-$M_\star$ plane and evolution of the main sequence in Section~\ref{sec:SFMS}. We find good agreement with previous measurements from \cite{DamenEvolutionSpecificStar2009,SpeagleHighlyConsistentFramework2014,LeeTurnoverGalaxyMain2015,DaviesGAMAHATLASmetaanalysis2016,LeslieVLACOSMOSGHzLarge2020}. 
    By combining measurements from GAMA with our new DEVILS measurements we see evidence of bending at the high mass end at low redshift ($z<0.45$) which is not evident using GAMA or DEVILS alone.
    We also find no evidence of a turn-over in the mass range of our data at high redshift ($z > 2.6$) suggesting that the shape of the main sequence evolves with redshift, where bending only occurs at low redshift.
    The cause of this will be further explored in Thorne et al. (in prep.)
\end{itemize}

\section{Data Availability}\label{sec:DataAvailability}
The data products described in this paper are currently available for internal DEVILS team use for proprietary science in the D10\_ProSpectCat data management unit (DMU). 
This DMU will be made public with subsequent DEVILS data releases via data central.
The fit values to the galaxy stellar mass function and SFR-$M_\star$ main sequence are presented in Appendices~\ref{App:Vals} and are available at MNRAS online. 

\section*{Acknowledgements}
We thank the anonymous referee, whose comments improved the paper.
JET is supported by the Australian
Government Research Training Program (RTP) Scholarship.
ASGR and LJMD acknowledge support from the \textit{Australian Research Council's} Future Fellowship scheme (FT200100375 and FT200100055 respectively). 
SB and SPD acknowledge support from the \textit{Australian Research Council's} Discovery Project scheme (DP180103740). 
MS has been supported by the European Union's  Horizon 2020 research and innovation programme under the Maria Skłodowska-Curie (grant agreement No 754510), the National Science Centre of Poland (grant UMO-2016/23/N/ST9/02963) and by the Spanish Ministry of Science and Innovation through Juan de la Cierva-formacion program (reference FJC2018-038792-I).
AHW is supported by the European Research Council (Grant No. 770935).

DEVILS is an Australian project based around a spectroscopic campaign using the Anglo-Australian Telescope. The DEVILS input catalogue is generated from data taken as part of the ESO VISTA-VIDEO \citep{JarvisVISTADeepExtragalactic2013} and UltraVISTA \citep{McCrackenUltraVISTAnewultradeep2012} surveys. DEVILS is part funded via Discovery Programs by the Australian Research Council and the participating institutions. The DEVILS website is \url{https://devilsurvey.org}. The DEVILS data is hosted and provided by AAO Data Central (\url{https://datacentral.org.au/}).

GAMA is a joint European-Australasian project based around a spectroscopic campaign using the Anglo- Australian Telescope. The GAMA input catalogue is based on data taken from the Sloan Digital Sky Survey and the UKIRT Infrared Deep Sky Survey. Complementary imaging of the GAMA regions is being obtained by a number of in-dependent survey programmes including GALEX MIS, VST KiDS, VISTA VIKING, WISE, Herschel-ATLAS, GMRT and ASKAP providing UV to radio coverage. GAMA is funded by the STFC (UK), the ARC (Australia), the AAO, and the participating institutions. The GAMA website is \url{http://www.gama-survey.org/}.

This work was supported by resources provided by the Pawsey Supercomputing Centre with funding from the Australian Government and the Government of Western Australia. We gratefully acknowledge DUG Technology for their support and HPC services.

All of the work presented here was made possible by the free and open R software environment \citep{RCoreTeamLanguageEnvironmentStatistical2020}. All figures in this paper were made using the R \textsc{magicaxis} package \citep{RobothammagicaxisPrettyscientific2016}. This work also makes use of the \textsc{celestial} package \citep{RobothamCelestialCommonastronomical2016} and \textsc{dftools} \citep{ObreschkowdftoolsDistributionfunction2018}. 




\bibliographystyle{mnras}
\bibliography{MyBib} 

\begin{thebibliography}{}
\makeatletter
\relax
\def\mn@urlcharsother{\let\do\@makeother \do\$\do\&\do\#\do\^\do\_\do\%\do\~}
\def\mn@doi{\begingroup\mn@urlcharsother \@ifnextchar [ {\mn@doi@}
  {\mn@doi@[]}}
\def\mn@doi@[#1]#2{\def\@tempa{#1}\ifx\@tempa\@empty \href
  {http://dx.doi.org/#2} {doi:#2}\else \href {http://dx.doi.org/#2} {#1}\fi
  \endgroup}
\def\mn@eprint#1#2{\mn@eprint@#1:#2::\@nil}
\def\mn@eprint@arXiv#1{\href {http://arxiv.org/abs/#1} {{\tt arXiv:#1}}}
\def\mn@eprint@dblp#1{\href {http://dblp.uni-trier.de/rec/bibtex/#1.xml}
  {dblp:#1}}
\def\mn@eprint@#1:#2:#3:#4\@nil{\def\@tempa {#1}\def\@tempb {#2}\def\@tempc
  {#3}\ifx \@tempc \@empty \let \@tempc \@tempb \let \@tempb \@tempa \fi \ifx
  \@tempb \@empty \def\@tempb {arXiv}\fi \@ifundefined
  {mn@eprint@\@tempb}{\@tempb:\@tempc}{\expandafter \expandafter \csname
  mn@eprint@\@tempb\endcsname \expandafter{\@tempc}}}

\bibitem[\protect\citeauthoryear{Adams, Bowler, Jarvis, H{\"a}u{\ss}ler,
  McLure, Bunker, Dunlop  \& Verma}{Adams
  et~al.}{2020}]{AdamsrestframeUVluminosity2020}
Adams N.~J.,  Bowler R. A.~A.,  Jarvis M.~J.,  H{\"a}u{\ss}ler B.,  McLure
  R.~J.,  Bunker A.,  Dunlop J.~S.,   Verma A.,  2020, \mn@doi [MNRAS]
  {10.1093/mnras/staa687}, 494, 1771

\bibitem[\protect\citeauthoryear{Aihara et~al.,}{Aihara
  et~al.}{2019}]{AiharaSeconddatarelease2019}
Aihara H.,  et~al., 2019, \mn@doi [Publications of the Astronomical Society of
  Japan] {10.1093/pasj/psz103}, 71, 114

\bibitem[\protect\citeauthoryear{Alarcon et~al.,}{Alarcon
  et~al.}{2021}]{AlarconPAUSurveyimproved2021}
Alarcon A.,  et~al., 2021, \mn@doi [MNRAS] {10.1093/mnras/staa3659}, 501, 6103

\bibitem[\protect\citeauthoryear{Allard \& Hauschildt}{Allard \&
  Hauschildt}{1995}]{AllardModelatmospheressub1995}
Allard F.,  Hauschildt P.~H.,  1995, \mn@doi [ApJ] {10.1086/175708}, 445, 433

\bibitem[\protect\citeauthoryear{Alongi, Bertelli, Bressan, Chiosi, Fagotto,
  Greggio  \& Nasi}{Alongi
  et~al.}{1993}]{AlongiEvolutionarysequencesstellar1993}
Alongi M.,  Bertelli G.,  Bressan A.,  Chiosi C.,  Fagotto F.,  Greggio L.,
  Nasi E.,  1993, A\&AS, 97, 851

\bibitem[\protect\citeauthoryear{Andrews, Driver, Davies, Kafle, Robotham  \&
  Wright}{Andrews et~al.}{2017}]{AndrewsG10COSMOS382017}
Andrews S.~K.,  Driver S.~P.,  Davies L. J.~M.,  Kafle P.~R.,  Robotham A.
  S.~G.,   Wright A.~H.,  2017, \mn@doi [MNRAS] {10.1093/mnras/stw2395}, 464,
  1569

\bibitem[\protect\citeauthoryear{Andrews, Driver, Davies, Lagos  \&
  Robotham}{Andrews et~al.}{2018}]{AndrewsModellingcosmicspectral2018}
Andrews S.~K.,  Driver S.~P.,  Davies L. J.~M.,  Lagos C. d.~P.,   Robotham A.
  S.~G.,  2018, \mn@doi [MNRAS] {10.1093/mnras/stx2843}, 474, 898

\bibitem[\protect\citeauthoryear{Arnouts \& Ilbert}{Arnouts \&
  Ilbert}{2011}]{ArnoutsLePHAREPhotometricAnalysis2011}
Arnouts S.,  Ilbert O.,  2011, Astrophysics Source Code Library, p.
  ascl:1108.009

\bibitem[\protect\citeauthoryear{Arnouts et~al.,}{Arnouts
  et~al.}{2013}]{ArnoutsEncodinginfraredexcess2013}
Arnouts S.,  et~al., 2013, \mn@doi [A\&A] {10.1051/0004-6361/201321768}, 558,
  A67

\bibitem[\protect\citeauthoryear{Baldry, Glazebrook  \& Driver}{Baldry
  et~al.}{2008}]{Baldrygalaxystellarmass2008}
Baldry I.~K.,  Glazebrook K.,   Driver S.~P.,  2008, \mn@doi [MNRAS]
  {10.1111/j.1365-2966.2008.13348.x}, 388, 945

\bibitem[\protect\citeauthoryear{Baldry et~al.,}{Baldry
  et~al.}{2012}]{BaldryGalaxyMassAssembly2012}
Baldry I.~K.,  et~al., 2012, \mn@doi [MNRAS]
  {10.1111/j.1365-2966.2012.20340.x}, 421, 621

\bibitem[\protect\citeauthoryear{Baraffe, Chabrier, Allard  \&
  Hauschildt}{Baraffe et~al.}{1998}]{BaraffeEvolutionarymodelssolar1998}
Baraffe I.,  Chabrier G.,  Allard F.,   Hauschildt P.~H.,  1998, A\&A, 337, 403

\bibitem[\protect\citeauthoryear{Bell, McIntosh, Katz  \& Weinberg}{Bell
  et~al.}{2003}]{BellOpticalNearInfraredProperties2003}
Bell E.~F.,  McIntosh D.~H.,  Katz N.,   Weinberg M.~D.,  2003, \mn@doi [ApJS]
  {10.1086/378847}, 149, 289

\bibitem[\protect\citeauthoryear{Bellagamba, Maturi, Hamana, Meneghetti,
  Miyazaki  \& Moscardini}{Bellagamba
  et~al.}{2011}]{BellagambaOptimalfilteringoptical2011}
Bellagamba F.,  Maturi M.,  Hamana T.,  Meneghetti M.,  Miyazaki S.,
  Moscardini L.,  2011, \mn@doi [MNRAS] {10.1111/j.1365-2966.2011.18202.x},
  413, 1145

\bibitem[\protect\citeauthoryear{Bellstedt et~al.,}{Bellstedt
  et~al.}{2020a}]{BellstedtGalaxyMassAssembly2020a}
Bellstedt S.,  et~al., 2020a, \mn@doi [MNRAS] {10.1093/mnras/staa1466}, 496,
  3235

\bibitem[\protect\citeauthoryear{Bellstedt et~al.,}{Bellstedt
  et~al.}{2020b}]{BellstedtGalaxyMassAssembly2020b}
Bellstedt S.,  et~al., 2020b, \mn@doi [MNRAS] {10.1093/mnras/staa2620}, 498,
  5581

\bibitem[\protect\citeauthoryear{Bellstedt et~al.,}{Bellstedt
  et~al.}{2021}]{BellstedtGalaxymassassembly2021}
Bellstedt S.,  et~al., 2021, \mn@doi [MNRAS] {10.1093/mnras/stab550}, 503, 3309

\bibitem[\protect\citeauthoryear{Benaglia, Chauveau, Hunter  \& Young}{Benaglia
  et~al.}{2009}]{Benagliamixtoolspackageanalyzing2009}
Benaglia T.,  Chauveau D.,  Hunter D.~R.,   Young D.,  2009, J. Stat. Softw.,
  32, 1

\bibitem[\protect\citeauthoryear{Bertelli, Bressan, Chiosi, Fagotto  \&
  Nasi}{Bertelli et~al.}{1994}]{BertelliTheoreticalisochronesmodels1994}
Bertelli G.,  Bressan A.,  Chiosi C.,  Fagotto F.,   Nasi E.,  1994, A\&AS,
  106, 275

\bibitem[\protect\citeauthoryear{Bessell, Brett, Scholz  \& Wood}{Bessell
  et~al.}{1989}]{BessellColorsextendedstatic1989}
Bessell M.~S.,  Brett J.~M.,  Scholz M.,   Wood P.~R.,  1989, A\&AS, 77, 1

\bibitem[\protect\citeauthoryear{Bessell, Brett, Scholz  \& Wood}{Bessell
  et~al.}{1991}]{BessellColorsstratificationsextended1991}
Bessell M.~S.,  Brett J.~M.,  Scholz M.,   Wood P.~R.,  1991, A\&AS, 89, 335

\bibitem[\protect\citeauthoryear{Boquien, Burgarella, Roehlly, Buat, Ciesla,
  Corre, Inoue  \& Salas}{Boquien et~al.}{2019}]{BoquienCIGALEpythonCode2019}
Boquien M.,  Burgarella D.,  Roehlly Y.,  Buat V.,  Ciesla L.,  Corre D.,
  Inoue A.~K.,   Salas H.,  2019, \mn@doi [A\&A] {10.1051/0004-6361/201834156},
  622, A103

\bibitem[\protect\citeauthoryear{Bressan, Fagotto, Bertelli  \& Chiosi}{Bressan
  et~al.}{1993}]{BressanEvolutionarysequencesstellar1993}
Bressan A.,  Fagotto F.,  Bertelli G.,   Chiosi C.,  1993, A\&AS, 100, 647

\bibitem[\protect\citeauthoryear{Bressan, Marigo, Girardi, Salasnich, Dal~Cero,
  Rubele  \& Nanni}{Bressan et~al.}{2012}]{BressanPARSECstellartracks2012}
Bressan A.,  Marigo P.,  Girardi L.,  Salasnich B.,  Dal~Cero C.,  Rubele S.,
  Nanni A.,  2012, \mn@doi [MNRAS] {10.1111/j.1365-2966.2012.21948.x}, 427, 127

\bibitem[\protect\citeauthoryear{Brinchmann, Charlot, White, Tremonti,
  Kauffmann, Heckman  \& Brinkmann}{Brinchmann
  et~al.}{2004}]{Brinchmannphysicalpropertiesstarforming2004}
Brinchmann J.,  Charlot S.,  White S. D.~M.,  Tremonti C.,  Kauffmann G.,
  Heckman T.,   Brinkmann J.,  2004, \mn@doi [MNRAS]
  {10.1111/j.1365-2966.2004.07881.x}, 351, 1151

\bibitem[\protect\citeauthoryear{Bruzual \& Charlot}{Bruzual \&
  Charlot}{2003}]{BC03}
Bruzual G.,  Charlot S.,  2003, \mn@doi [MNRAS]
  {10.1046/j.1365-8711.2003.06897.x}, 344, 1000

\bibitem[\protect\citeauthoryear{Calzetti, Armus, Bohlin, Kinney, Koornneef  \&
  {Storchi-Bergmann}}{Calzetti et~al.}{2000}]{CalzettiDustContentOpacity2000}
Calzetti D.,  Armus L.,  Bohlin R.~C.,  Kinney A.~L.,  Koornneef J.,
  {Storchi-Bergmann} T.,  2000, \mn@doi [ApJ] {10.1086/308692}, 533, 682

\bibitem[\protect\citeauthoryear{Capak et~al.,}{Capak
  et~al.}{2007}]{CapakFirstReleaseCOSMOS2007}
Capak P.,  et~al., 2007, \mn@doi [ApJS] {10.1086/519081}, 172, 99

\bibitem[\protect\citeauthoryear{Cardelli, Clayton  \& Mathis}{Cardelli
  et~al.}{1989}]{Cardellirelationshipinfraredoptical1989}
Cardelli J.~A.,  Clayton G.~C.,   Mathis J.~S.,  1989, \mn@doi [ApJ]
  {10.1086/167900}, 345, 245

\bibitem[\protect\citeauthoryear{Carnall, McLure, Dunlop  \& Dav{\'e}}{Carnall
  et~al.}{2018}]{CarnallInferringstarformation2018}
Carnall A.~C.,  McLure R.~J.,  Dunlop J.~S.,   Dav{\'e} R.,  2018, \mn@doi
  [MNRAS] {10.1093/mnras/sty2169}, 480, 4379

\bibitem[\protect\citeauthoryear{Carnall, Leja, Johnson, McLure, Dunlop  \&
  Conroy}{Carnall et~al.}{2019}]{CarnallHowMeasureGalaxy2019}
Carnall A.~C.,  Leja J.,  Johnson B.~D.,  McLure R.~J.,  Dunlop J.~S.,   Conroy
  C.,  2019, \mn@doi [ApJ] {10.3847/1538-4357/ab04a2}, 873, 44

\bibitem[\protect\citeauthoryear{Casey}{Casey}{2012}]{CaseyFarinfraredspectralenergy2012}
Casey C.~M.,  2012, \mn@doi [MNRAS] {10.1111/j.1365-2966.2012.21455.x}, 425,
  3094

\bibitem[\protect\citeauthoryear{Cassisi, {degl'Innocenti}  \& Salaris}{Cassisi
  et~al.}{1997a}]{Cassisieffectdiffusionred1997}
Cassisi S.,  {degl'Innocenti} S.,   Salaris M.,  1997a, \mn@doi [MNRAS]
  {10.1093/mnras/290.3.515}, 290, 515

\bibitem[\protect\citeauthoryear{Cassisi, Castellani  \& Castellani}{Cassisi
  et~al.}{1997b}]{CassisiIntermediateagemetaldeficient1997}
Cassisi S.,  Castellani M.,   Castellani V.,  1997b, A\&A, 317, 108

\bibitem[\protect\citeauthoryear{Cassisi, Castellani, Ciarcelluti, Piotto  \&
  Zoccali}{Cassisi et~al.}{2000}]{CassisiGalacticglobularclusters2000}
Cassisi S.,  Castellani V.,  Ciarcelluti P.,  Piotto G.,   Zoccali M.,  2000,
  \mn@doi [MNRAS] {10.1046/j.1365-8711.2000.03457.x}, 315, 679

\bibitem[\protect\citeauthoryear{Cenarro, Cardiel, Gorgas, Peletier, Vazdekis
  \& Prada}{Cenarro et~al.}{2001}]{CenarroEmpiricalcalibrationnearinfrared2001}
Cenarro A.~J.,  Cardiel N.,  Gorgas J.,  Peletier R.~F.,  Vazdekis A.,   Prada
  F.,  2001, \mn@doi [MNRAS] {10.1046/j.1365-8711.2001.04688.x}, 326, 959

\bibitem[\protect\citeauthoryear{Chabrier}{Chabrier}{2003}]{ChabrierGalacticStellarSubstellar2003}
Chabrier G.,  2003, \mn@doi [PASP] {10.1086/376392}, 115, 763

\bibitem[\protect\citeauthoryear{Charbonnel, Meynet, Maeder  \&
  Schaerer}{Charbonnel et~al.}{1996}]{CharbonnelGridsstellarmodels1996}
Charbonnel C.,  Meynet G.,  Maeder A.,   Schaerer D.,  1996, A\&AS, 115, 339

\bibitem[\protect\citeauthoryear{Charbonnel, D{\"a}ppen, Schaerer, Bernasconi,
  Maeder, Meynet  \& Mowlavi}{Charbonnel
  et~al.}{1999}]{CharbonnelGridsstellarmodels1999}
Charbonnel C.,  D{\"a}ppen W.,  Schaerer D.,  Bernasconi P.~A.,  Maeder A.,
  Meynet G.,   Mowlavi N.,  1999, \mn@doi [A\&AS] {10.1051/aas:1999454}, 135,
  405

\bibitem[\protect\citeauthoryear{Charlot \& Fall}{Charlot \&
  Fall}{2000}]{CharlotSimpleModelAbsorption2000}
Charlot S.,  Fall S.~M.,  2000, \mn@doi [ApJ] {10.1086/309250}, 539, 718

\bibitem[\protect\citeauthoryear{Chevallard \& Charlot}{Chevallard \&
  Charlot}{2016}]{ChevallardModellinginterpretingspectral2016}
Chevallard J.,  Charlot S.,  2016, \mn@doi [MNRAS] {10.1093/mnras/stw1756},
  462, 1415

\bibitem[\protect\citeauthoryear{Choi, Dotter, Conroy, Cantiello, Paxton  \&
  Johnson}{Choi et~al.}{2016}]{ChoiMesaIsochronesStellar2016}
Choi J.,  Dotter A.,  Conroy C.,  Cantiello M.,  Paxton B.,   Johnson B.~D.,
  2016, \mn@doi [ApJ] {10.3847/0004-637X/823/2/102}, 823, 102

\bibitem[\protect\citeauthoryear{Cid~Fernandes, Mateus, Sodr{\'e},
  Stasi{\'n}ska  \& Gomes}{Cid~Fernandes
  et~al.}{2005}]{CidFernandesSemiempiricalanalysisSloan2005}
Cid~Fernandes R.,  Mateus A.,  Sodr{\'e} L.,  Stasi{\'n}ska G.,   Gomes J.~M.,
  2005, \mn@doi [MNRAS] {10.1111/j.1365-2966.2005.08752.x}, 358, 363

\bibitem[\protect\citeauthoryear{Comparat et~al.,}{Comparat
  et~al.}{2015}]{Comparat65evolutionbright2015}
Comparat J.,  et~al., 2015, \mn@doi [A\&A] {10.1051/0004-6361/201424767}, 575,
  A40

\bibitem[\protect\citeauthoryear{Conroy}{Conroy}{2013}]{ConroyModelingPanchromaticSpectral2013}
Conroy C.,  2013, \mn@doi [ARA\&A] {10.1146/annurev-astro-082812-141017}, 51,
  393

\bibitem[\protect\citeauthoryear{Conroy, Gunn  \& White}{Conroy
  et~al.}{2009}]{ConroyPropagationUncertaintiesStellar2009}
Conroy C.,  Gunn J.~E.,   White M.,  2009, \mn@doi [ApJ]
  {10.1088/0004-637X/699/1/486}, 699, 486

\bibitem[\protect\citeauthoryear{Cool et~al.,}{Cool
  et~al.}{2013}]{CoolPRIsmMUltiobjectSurvey2013}
Cool R.~J.,  et~al., 2013, \mn@doi [ApJ] {10.1088/0004-637X/767/2/118}, 767,
  118

\bibitem[\protect\citeauthoryear{Crain et~al.,}{Crain
  et~al.}{2015}]{CrainEAGLEsimulationsgalaxy2015}
Crain R.~A.,  et~al., 2015, \mn@doi [MNRAS] {10.1093/mnras/stv725}, 450, 1937

\bibitem[\protect\citeauthoryear{Da~Cunha, Charlot  \& Elbaz}{Da~Cunha
  et~al.}{2008}]{MAGPHYS}
Da~Cunha E.,  Charlot S.,   Elbaz D.,  2008, \mn@doi [MNRAS]
  {10.1111/j.1365-2966.2008.13535.x}, 388, 1595

\bibitem[\protect\citeauthoryear{Daddi et~al.,}{Daddi
  et~al.}{2007}]{DaddiMultiwavelengthStudyMassive2007}
Daddi E.,  et~al., 2007, \mn@doi [ApJ] {10.1086/521818}, 670, 156

\bibitem[\protect\citeauthoryear{Dale, Helou, Magdis, Armus, {D{\'i}az-Santos}
  \& Shi}{Dale et~al.}{2014}]{DaleTwoParameterModelInfrared2014}
Dale D.~A.,  Helou G.,  Magdis G.~E.,  Armus L.,  {D{\'i}az-Santos} T.,   Shi
  Y.,  2014, \mn@doi [ApJ] {10.1088/0004-637X/784/1/83}, 784, 83

\bibitem[\protect\citeauthoryear{Damen, Labb{\'e}, Franx, {van Dokkum}, Taylor
  \& Gawiser}{Damen et~al.}{2009}]{DamenEvolutionSpecificStar2009}
Damen M.,  Labb{\'e} I.,  Franx M.,  {van Dokkum} P.~G.,  Taylor E.~N.,
  Gawiser E.~J.,  2009, \mn@doi [ApJ] {10.1088/0004-637X/690/1/937}, 690, 937

\bibitem[\protect\citeauthoryear{Damjanov, Zahid, Geller, Fabricant  \&
  Hwang}{Damjanov et~al.}{2018}]{DamjanovhCOSMOSDenseSpectroscopic2018}
Damjanov I.,  Zahid H.~J.,  Geller M.~J.,  Fabricant D.~G.,   Hwang H.~S.,
  2018, \mn@doi [ApJS] {10.3847/1538-4365/aaa01c}, 234, 21

\bibitem[\protect\citeauthoryear{Davidzon et~al.,}{Davidzon
  et~al.}{2017}]{DavidzonCOSMOS2015galaxystellar2017}
Davidzon I.,  et~al., 2017, \mn@doi [A\&A] {10.1051/0004-6361/201730419}, 605,
  A70

\bibitem[\protect\citeauthoryear{Davies et~al.,}{Davies
  et~al.}{2015}]{DaviesGalaxyMassAssembly2015}
Davies L. J.~M.,  et~al., 2015, \mn@doi [MNRAS] {10.1093/mnras/stu2515}, 447,
  1014

\bibitem[\protect\citeauthoryear{Davies et~al.,}{Davies
  et~al.}{2016}]{DaviesGAMAHATLASmetaanalysis2016}
Davies L. J.~M.,  et~al., 2016, \mn@doi [MNRAS] {10.1093/mnras/stw1342}, 461,
  458

\bibitem[\protect\citeauthoryear{Davies et~al.,}{Davies
  et~al.}{2018}]{DaviesDeepExtragalacticVIsible2018}
Davies L. J.~M.,  et~al., 2018, \mn@doi [MNRAS] {10.1093/mnras/sty1553}, 480,
  768

\bibitem[\protect\citeauthoryear{Davies et~al.,}{Davies
  et~al.}{2019}]{DaviesStarFormingMainSequence2019}
Davies L. J.~M.,  et~al., 2019, \mn@doi [MNRAS] {10.1093/mnras/sty2957}, 483,
  1881

\bibitem[\protect\citeauthoryear{Dotter}{Dotter}{2016}]{DotterMESAIsochronesStellar2016}
Dotter A.,  2016, \mn@doi [ApJS] {10.3847/0067-0049/222/1/8}, 222, 8

\bibitem[\protect\citeauthoryear{Draine}{Draine}{2003}]{DraineInterstellarDustGrains2003}
Draine B.~T.,  2003, \mn@doi [ARA\&A] {10.1146/annurev.astro.41.011802.094840},
  41, 241

\bibitem[\protect\citeauthoryear{Draine \& Li}{Draine \&
  Li}{2007}]{DraineInfraredEmissionInterstellar2007}
Draine B.~T.,  Li A.,  2007, \mn@doi [ApJ] {10.1086/511055}, 657, 810

\bibitem[\protect\citeauthoryear{Draine et~al.,}{Draine
  et~al.}{2007}]{DraineDustMassesPAH2007}
Draine B.~T.,  et~al., 2007, \mn@doi [ApJ] {10.1086/518306}, 663, 866

\bibitem[\protect\citeauthoryear{Driver et~al.,}{Driver
  et~al.}{2018}]{GAMA_MAGPHYS}
Driver S.~P.,  et~al., 2018, \mn@doi [MNRAS] {10.1093/mnras/stx2728}, 475, 2891

\bibitem[\protect\citeauthoryear{Eggleton}{Eggleton}{1971}]{Eggletonevolutionlowmass1971}
Eggleton P.~P.,  1971, \mn@doi [MNRAS] {10.1093/mnras/151.3.351}, 151, 351

\bibitem[\protect\citeauthoryear{Elbaz et~al.,}{Elbaz
  et~al.}{2007}]{Elbazreversalstarformationdensity2007}
Elbaz D.,  et~al., 2007, \mn@doi [A\&A] {10.1051/0004-6361:20077525}, 468, 33

\bibitem[\protect\citeauthoryear{Eldridge \& Stanway}{Eldridge \&
  Stanway}{2009}]{EldridgeSpectralpopulationsynthesis2009}
Eldridge J.~J.,  Stanway E.~R.,  2009, \mn@doi [MNRAS]
  {10.1111/j.1365-2966.2009.15514.x}, 400, 1019

\bibitem[\protect\citeauthoryear{Eldridge \& Stanway}{Eldridge \&
  Stanway}{2019}]{EldridgePopulationspectralsynthesis2019}
Eldridge J.~J.,  Stanway E.~R.,  2019, \mn@doi [ArXiv200511883 Astro-Ph]
  {10.1017/9781108553070.023}, pp 307--328

\bibitem[\protect\citeauthoryear{Eldridge \& Tout}{Eldridge \&
  Tout}{2004}]{Eldridgeprogenitorscorecollapsesupernovae2004}
Eldridge J.~J.,  Tout C.~A.,  2004, \mn@doi [MNRAS]
  {10.1111/j.1365-2966.2004.08041.x}, 353, 87

\bibitem[\protect\citeauthoryear{Eldridge, Stanway, Xiao, McClelland, Taylor,
  Ng, Greis  \& Bray}{Eldridge
  et~al.}{2017}]{EldridgeBinaryPopulationSpectral2017}
Eldridge J.~J.,  Stanway E.~R.,  Xiao L.,  McClelland L. A.~S.,  Taylor G.,  Ng
  M.,  Greis S. M.~L.,   Bray J.~C.,  2017, \mn@doi [PASA]
  {10.1017/pasa.2017.51}, 34, e058

\bibitem[\protect\citeauthoryear{Fagotto, Bressan, Bertelli  \& Chiosi}{Fagotto
  et~al.}{1994a}]{FagottoEvolutionarysequencesstellar1994a}
Fagotto F.,  Bressan A.,  Bertelli G.,   Chiosi C.,  1994a, A\&AS, 104, 365

\bibitem[\protect\citeauthoryear{Fagotto, Bressan, Bertelli  \& Chiosi}{Fagotto
  et~al.}{1994b}]{FagottoEvolutionarysequencesstellar1994}
Fagotto F.,  Bressan A.,  Bertelli G.,   Chiosi C.,  1994b, A\&AS, 105, 29

\bibitem[\protect\citeauthoryear{Fanelli, O'Connell, Burstein  \& Wu}{Fanelli
  et~al.}{1992}]{FanelliSpectralsynthesisultraviolet1992}
Fanelli M.~N.,  O'Connell R.~W.,  Burstein D.,   Wu C.-C.,  1992, \mn@doi
  [ApJS] {10.1086/191714}, 82, 197

\bibitem[\protect\citeauthoryear{Ferland, Korista, Verner, Ferguson, Kingdon
  \& Verner}{Ferland et~al.}{1998}]{FerlandCLOUDY90Numerical1998}
Ferland G.~J.,  Korista K.~T.,  Verner D.~A.,  Ferguson J.~W.,  Kingdon J.~B.,
   Verner E.~M.,  1998, \mn@doi [PASP] {10.1086/316190}, 110, 761

\bibitem[\protect\citeauthoryear{Ferland et~al.,}{Ferland
  et~al.}{2013}]{Ferland2013ReleaseCloudy2013}
Ferland G.~J.,  et~al., 2013, Revista Mexicana de Astronomia y Astrofisica, 49,
  137

\bibitem[\protect\citeauthoryear{Finoguenov et~al.,}{Finoguenov
  et~al.}{2007}]{FinoguenovXMMNewtonWideFieldSurvey2007}
Finoguenov A.,  et~al., 2007, \mn@doi [ApJS] {10.1086/516577}, 172, 182

\bibitem[\protect\citeauthoryear{Fluks, Plez, The, {de Winter}, Westerlund  \&
  Steenman}{Fluks et~al.}{1994}]{FluksspectraphotometryMgiant1994}
Fluks M.~A.,  Plez B.,  The P.~S.,  {de Winter} D.,  Westerlund B.~E.,
  Steenman H.~C.,  1994, A\&AS, 105, 311

\bibitem[\protect\citeauthoryear{Fritz, Franceschini  \& Hatziminaoglou}{Fritz
  et~al.}{2006}]{FritzRevisitinginfraredspectra2006}
Fritz J.,  Franceschini A.,   Hatziminaoglou E.,  2006, \mn@doi [MNRAS]
  {10.1111/j.1365-2966.2006.09866.x}, 366, 767

\bibitem[\protect\citeauthoryear{Furlong et~al.,}{Furlong
  et~al.}{2015}]{FurlongEvolutiongalaxystellar2015}
Furlong M.,  et~al., 2015, \mn@doi [MNRAS] {10.1093/mnras/stv852}, 450, 4486

\bibitem[\protect\citeauthoryear{Girardi, Bressan, Chiosi, Bertelli  \&
  Nasi}{Girardi et~al.}{1996}]{GirardiEvolutionarysequencesstellar1996}
Girardi L.,  Bressan A.,  Chiosi C.,  Bertelli G.,   Nasi E.,  1996, A\&AS,
  117, 113

\bibitem[\protect\citeauthoryear{Girardi, Bressan, Bertelli  \& Chiosi}{Girardi
  et~al.}{2000}]{GirardiEvolutionarytracksisochrones2000}
Girardi L.,  Bressan A.,  Bertelli G.,   Chiosi C.,  2000, \mn@doi [A\&AS]
  {10.1051/aas:2000126}, 141, 371

\bibitem[\protect\citeauthoryear{Gr{\"a}fener, Owocki  \& Vink}{Gr{\"a}fener
  et~al.}{2012}]{GrafenerStellarenvelopeinflation2012}
Gr{\"a}fener G.,  Owocki S.~P.,   Vink J.~S.,  2012, \mn@doi [A\&A]
  {10.1051/0004-6361/201117497}, 538, A40

\bibitem[\protect\citeauthoryear{Gregg et~al.,}{Gregg
  et~al.}{2006}]{2006hstc.conf..209G}
Gregg M.~D.,  et~al., 2006, in Koekemoer A.~M.,  Goudfrooij P.,   Dressel
  L.~L.,  eds, The 2005 {{HST}} Calibration Workshop: {{Hubble}} after the
  Transition to Two-Gyro Mode. p.~209

\bibitem[\protect\citeauthoryear{Guo, Zheng, Wang  \& Fu}{Guo
  et~al.}{2015}]{GuoStarFormationMain2015}
Guo K.,  Zheng X.~Z.,  Wang T.,   Fu H.,  2015, \mn@doi [ApJL]
  {10.1088/2041-8205/808/2/L49}, 808, L49

\bibitem[\protect\citeauthoryear{Han \& Han}{Han \&
  Han}{2012}]{HanDecodingSpectralEnergy2012}
Han Y.,  Han Z.,  2012, \mn@doi [ApJ] {10.1088/0004-637X/749/2/123}, 749, 123

\bibitem[\protect\citeauthoryear{Han \& Han}{Han \&
  Han}{2014}]{HanBayeSEDGeneralApproach2014}
Han Y.,  Han Z.,  2014, \mn@doi [ApJS] {10.1088/0067-0049/215/1/2}, 215, 2

\bibitem[\protect\citeauthoryear{Han \& Han}{Han \&
  Han}{2019}]{HanComprehensiveBayesianDiscrimination2019}
Han Y.,  Han Z.,  2019, \mn@doi [ApJS] {10.3847/1538-4365/aaeffa}, 240, 3

\bibitem[\protect\citeauthoryear{Hasinger et~al.,}{Hasinger
  et~al.}{2018}]{HasingerDEIMOS10KSpectroscopic2018}
Hasinger G.,  et~al., 2018, \mn@doi [ApJ] {10.3847/1538-4357/aabacf}, 858, 77

\bibitem[\protect\citeauthoryear{Hsieh, Wang, Hsieh, Lin, Yan, Lim  \&
  Ho}{Hsieh et~al.}{2012}]{HsiehTaiwanECDFSNearInfrared2012}
Hsieh B.-C.,  Wang W.-H.,  Hsieh C.-C.,  Lin L.,  Yan H.,  Lim J.,   Ho P.
  T.~P.,  2012, \mn@doi [ApJS] {10.1088/0067-0049/203/2/23}, 203, 23

\bibitem[\protect\citeauthoryear{Ilbert et~al.,}{Ilbert
  et~al.}{2009}]{IlbertCosmosPhotometricRedshifts2009}
Ilbert O.,  et~al., 2009, \mn@doi [ApJ] {10.1088/0004-637X/690/2/1236}, 690,
  1236

\bibitem[\protect\citeauthoryear{Ivanov, Coccato, Neeser, Selman, Pizzella,
  Dalla~Bont{\`a}, Corsini  \& Morelli}{Ivanov
  et~al.}{2019}]{IvanovMUSElibrarystellar2019}
Ivanov V.~D.,  Coccato L.,  Neeser M.~J.,  Selman F.,  Pizzella A.,
  Dalla~Bont{\`a} E.,  Corsini E.~M.,   Morelli L.,  2019, \mn@doi [A\&A]
  {10.1051/0004-6361/201936178}, 629, A100

\bibitem[\protect\citeauthoryear{Jarvis et~al.,}{Jarvis
  et~al.}{2013}]{JarvisVISTADeepExtragalactic2013}
Jarvis M.~J.,  et~al., 2013, \mn@doi [MNRAS] {10.1093/mnras/sts118}, 428, 1281

\bibitem[\protect\citeauthoryear{Johnson, Leja, Conroy  \& Speagle}{Johnson
  et~al.}{2020}]{JohnsonStellarPopulationInference2020}
Johnson B.~D.,  Leja J.,  Conroy C.,   Speagle J.~S.,  2020, arXiv e-prints,
  2012, arXiv:2012.01426

\bibitem[\protect\citeauthoryear{Kashino et~al.,}{Kashino
  et~al.}{2019}]{KashinoFMOSCOSMOSSurveyStarforming2019}
Kashino D.,  et~al., 2019, \mn@doi [ApJS] {10.3847/1538-4365/ab06c4}, 241, 10

\bibitem[\protect\citeauthoryear{Katsianis et~al.,}{Katsianis
  et~al.}{2019}]{KatsianisEvolvingMassdependentssSFRM2019}
Katsianis A.,  et~al., 2019, \mn@doi [ApJ] {10.3847/1538-4357/ab1f8d}, 879, 11

\bibitem[\protect\citeauthoryear{Kennicutt \& Evans}{Kennicutt \&
  Evans}{2012}]{KennicuttStarFormationMilky2012}
Kennicutt R.~C.,  Evans N.~J.,  2012, \mn@doi [ARA\&A]
  {10.1146/annurev-astro-081811-125610}, 50, 531

\bibitem[\protect\citeauthoryear{Kriek et~al.,}{Kriek
  et~al.}{2015}]{KriekMOSFIREDeepEvolution2015}
Kriek M.,  et~al., 2015, \mn@doi [ApJS] {10.1088/0067-0049/218/2/15}, 218, 15

\bibitem[\protect\citeauthoryear{Kroupa}{Kroupa}{2001}]{Kroupavariationinitialmass2001}
Kroupa P.,  2001, \mn@doi [MNRAS] {10.1046/j.1365-8711.2001.04022.x}, 322, 231

\bibitem[\protect\citeauthoryear{Kroupa \& Boily}{Kroupa \&
  Boily}{2002}]{Kroupamassfunctionstar2002}
Kroupa P.,  Boily C.~M.,  2002, \mn@doi [MNRAS]
  {10.1046/j.1365-8711.2002.05848.x}, 336, 1188

\bibitem[\protect\citeauthoryear{Kurucz}{Kurucz}{1992}]{1992IAUS..149..225K}
Kurucz R.~L.,  1992, in Barbuy B.,  Renzini A.,  eds, ~{{IAUS}} Vol. 149, The
  Stellar Populations of Galaxies. p.~225

\bibitem[\protect\citeauthoryear{Lacey et~al.,}{Lacey
  et~al.}{2016}]{Laceyunifiedmultiwavelengthmodel2016}
Lacey C.~G.,  et~al., 2016, \mn@doi [MNRAS] {10.1093/mnras/stw1888}, 462, 3854

\bibitem[\protect\citeauthoryear{Lagos, Tobar, Robotham, Obreschkow, Mitchell,
  Power  \& Elahi}{Lagos et~al.}{2018}]{LagosSharkintroducingopen2018}
Lagos C. d.~P.,  Tobar R.~J.,  Robotham A. S.~G.,  Obreschkow D.,  Mitchell
  P.~D.,  Power C.,   Elahi P.~J.,  2018, \mn@doi [MNRAS]
  {10.1093/mnras/sty2440}, 481, 3573

\bibitem[\protect\citeauthoryear{Laigle et~al.,}{Laigle
  et~al.}{2016}]{LaigleCOSMOS2015CATALOGEXPLORING2016}
Laigle C.,  et~al., 2016, \mn@doi [ApJS] {10.3847/0067-0049/224/2/24}, 224, 24

\bibitem[\protect\citeauthoryear{Lan{\c c}on \& Mouhcine}{Lan{\c c}on \&
  Mouhcine}{2002}]{Lanconmodellingintermediateagestellar2002}
Lan{\c c}on A.,  Mouhcine M.,  2002, \mn@doi [A\&A]
  {10.1051/0004-6361:20020585}, 393, 167

\bibitem[\protect\citeauthoryear{{Lara-L{\'o}pez} et~al.,}{{Lara-L{\'o}pez}
  et~al.}{2013}]{Lara-LopezGalaxyMassAssembly2013}
{Lara-L{\'o}pez} M.~A.,  et~al., 2013, \mn@doi [MNRAS] {10.1093/mnras/stt1031},
  434, 451

\bibitem[\protect\citeauthoryear{Le~Bertre}{Le~Bertre}{1997}]{LeBertreOpticalinfraredobservations1997}
Le~Bertre T.,  1997, A\&A, 324, 1059

\bibitem[\protect\citeauthoryear{Le~Borgne et~al.,}{Le~Borgne
  et~al.}{2003}]{LeBorgneSTELIBlibrarystellar2003}
Le~Borgne J.-F.,  et~al., 2003, \mn@doi [A\&A] {10.1051/0004-6361:20030243},
  402, 433

\bibitem[\protect\citeauthoryear{Le~F{\`e}vre et~al.,}{Le~F{\`e}vre
  et~al.}{2013}]{LeFevreVIMOSVLTDeep2013}
Le~F{\`e}vre O.,  et~al., 2013, \mn@doi [A\&A] {10.1051/0004-6361/201322179},
  559, A14

\bibitem[\protect\citeauthoryear{Le~F{\`e}vre et~al.,}{Le~F{\`e}vre
  et~al.}{2015}]{LeFevreVIMOSUltraDeepSurvey2015}
Le~F{\`e}vre O.,  et~al., 2015, \mn@doi [A\&A] {10.1051/0004-6361/201423829},
  576, A79

\bibitem[\protect\citeauthoryear{Le~Floc'h et~al.,}{Le~Floc'h
  et~al.}{2009}]{LeFlochDeepSpitzer242009}
Le~Floc'h E.,  et~al., 2009, \mn@doi [ApJ] {10.1088/0004-637X/703/1/222}, 703,
  222

\bibitem[\protect\citeauthoryear{Le~Sidaner \& Le~Bertre}{Le~Sidaner \&
  Le~Bertre}{1996}]{LeSidanerOpticalinfraredobservations1996}
Le~Sidaner P.,  Le~Bertre T.,  1996, A\&A, 314, 896

\bibitem[\protect\citeauthoryear{Lee, Idzi, Ferguson, Somerville, Wiklind  \&
  Giavalisco}{Lee et~al.}{2009}]{LeeBiasesUncertaintiesPhysical2009}
Lee S.-K.,  Idzi R.,  Ferguson H.~C.,  Somerville R.~S.,  Wiklind T.,
  Giavalisco M.,  2009, \mn@doi [ApJS] {10.1088/0067-0049/184/1/100}, 184, 100

\bibitem[\protect\citeauthoryear{Lee, Ferguson, Somerville, Wiklind  \&
  Giavalisco}{Lee et~al.}{2010}]{LeeEstimationStarFormation2010}
Lee S.-K.,  Ferguson H.~C.,  Somerville R.~S.,  Wiklind T.,   Giavalisco M.,
  2010, \mn@doi [ApJ] {10.1088/0004-637X/725/2/1644}, 725, 1644

\bibitem[\protect\citeauthoryear{Lee et~al.,}{Lee
  et~al.}{2015}]{LeeTurnoverGalaxyMain2015}
Lee N.,  et~al., 2015, \mn@doi [ApJ] {10.1088/0004-637X/801/2/80}, 801, 80

\bibitem[\protect\citeauthoryear{Lee et~al.,}{Lee
  et~al.}{2018}]{LeeFirstDataRelease2018}
Lee K.-G.,  et~al., 2018, \mn@doi [ApJS] {10.3847/1538-4365/aace58}, 237, 31

\bibitem[\protect\citeauthoryear{Leja, Johnson, Conroy, {van Dokkum}  \&
  Byler}{Leja et~al.}{2017}]{Prospector}
Leja J.,  Johnson B.~D.,  Conroy C.,  {van Dokkum} P.~G.,   Byler N.,  2017,
  \mn@doi [ApJ] {10.3847/1538-4357/aa5ffe}, 837, 170

\bibitem[\protect\citeauthoryear{Leja, Carnall, Johnson, Conroy  \&
  Speagle}{Leja et~al.}{2019}]{LejaHowMeasureGalaxy2019}
Leja J.,  Carnall A.~C.,  Johnson B.~D.,  Conroy C.,   Speagle J.~S.,  2019,
  \mn@doi [ApJ] {10.3847/1538-4357/ab133c}, 876, 3

\bibitem[\protect\citeauthoryear{Leja, Speagle, Johnson, Conroy, {van Dokkum}
  \& Franx}{Leja et~al.}{2020}]{LejaNewCensusUniverse2020}
Leja J.,  Speagle J.~S.,  Johnson B.~D.,  Conroy C.,  {van Dokkum} P.,   Franx
  M.,  2020, \mn@doi [ApJ] {10.3847/1538-4357/ab7e27}, 893, 111

\bibitem[\protect\citeauthoryear{Leslie et~al.,}{Leslie
  et~al.}{2020}]{LeslieVLACOSMOSGHzLarge2020}
Leslie S.~K.,  et~al., 2020, \mn@doi [ApJ] {10.3847/1538-4357/aba044}, 899, 58

\bibitem[\protect\citeauthoryear{Levesque, Kewley  \& Larson}{Levesque
  et~al.}{2010}]{LevesqueTheoreticalModelingStarForming2010}
Levesque E.~M.,  Kewley L.~J.,   Larson K.~L.,  2010, \mn@doi [AJ]
  {10.1088/0004-6256/139/2/712}, 139, 712

\bibitem[\protect\citeauthoryear{Lilly et~al.,}{Lilly
  et~al.}{2009}]{LillyzCOSMOS10kBrightSpectroscopic2009}
Lilly S.~J.,  et~al., 2009, \mn@doi [ApJS] {10.1088/0067-0049/184/2/218}, 184,
  218

\bibitem[\protect\citeauthoryear{Lower, Narayanan, Leja, Johnson, Conroy  \&
  Dav{\'e}}{Lower et~al.}{2020}]{LowerHowWellCan2020}
Lower S.,  Narayanan D.,  Leja J.,  Johnson B.~D.,  Conroy C.,   Dav{\'e} R.,
  2020, \mn@doi [ApJ] {10.3847/1538-4357/abbfa7}, 904, 33

\bibitem[\protect\citeauthoryear{Lutz et~al.,}{Lutz
  et~al.}{2011}]{LutzPACSEvolutionaryProbe2011}
Lutz D.,  et~al., 2011, \mn@doi [A\&A] {10.1051/0004-6361/201117107}, 532, A90

\bibitem[\protect\citeauthoryear{Madau \& Dickinson}{Madau \&
  Dickinson}{2014}]{MadauCosmicStarFormationHistory2014}
Madau P.,  Dickinson M.,  2014, \mn@doi [ARA\&A]
  {10.1146/annurev-astro-081811-125615}, 52, 415

\bibitem[\protect\citeauthoryear{Maiolino \& Mannucci}{Maiolino \&
  Mannucci}{2019}]{Maiolinoremetallicacosmic2019}
Maiolino R.,  Mannucci F.,  2019, \mn@doi [A\&AR] {10.1007/s00159-018-0112-2},
  27, 3

\bibitem[\protect\citeauthoryear{Maraston}{Maraston}{2005}]{MarastonEvolutionarypopulationsynthesis2005}
Maraston C.,  2005, \mn@doi [MNRAS] {10.1111/j.1365-2966.2005.09270.x}, 362,
  799

\bibitem[\protect\citeauthoryear{Maraston, Pforr, Renzini, Daddi, Dickinson,
  Cimatti  \& Tonini}{Maraston et~al.}{2010}]{MarastonStarformationrates2010}
Maraston C.,  Pforr J.,  Renzini A.,  Daddi E.,  Dickinson M.,  Cimatti A.,
  Tonini C.,  2010, \mn@doi [MNRAS] {10.1111/j.1365-2966.2010.16973.x}, 407,
  830

\bibitem[\protect\citeauthoryear{Marchesini, {van Dokkum},
  F{\"o}rster~Schreiber, Franx, Labb{\'e}  \& Wuyts}{Marchesini
  et~al.}{2009}]{MarchesiniEvolutionStellarMass2009}
Marchesini D.,  {van Dokkum} P.~G.,  F{\"o}rster~Schreiber N.~M.,  Franx M.,
  Labb{\'e} I.,   Wuyts S.,  2009, \mn@doi [ApJ]
  {10.1088/0004-637X/701/2/1765}, 701, 1765

\bibitem[\protect\citeauthoryear{Masters, Stern, Cohen, Capak, Rhodes,
  Castander  \& Paltani}{Masters
  et~al.}{2017}]{MastersCompleteCalibrationColorRedshift2017}
Masters D.~C.,  Stern D.~K.,  Cohen J.~G.,  Capak P.~L.,  Rhodes J.~D.,
  Castander F.~J.,   Paltani S.,  2017, \mn@doi [ApJ]
  {10.3847/1538-4357/aa6f08}, 841, 111

\bibitem[\protect\citeauthoryear{Masters et~al.,}{Masters
  et~al.}{2019}]{MastersCompleteCalibrationColorRedshift2019}
Masters D.~C.,  et~al., 2019, \mn@doi [ApJ] {10.3847/1538-4357/ab184d}, 877, 81

\bibitem[\protect\citeauthoryear{McCracken et~al.,}{McCracken
  et~al.}{2012}]{McCrackenUltraVISTAnewultradeep2012}
McCracken H.~J.,  et~al., 2012, \mn@doi [A\&A] {10.1051/0004-6361/201219507},
  544, A156

\bibitem[\protect\citeauthoryear{Momcheva et~al.,}{Momcheva
  et~al.}{2016}]{Momcheva3DHSTSurveyHubble2016}
Momcheva I.~G.,  et~al., 2016, \mn@doi [ApJS] {10.3847/0067-0049/225/2/27},
  225, 27

\bibitem[\protect\citeauthoryear{Muzzin et~al.,}{Muzzin
  et~al.}{2013}]{MuzzinEvolutionStellarMass2013}
Muzzin A.,  et~al., 2013, \mn@doi [ApJ] {10.1088/0004-637X/777/1/18}, 777, 18

\bibitem[\protect\citeauthoryear{Nagamine, Fukugita, Cen  \& Ostriker}{Nagamine
  et~al.}{2001}]{NagamineStarFormationHistory2001}
Nagamine K.,  Fukugita M.,  Cen R.,   Ostriker J.~P.,  2001, \mn@doi [ApJ]
  {10.1086/322293}, 558, 497

\bibitem[\protect\citeauthoryear{Nenkova, Sirocky, Nikutta, Ivezi{\'c}  \&
  Elitzur}{Nenkova et~al.}{2008}]{NenkovaAGNDustyTori2008}
Nenkova M.,  Sirocky M.~M.,  Nikutta R.,  Ivezi{\'c} {\v Z}.,   Elitzur M.,
  2008, \mn@doi [ApJ] {10.1086/590483}, 685, 160

\bibitem[\protect\citeauthoryear{Noeske et~al.,}{Noeske
  et~al.}{2007}]{NoeskeStarFormationAEGIS2007}
Noeske K.~G.,  et~al., 2007, \mn@doi [ApJL] {10.1086/517926}, 660, L43

\bibitem[\protect\citeauthoryear{Noll, Burgarella, Giovannoli, Buat, Marcillac
  \& {Mu{\~n}oz-Mateos}}{Noll et~al.}{2009}]{CIGALE}
Noll S.,  Burgarella D.,  Giovannoli E.,  Buat V.,  Marcillac D.,
  {Mu{\~n}oz-Mateos} J.~C.,  2009, \mn@doi [A \& A]
  {10.1051/0004-6361/200912497}, 507, 1793

\bibitem[\protect\citeauthoryear{Nomoto, Kobayashi  \& Tominaga}{Nomoto
  et~al.}{2013}]{NomotoNucleosynthesisStarsChemical2013}
Nomoto K.,  Kobayashi C.,   Tominaga N.,  2013, \mn@doi [ARA\&A]
  {10.1146/annurev-astro-082812-140956}, 51, 457

\bibitem[\protect\citeauthoryear{Obreschkow}{Obreschkow}{2018}]{ObreschkowdftoolsDistributionfunction2018}
Obreschkow D.,  2018, Astrophysics Source Code Library, p. ascl:1805.002

\bibitem[\protect\citeauthoryear{Obreschkow, Murray, Robotham  \&
  Westmeier}{Obreschkow et~al.}{2018}]{ObreschkowEddingtondemoninferring2018}
Obreschkow D.,  Murray S.~G.,  Robotham A. S.~G.,   Westmeier T.,  2018,
  \mn@doi [MNRAS] {10.1093/mnras/stx3155}, 474, 5500

\bibitem[\protect\citeauthoryear{Oesch et~al.,}{Oesch
  et~al.}{2016}]{OeschREMARKABLYLUMINOUSGALAXY2016}
Oesch P.~A.,  et~al., 2016, \mn@doi [ApJ] {10.3847/0004-637X/819/2/129}, 819,
  129

\bibitem[\protect\citeauthoryear{Oliver et~al.,}{Oliver
  et~al.}{2012}]{OliverHerschelMultitieredExtragalactic2012}
Oliver S.~J.,  et~al., 2012, \mn@doi [MNRAS]
  {10.1111/j.1365-2966.2012.20912.x}, 424, 1614

\bibitem[\protect\citeauthoryear{Papovich, Dickinson  \& Ferguson}{Papovich
  et~al.}{2001}]{PapovichStellarPopulationsEvolution2001}
Papovich C.,  Dickinson M.,   Ferguson H.~C.,  2001, \mn@doi [ApJ]
  {10.1086/322412}, 559, 620

\bibitem[\protect\citeauthoryear{{Paulino-Afonso}, Sobral, Darvish, Ribeiro,
  Stroe, Best, Afonso  \& Matsuda}{{Paulino-Afonso}
  et~al.}{2018}]{Paulino-AfonsoVIS3COSSurveyoverview2018}
{Paulino-Afonso} A.,  Sobral D.,  Darvish B.,  Ribeiro B.,  Stroe A.,  Best P.,
   Afonso J.,   Matsuda Y.,  2018, \mn@doi [A\&A]
  {10.1051/0004-6361/201832688}, 620, A186

\bibitem[\protect\citeauthoryear{Paxton, Bildsten, Dotter, Herwig, Lesaffre  \&
  Timmes}{Paxton et~al.}{2011}]{PaxtonModulesExperimentsStellar2011}
Paxton B.,  Bildsten L.,  Dotter A.,  Herwig F.,  Lesaffre P.,   Timmes F.,
  2011, \mn@doi [ApJS] {10.1088/0067-0049/192/1/3}, 192, 3

\bibitem[\protect\citeauthoryear{Paxton et~al.,}{Paxton
  et~al.}{2013}]{PaxtonModulesExperimentsStellar2013}
Paxton B.,  et~al., 2013, \mn@doi [ApJS] {10.1088/0067-0049/208/1/4}, 208, 4

\bibitem[\protect\citeauthoryear{Paxton et~al.,}{Paxton
  et~al.}{2015}]{PaxtonModulesExperimentsStellar2015}
Paxton B.,  et~al., 2015, \mn@doi [ApJS] {10.1088/0067-0049/220/1/15}, 220, 15

\bibitem[\protect\citeauthoryear{Pearson et~al.,}{Pearson
  et~al.}{2018}]{PearsonMainsequencestar2018}
Pearson W.~J.,  et~al., 2018, \mn@doi [A\&A] {10.1051/0004-6361/201832821},
  615, A146

\bibitem[\protect\citeauthoryear{Pei \& Fall}{Pei \&
  Fall}{1995}]{PeiCosmicChemicalEvolution1995}
Pei Y.~C.,  Fall S.~M.,  1995, \mn@doi [ApJ] {10.1086/176466}, 454, 69

\bibitem[\protect\citeauthoryear{Pforr, Maraston  \& Tonini}{Pforr
  et~al.}{2012}]{PforrRecoveringgalaxystellar2012}
Pforr J.,  Maraston C.,   Tonini C.,  2012, \mn@doi [MNRAS]
  {10.1111/j.1365-2966.2012.20848.x}, 422, 3285

\bibitem[\protect\citeauthoryear{Pickles}{Pickles}{1998}]{PicklesStellarSpectralFlux1998}
Pickles A.~J.,  1998, \mn@doi [PASP] {10.1086/316197}, 110, 863

\bibitem[\protect\citeauthoryear{Pietrinferni, Cassisi, Salaris  \&
  Castelli}{Pietrinferni et~al.}{2004}]{PietrinferniLargeStellarEvolution2004}
Pietrinferni A.,  Cassisi S.,  Salaris M.,   Castelli F.,  2004, \mn@doi [ApJ]
  {10.1086/422498}, 612, 168

\bibitem[\protect\citeauthoryear{Pietrinferni, Cassisi, Salaris  \&
  Hidalgo}{Pietrinferni et~al.}{2013}]{PietrinferniBaSTIStellarEvolution2013}
Pietrinferni A.,  Cassisi S.,  Salaris M.,   Hidalgo S.,  2013, \mn@doi [A\&A]
  {10.1051/0004-6361/201321950}, 558, A46

\bibitem[\protect\citeauthoryear{Pillepich et~al.,}{Pillepich
  et~al.}{2018}]{PillepichFirstresultsIllustrisTNG2018}
Pillepich A.,  et~al., 2018, \mn@doi [MNRAS] {10.1093/mnras/stx3112}, 475, 648

\bibitem[\protect\citeauthoryear{{Planck Collaboration} et~al.,}{{Planck
  Collaboration} et~al.}{2016}]{PlanckCollaborationPlanck2015results2016}
{Planck Collaboration} et~al., 2016, \mn@doi [A\&A]
  {10.1051/0004-6361/201525830}, 594, A13

\bibitem[\protect\citeauthoryear{Pols, Tout, Eggleton  \& Han}{Pols
  et~al.}{1995}]{PolsApproximateinputphysics1995}
Pols O.~R.,  Tout C.~A.,  Eggleton P.~P.,   Han Z.,  1995, \mn@doi [MNRAS]
  {10.1093/mnras/274.3.964}, 274, 964

\bibitem[\protect\citeauthoryear{{R Core Team}}{{R Core
  Team}}{2020}]{RCoreTeamLanguageEnvironmentStatistical2020}
{R Core Team} 2020, R: {{A Language}} and {{Environment}} for {{Statistical}}
  {{Computing}}, R Foundation for Statistical Computing

\bibitem[\protect\citeauthoryear{Rauch}{Rauch}{2002}]{2002RMxAC..12..150R}
Rauch T.,  2002, in Henney W.~J.,  Franco J.,   Martos M.,  eds,  Revista
  Mexicana de Astronomia y Astrofisica Conference Series Vol. 12, Revista
  Mexicana de Astronomia y Astrofisica Conference Series. pp 150--151

\bibitem[\protect\citeauthoryear{Robotham}{Robotham}{2016a}]{RobothamCelestialCommonastronomical2016}
Robotham A. S.~G.,  2016a, Astrophysics Source Code Library, p. ascl:1602.011

\bibitem[\protect\citeauthoryear{Robotham}{Robotham}{2016b}]{RobothammagicaxisPrettyscientific2016}
Robotham A. S.~G.,  2016b, Astrophysics Source Code Library, p. ascl:1604.004

\bibitem[\protect\citeauthoryear{Robotham, Davies, Driver, Koushan, Taranu,
  Casura  \& Liske}{Robotham
  et~al.}{2018}]{RobothamProFoundSourceExtraction2018}
Robotham A. S.~G.,  Davies L. J.~M.,  Driver S.~P.,  Koushan S.,  Taranu D.~S.,
   Casura S.,   Liske J.,  2018, \mn@doi [MNRAS] {10.1093/mnras/sty440}, 476,
  3137

\bibitem[\protect\citeauthoryear{Robotham, Bellstedt, Lagos, Thorne, Davies,
  Driver  \& Bravo}{Robotham
  et~al.}{2020}]{RobothamProSpectgeneratingspectral2020}
Robotham A. S.~G.,  Bellstedt S.,  Lagos C. d.~P.,  Thorne J.~E.,  Davies
  L.~J.,  Driver S.~P.,   Bravo M.,  2020, \mn@doi [MNRAS]
  {10.1093/mnras/staa1116}, 495, 905

\bibitem[\protect\citeauthoryear{R{\"o}ck, Vazdekis, Peletier, Knapen  \&
  {Falc{\'o}n-Barroso}}{R{\"o}ck
  et~al.}{2015}]{RockStellarpopulationsynthesis2015}
R{\"o}ck B.,  Vazdekis A.,  Peletier R.~F.,  Knapen J.~H.,
  {Falc{\'o}n-Barroso} J.,  2015, \mn@doi [MNRAS] {10.1093/mnras/stv503}, 449,
  2853

\bibitem[\protect\citeauthoryear{R{\"o}ck, Vazdekis, Ricciardelli, Peletier,
  Knapen  \& {Falc{\'o}n-Barroso}}{R{\"o}ck
  et~al.}{2016}]{RockMILESextendedStellar2016}
R{\"o}ck B.,  Vazdekis A.,  Ricciardelli E.,  Peletier R.~F.,  Knapen J.~H.,
  {Falc{\'o}n-Barroso} J.,  2016, \mn@doi [A\&A] {10.1051/0004-6361/201527570},
  589, A73

\bibitem[\protect\citeauthoryear{Salim et~al.,}{Salim
  et~al.}{2007}]{SalimUVStarFormation2007}
Salim S.,  et~al., 2007, \mn@doi [ApJS] {10.1086/519218}, 173, 267

\bibitem[\protect\citeauthoryear{Salim, Boquien  \& Lee}{Salim
  et~al.}{2018}]{SalimDustAttenuationCurves2018}
Salim S.,  Boquien M.,   Lee J.~C.,  2018, \mn@doi [ApJ]
  {10.3847/1538-4357/aabf3c}, 859, 11

\bibitem[\protect\citeauthoryear{Salpeter}{Salpeter}{1955}]{SalpeterLuminosityFunctionStellar1955}
Salpeter E.~E.,  1955, \mn@doi [ApJ] {10.1086/145971}, 121, 161

\bibitem[\protect\citeauthoryear{{S{\'a}nchez-Bl{\'a}zquez}
  et~al.,}{{S{\'a}nchez-Bl{\'a}zquez}
  et~al.}{2006}]{Sanchez-BlazquezMediumresolutionIsaacNewton2006}
{S{\'a}nchez-Bl{\'a}zquez} P.,  et~al., 2006, \mn@doi [MNRAS]
  {10.1111/j.1365-2966.2006.10699.x}, 371, 703

\bibitem[\protect\citeauthoryear{Sanders et~al.,}{Sanders
  et~al.}{2007}]{SandersSCOSMOSSpitzerLegacy2007}
Sanders D.~B.,  et~al., 2007, \mn@doi [ApJSS] {10.1086/517885}, 172, 86

\bibitem[\protect\citeauthoryear{Schaller, Schaerer, Meynet  \&
  Maeder}{Schaller et~al.}{1992}]{SchallerNewgridsstellar1992}
Schaller G.,  Schaerer D.,  Meynet G.,   Maeder A.,  1992, A\&AS, 96, 269

\bibitem[\protect\citeauthoryear{Schaye et~al.,}{Schaye
  et~al.}{2015}]{SchayeEAGLEprojectsimulating2015}
Schaye J.,  et~al., 2015, \mn@doi [MNRAS] {10.1093/mnras/stu2058}, 446, 521

\bibitem[\protect\citeauthoryear{Schechter}{Schechter}{1976}]{Schechteranalyticexpressionluminosity1976}
Schechter P.,  1976, \mn@doi [ApJ] {10.1086/154079}, 203, 297

\bibitem[\protect\citeauthoryear{Schoenberner}{Schoenberner}{1983}]{SchoenbernerLatestagesstellar1983}
Schoenberner D.,  1983, \mn@doi [ApJ] {10.1086/161333}, 272, 708

\bibitem[\protect\citeauthoryear{Schreiber et~al.,}{Schreiber
  et~al.}{2015}]{SchreiberHerschelviewdominant2015}
Schreiber C.,  et~al., 2015, \mn@doi [A\&A] {10.1051/0004-6361/201425017}, 575,
  A74

\bibitem[\protect\citeauthoryear{Skelton et~al.,}{Skelton
  et~al.}{2014}]{Skelton3DHSTWFC3selectedPhotometric2014}
Skelton R.~E.,  et~al., 2014, \mn@doi [ApJS] {10.1088/0067-0049/214/2/24}, 214,
  24

\bibitem[\protect\citeauthoryear{Smethurst et~al.,}{Smethurst
  et~al.}{2015}]{SmethurstGalaxyZooevidence2015}
Smethurst R.~J.,  et~al., 2015, \mn@doi [MNRAS] {10.1093/mnras/stv161}, 450,
  435

\bibitem[\protect\citeauthoryear{Smith, Norris  \& Crowther}{Smith
  et~al.}{2002}]{SmithRealisticionizingfluxes2002}
Smith L.~J.,  Norris R. P.~F.,   Crowther P.~A.,  2002, \mn@doi [MNRAS]
  {10.1046/j.1365-8711.2002.06042.x}, 337, 1309

\bibitem[\protect\citeauthoryear{Somerville \& Primack}{Somerville \&
  Primack}{1999}]{SomervilleSemianalyticmodellinggalaxy1999}
Somerville R.~S.,  Primack J.~R.,  1999, \mn@doi [MNRAS]
  {10.1046/j.1365-8711.1999.03032.x}, 310, 1087

\bibitem[\protect\citeauthoryear{Speagle, Steinhardt, Capak  \&
  Silverman}{Speagle et~al.}{2014}]{SpeagleHighlyConsistentFramework2014}
Speagle J.~S.,  Steinhardt C.~L.,  Capak P.~L.,   Silverman J.~D.,  2014,
  \mn@doi [ApJS] {10.1088/0067-0049/214/2/15}, 214, 15

\bibitem[\protect\citeauthoryear{Straatman et~al.,}{Straatman
  et~al.}{2018}]{StraatmanLargeEarlyGalaxy2018}
Straatman C. M.~S.,  et~al., 2018, \mn@doi [ApJS] {10.3847/1538-4365/aae37a},
  239, 27

\bibitem[\protect\citeauthoryear{Taylor et~al.,}{Taylor
  et~al.}{2015}]{TaylorGalaxyMassAssembly2015}
Taylor E.~N.,  et~al., 2015, \mn@doi [MNRAS] {10.1093/mnras/stu1900}, 446, 2144

\bibitem[\protect\citeauthoryear{Tomczak et~al.,}{Tomczak
  et~al.}{2014}]{TomczakGalaxyStellarMass2014}
Tomczak A.~R.,  et~al., 2014, \mn@doi [ApJ] {10.1088/0004-637X/783/2/85}, 783,
  85

\bibitem[\protect\citeauthoryear{Tomczak et~al.,}{Tomczak
  et~al.}{2016}]{TomczakSFRMRelationEmpirical2016}
Tomczak A.~R.,  et~al., 2016, \mn@doi [ApJ] {10.3847/0004-637X/817/2/118}, 817,
  118

\bibitem[\protect\citeauthoryear{Trayford, Lagos, Robotham  \&
  Obreschkow}{Trayford et~al.}{2020}]{TrayfordFadegreysystematic2020}
Trayford J.~W.,  Lagos C. d.~P.,  Robotham A. S.~G.,   Obreschkow D.,  2020,
  \mn@doi [MNRAS] {10.1093/mnras/stz3234}, 491, 3937

\bibitem[\protect\citeauthoryear{Valdes, Gupta, Rose, Singh  \& Bell}{Valdes
  et~al.}{2004}]{ValdesIndoUSLibraryCoude2004}
Valdes F.,  Gupta R.,  Rose J.~A.,  Singh H.~P.,   Bell D.~J.,  2004, \mn@doi
  [ApJS] {10.1086/386343}, 152, 251

\bibitem[\protect\citeauthoryear{Vassiliadis \& Wood}{Vassiliadis \&
  Wood}{1993}]{VassiliadisEvolutionlowintermediatemass1993}
Vassiliadis E.,  Wood P.~R.,  1993, \mn@doi [ApJ] {10.1086/173033}, 413, 641

\bibitem[\protect\citeauthoryear{Vassiliadis \& Wood}{Vassiliadis \&
  Wood}{1994}]{VassiliadisPostasymptoticgiantbranch1994}
Vassiliadis E.,  Wood P.~R.,  1994, \mn@doi [ApJS] {10.1086/191962}, 92, 125

\bibitem[\protect\citeauthoryear{Vazdekis, Koleva, Ricciardelli, R{\"o}ck  \&
  {Falc{\'o}n-Barroso}}{Vazdekis
  et~al.}{2016}]{VazdekisUVextendedEMILESstellar2016}
Vazdekis A.,  Koleva M.,  Ricciardelli E.,  R{\"o}ck B.,   {Falc{\'o}n-Barroso}
  J.,  2016, \mn@doi [MNRAS] {10.1093/mnras/stw2231}, 463, 3409

\bibitem[\protect\citeauthoryear{Walcher, Groves, Budavari  \& Dale}{Walcher
  et~al.}{2011}]{WalcherFittingintegratedSpectral2011}
Walcher C.~J.,  Groves B.,  Budavari T.,   Dale D.,  2011, \mn@doi [Ap\&SS]
  {10.1007/s10509-010-0458-z}, 331, 1

\bibitem[\protect\citeauthoryear{Whitaker, {van Dokkum}, Brammer  \&
  Franx}{Whitaker et~al.}{2012}]{WhitakerStarFormationMass2012}
Whitaker K.~E.,  {van Dokkum} P.~G.,  Brammer G.,   Franx M.,  2012, \mn@doi
  [ApJL] {10.1088/2041-8205/754/2/L29}, 754, L29

\bibitem[\protect\citeauthoryear{Whitaker et~al.,}{Whitaker
  et~al.}{2014}]{WhitakerConstrainingLowmassSlope2014}
Whitaker K.~E.,  et~al., 2014, \mn@doi [ApJ] {10.1088/0004-637X/795/2/104},
  795, 104

\bibitem[\protect\citeauthoryear{Winget, Hansen, Liebert, {van Horn}, Fontaine,
  Nather, Kepler  \& Lamb}{Winget
  et~al.}{1987}]{Wingetindependentmethoddetermining1987}
Winget D.~E.,  Hansen C.~J.,  Liebert J.,  {van Horn} H.~M.,  Fontaine G.,
  Nather R.~E.,  Kepler S.~O.,   Lamb D.~Q.,  1987, \mn@doi [ApJL]
  {10.1086/184864}, 315, L77

\bibitem[\protect\citeauthoryear{Worthey}{Worthey}{1994}]{WortheyComprehensivestellarpopulation1994}
Worthey G.,  1994, \mn@doi [ApJS] {10.1086/192096}, 95, 107

\bibitem[\protect\citeauthoryear{Wright et~al.,}{Wright
  et~al.}{2016}]{WrightGalaxyMassAssembly2016}
Wright A.~H.,  et~al., 2016, \mn@doi [MNRAS] {10.1093/mnras/stw832}, 460, 765

\bibitem[\protect\citeauthoryear{Wright et~al.,}{Wright
  et~al.}{2017}]{WrightGalaxyMassAssembly2017}
Wright A.~H.,  et~al., 2017, \mn@doi [MNRAS] {10.1093/mnras/stx1149}, 470, 283

\bibitem[\protect\citeauthoryear{Wright, Driver  \& Robotham}{Wright
  et~al.}{2018}]{WrightGAMAG10COSMOS3DHST2018}
Wright A.~H.,  Driver S.~P.,   Robotham A. S.~G.,  2018, \mn@doi [MNRAS]
  {10.1093/mnras/sty2136}, 480, 3491

\bibitem[\protect\citeauthoryear{Wuyts et~al.,}{Wuyts
  et~al.}{2011}]{WuytsGalaxyStructureMode2011}
Wuyts S.,  et~al., 2011, \mn@doi [ApJ] {10.1088/0004-637X/742/2/96}, 742, 96

\bibitem[\protect\citeauthoryear{Yang et~al.,}{Yang
  et~al.}{2020}]{YangxcigalefittingAGN2020}
Yang G.,  et~al., 2020, \mn@doi [MNRAS] {10.1093/mnras/stz3001}, 491, 740

\bibitem[\protect\citeauthoryear{Zamojski et~al.,}{Zamojski
  et~al.}{2007}]{ZamojskiDeepGALEXImaging2007}
Zamojski M.~A.,  et~al., 2007, \mn@doi [ApJSS] {10.1086/516593}, 172, 468

\bibitem[\protect\citeauthoryear{{van Dokkum}}{{van
  Dokkum}}{2008}]{vanDokkumEvidenceCosmicEvolution2008}
{van Dokkum} P.~G.,  2008, \mn@doi [ApJ] {10.1086/525014}, 674, 29

\makeatother
\end{thebibliography}




\appendix
\section{Tabular form of Figure 1}

\definecolor{Gray}{gray}{0.9}
\definecolor{ps}{RGB}{238,34,12}
\definecolor{magphys}{RGB}{254,174,0}
\definecolor{beagle}{RGB}{190,168,2}
\definecolor{BayeSED}{RGB}{222,105,165}
\definecolor{BAGPIPES}{RGB}{0,118,186}
\definecolor{Prospector}{RGB}{29,177,0}
\definecolor{CIGALE}{RGB}{151,14,83}

\begin{table*}
    \centering
    \caption{The tabular form of Figure~\ref{fig:SED_Diagram}. The SED fitting codes are each represented by a column and a cell is shaded if that SED fitting code uses that particular template. 
    BPASS is \citet{EldridgeSpectralpopulationsynthesis2009,EldridgeBinaryPopulationSpectral2017}, M05 is \citet{MarastonEvolutionarypopulationsynthesis2005}, E-MILES is \citet{VazdekisUVextendedEMILESstellar2016}, BC03 is \citet{BC03}, and FSPS is \citet{ConroyPropagationUncertaintiesStellar2009}.
    The MAPPINGS-III tables are presented in \citet{LevesqueTheoreticalModelingStarForming2010} and CLOUDY is described in \citet{FerlandCLOUDY90Numerical1998,Ferland2013ReleaseCloudy2013}.}
    \label{tab:SED_Diagram}
    \begin{tabular}{ c  c  c  c  c  c  c  c  c  }
    \cline{3-9}
    \multicolumn{1}{l}{}  &  & \multicolumn{7}{c}{SED Fitting Code} \\
    \hline
	Template Type	&	Template	&	\textsc{ProSpect}	&	\textsc{magphys}	&	\textsc{beagle}	&	\textsc{prospector}	&	\textsc{bagpipes}	&	\textsc{cigale}	&	\textsc{BayeSED}	\\
\hline	
\multirow{5}{*}{Stellar Templates}	&	BPASS	&	\cellcolor{ps}	&		&		&		&		&		&		\\
	&	M05	&		&		&		&		&		&	\cellcolor{CIGALE}	&	\cellcolor{BayeSED}	\\
	&	E-MILES	&	\cellcolor{ps}	&		&		&		&		&		&		\\
	&	BC03	&	\cellcolor{ps}	&	\cellcolor{magphys}	&	\cellcolor{beagle}	&		&	\cellcolor{BAGPIPES}	&	\cellcolor{CIGALE}	&	\cellcolor{BayeSED}	\\
	&	FSPS	&		&		&		&	\cellcolor{Prospector}	&		&		&		\\
\hline																	
\multirow{4}{*}{Dust Attenuation}	&	\citet{Cardellirelationshipinfraredoptical1989}	&		&		&		&	\cellcolor{Prospector}	&	\cellcolor{BAGPIPES}	&		&		\\
	&	\citet{CharlotSimpleModelAbsorption2000}	&	\cellcolor{ps}	&	\cellcolor{magphys}	&	\cellcolor{beagle}	&	\cellcolor{Prospector}	&	\cellcolor{BAGPIPES}	&	\cellcolor{CIGALE}	&		\\
	&	\citet{SalimDustAttenuationCurves2018}	&		&		&		&		&	\cellcolor{BAGPIPES}	&		&		\\
	&	\citet{CalzettiDustContentOpacity2000}	&		&		&	\cellcolor{beagle}	&	\cellcolor{Prospector}	&	\cellcolor{BAGPIPES}	&	\cellcolor{CIGALE}	&	\cellcolor{BayeSED}	\\
\hline																	
\multirow{4}{*}{Dust Emission}	&	\citet{DaleTwoParameterModelInfrared2014}	&	\cellcolor{ps}	&		&	\cellcolor{beagle}	&		&		&	\cellcolor{CIGALE}	&		\\
	&	\citet{DraineInfraredEmissionInterstellar2007}	&		&		&		&	\cellcolor{Prospector}	&	\cellcolor{BAGPIPES}	&	\cellcolor{CIGALE}	&		\\
	&	\citet{CaseyFarinfraredspectralenergy2012}	&		&		&		&		&		&	\cellcolor{CIGALE}	&		\\
	&	Grey-body	&		&	\cellcolor{magphys}	&		&		&		&		&		\\
\hline																	
\multirow{4}{*}{Initial Mass Function}	&	\citet{ChabrierGalacticStellarSubstellar2003}	&	\cellcolor{ps}	&	\cellcolor{magphys}	&	\cellcolor{beagle}	&	\cellcolor{Prospector}	&		&	\cellcolor{CIGALE}	&	\cellcolor{BayeSED}	\\
	&	\citet{SalpeterLuminosityFunctionStellar1955}	&		&		&		&	\cellcolor{Prospector}	&		&	\cellcolor{CIGALE}	&	\cellcolor{BayeSED}	\\
	&	\citet{Kroupavariationinitialmass2001}	&		&		&		&	\cellcolor{Prospector}	&		&	\cellcolor{CIGALE}	&	\cellcolor{BayeSED}	\\
	&	\citet{Kroupamassfunctionstar2002}	&		&		&		&		&	\cellcolor{BAGPIPES}	&		&		\\
\hline																	
\multirow{2}{*}{Emission Lines}	&	MAPPINGS-III	&	\cellcolor{ps}	&		&		&		&		&		&		\\
	&	CLOUDY &		&		&	\cellcolor{beagle}	&	\cellcolor{Prospector}	&	\cellcolor{BAGPIPES}	&	\cellcolor{CIGALE}	&		\\
\hline																	
\multirow{2}{*}{Star Formation Histories}	&	Parametric	&	\cellcolor{ps}	&	\cellcolor{magphys}	&	\cellcolor{beagle}	&	\cellcolor{Prospector}	&	\cellcolor{BAGPIPES}	&	\cellcolor{CIGALE}	&	\cellcolor{BayeSED}	\\
	&	Non-parametric	&	\cellcolor{ps}	&		&	\cellcolor{beagle}	&	\cellcolor{Prospector}	&	\cellcolor{BAGPIPES}	&		&		\\
\hline																	
\multirow{2}{*}{Metallicity}	&	Constant but free	&	\cellcolor{ps}	&	\cellcolor{magphys}	&		&	\cellcolor{Prospector}	&		&	\cellcolor{CIGALE}	&	\cellcolor{BayeSED}	\\
	&	Evolving	&	\cellcolor{ps}	&		&	\cellcolor{beagle}	&	\cellcolor{Prospector}	&		&		&		\\
\hline																	
\multirow{5}{*}{AGN Templates}	&	\citet{FritzRevisitinginfraredspectra2006}	&	\cellcolor{ps}	&		&		&		&		&	\cellcolor{CIGALE}	&		\\
	&	\citet{AndrewsModellingcosmicspectral2018}	&	\cellcolor{ps}	&		&		&		&		&		&		\\
	&	\citet{CaseyFarinfraredspectralenergy2012}	&		&		&		&		&		&	\cellcolor{CIGALE}	&		\\
	&	\citet{DaleTwoParameterModelInfrared2014}	&	\cellcolor{ps}	&		&		&		&		&	\cellcolor{CIGALE}	&		\\
	&	\citet{NenkovaAGNDustyTori2008}	&		&		&		&	\cellcolor{Prospector}	&		&		&		\\
\hline																	
\multirow{3}{*}{Other}	&	Radio Extension	&	\cellcolor{ps}	&		&		&		&		&	\cellcolor{CIGALE}	&		\\
	&	Spectral Fitting 	&	\cellcolor{ps}	&		&	\cellcolor{beagle}	&	\cellcolor{Prospector}	&	\cellcolor{BAGPIPES}	&		&		\\
	&	X-ray Extension	&		&		&		&		&		&	\cellcolor{CIGALE}	&		\\
	\hline
    \end{tabular}
\end{table*}

\section{Stellar Template Schematic}
Here we show the stellar template counterpart to Figure~\ref{fig:SED_Diagram} and Table~\ref{tab:SED_Diagram}. Interactive versions of both diagrams are available at \url{https://jethorne.github.io/}.

\begin{figure*}
    \centering
    \includegraphics[width = \linewidth]{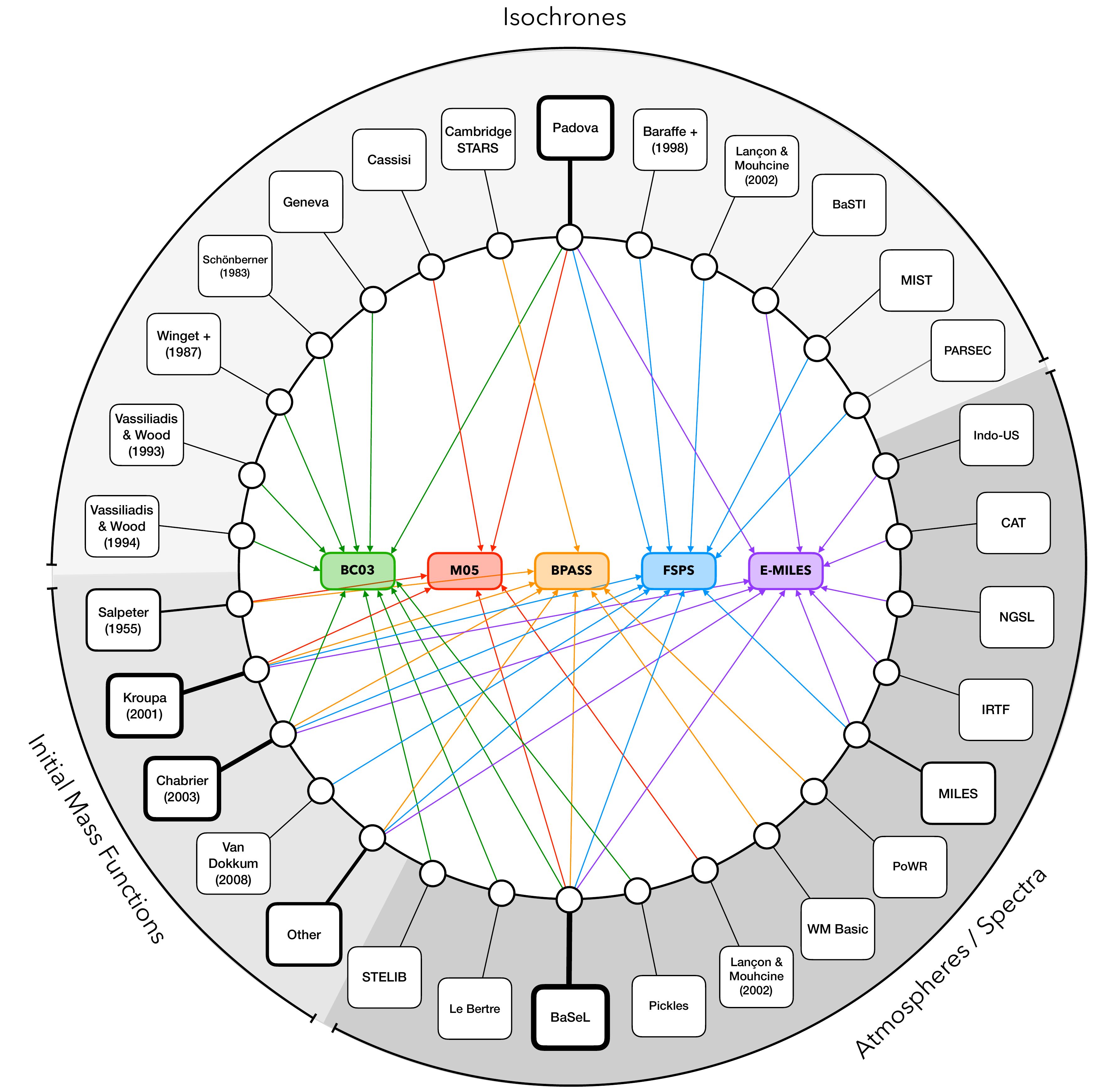}
    \caption{Schematic depicting some of the most popular stellar templates and the input isochrones, initial mass functions and atmospheres.
    BC03 \citep{BC03} is shown in green, M05 \citep{MarastonEvolutionarypopulationsynthesis2005} is shown in red, BPASS \citep{EldridgeSpectralpopulationsynthesis2009,EldridgeBinaryPopulationSpectral2017} is shown in orange, FSPS \citep{ConroyPropagationUncertaintiesStellar2009} is shown in blue and E-MILES \citep{Sanchez-BlazquezMediumresolutionIsaacNewton2006,VazdekisUVextendedEMILESstellar2016} is shown in purple.
    Additional isochrone references - \citet{SchoenbernerLatestagesstellar1983,Wingetindependentmethoddetermining1987,VassiliadisEvolutionlowintermediatemass1993,VassiliadisPostasymptoticgiantbranch1994,BaraffeEvolutionarymodelssolar1998,Lanconmodellingintermediateagestellar2002}; GENEVA: \citet{SchallerNewgridsstellar1992,CharbonnelGridsstellarmodels1996,CharbonnelGridsstellarmodels1999}, Padova: \citet{AlongiEvolutionarysequencesstellar1993,BressanEvolutionarysequencesstellar1993,FagottoEvolutionarysequencesstellar1994,FagottoEvolutionarysequencesstellar1994a,GirardiEvolutionarysequencesstellar1996,GirardiEvolutionarytracksisochrones2000}, Cassisi: \citet{Cassisieffectdiffusionred1997,CassisiIntermediateagemetaldeficient1997, CassisiGalacticglobularclusters2000}, Cambridge STARS: \citet{Eggletonevolutionlowmass1971,PolsApproximateinputphysics1995,Eldridgeprogenitorscorecollapsesupernovae2004}, BaSTI: \citet{PietrinferniLargeStellarEvolution2004,PietrinferniBaSTIStellarEvolution2013}, MIST: \citet{PaxtonModulesExperimentsStellar2011,PaxtonModulesExperimentsStellar2013,PaxtonModulesExperimentsStellar2015,ChoiMesaIsochronesStellar2016,DotterMESAIsochronesStellar2016}, PARSEC: \citet{BressanPARSECstellartracks2012}.
    Initial Mass Functions -  \citet{SalpeterLuminosityFunctionStellar1955,Kroupavariationinitialmass2001,ChabrierGalacticStellarSubstellar2003,vanDokkumEvidenceCosmicEvolution2008}
    Atmospheres/Spectra - STELIB: \citet{LeBorgneSTELIBlibrarystellar2003}, Le Bertre: \citet{LeBertreOpticalinfraredobservations1997,LeSidanerOpticalinfraredobservations1996}, BaSeL: \citet{BessellColorsextendedstatic1989,BessellColorsstratificationsextended1991,FluksspectraphotometryMgiant1994,AllardModelatmospheressub1995,2002RMxAC..12..150R}, Pickles: \citet{FanelliSpectralsynthesisultraviolet1992,PicklesStellarSpectralFlux1998}, WM Basic: \citet{SmithRealisticionizingfluxes2002}, PoWR: \citet{GrafenerStellarenvelopeinflation2012}, MILES: \citet{Sanchez-BlazquezMediumresolutionIsaacNewton2006}, IRTF: \citet{RockStellarpopulationsynthesis2015,RockMILESextendedStellar2016}, NGSL: \citet{2006hstc.conf..209G}, CAT:\citet{CenarroEmpiricalcalibrationnearinfrared2001}, Indo-US: \citet{ValdesIndoUSLibraryCoude2004}.
    }
    \label{fig:SPS}
\end{figure*}

\definecolor{Gray}{gray}{0.9}
\definecolor{BC03}{RGB}{0,143,0}
\definecolor{M05}{RGB}{255,38,0}
\definecolor{BPASS}{RGB}{255,147,0}
\definecolor{FSPS}{RGB}{0,150,255}
\definecolor{EMILES}{RGB}{148,55,255}

\begin{table*}
    \centering
    \caption{The tabular form of Figure~\ref{fig:SPS} where the stellar templates are represented as separate columns. As in Table~\ref{tab:SED_Diagram}, a cell is shaded if the stellar template makes use of the various input templates or models. The various references are included in the caption of Figure~\ref{fig:SPS}} 
    \label{tab:SPS}
    \begin{tabular}{ c  c  c  c  c  c  c   }
    \cline{3-7}
    \multicolumn{1}{l}{}  &  & \multicolumn{5}{c}{Stellar Template} \\
    \hline
	Template Type &	Template	&	M05	&	BPASS	&	BC03	&	FSPS	&	E-MILES	\\
	\hline
\multirow{13}{*}{Isochrones}	&	\citet{VassiliadisPostasymptoticgiantbranch1994}	&		&		&	\cellcolor{BC03}	&		&		\\
	&	\citet{VassiliadisEvolutionlowintermediatemass1993}	&		&		&	\cellcolor{BC03}	&		&		\\
	&	\citet{Wingetindependentmethoddetermining1987}	&		&		&	\cellcolor{BC03}	&		&		\\
	&	\citet{SchoenbernerLatestagesstellar1983}	&		&		&	\cellcolor{BC03}	&		&		\\
	&	Geneva	&		&		&	\cellcolor{BC03}	&		&		\\
	&	Cassisi	&	\cellcolor{M05}	&		&		&		&		\\
	&	Cambridge STARS	&		&	\cellcolor{BPASS}	&		&		&		\\
	&	Padova	&	\cellcolor{M05}	&		&	\cellcolor{BC03}	&	\cellcolor{FSPS}	&	\cellcolor{EMILES}	\\
	&	\citet{BaraffeEvolutionarymodelssolar1998}	&		&		&		&	\cellcolor{FSPS}	&		\\
	&	\citet{Lanconmodellingintermediateagestellar2002}	&		&		&		&	\cellcolor{FSPS}	&		\\
	&	BaSTI	&		&		&		&		&	\cellcolor{EMILES}	\\
	&	MIST	&		&		&		&	\cellcolor{FSPS}	&		\\
	&	PARSEC	&		&		&		&	\cellcolor{FSPS}	&		\\
	\hline
\multirow{12}{*}{Atmospheres / Spectra}	&	Indo-US	&		&		&		&		&	\cellcolor{EMILES}	\\
	&	CAT	&		&		&		&		&	\cellcolor{EMILES}	\\
	&	NGSL	&		&		&		&		&	\cellcolor{EMILES}	\\
	&	IRTF	&		&		&		&		&	\cellcolor{EMILES}	\\
	&	MILES	&		&		&		&	\cellcolor{FSPS}	&	\cellcolor{EMILES}	\\
	&	PoWR	&		&	\cellcolor{BPASS}	&		&		&		\\
	&	WM Basic	&		&	\cellcolor{BPASS}	&		&		&		\\
	&	\citet{Lanconmodellingintermediateagestellar2002}	&	\cellcolor{M05}	&		&		&		&		\\
	&	Pickles	&		&		&	\cellcolor{BC03}	&		&		\\
	&	BaSeL	&	\cellcolor{M05}	&	\cellcolor{BPASS}	&	\cellcolor{BC03}	&	\cellcolor{FSPS}	&	\cellcolor{EMILES}	\\
	&	Le Bertre	&		&		&	\cellcolor{BC03}	&		&		\\
	&	STELIB	&		&		&	\cellcolor{BC03}	&		&		\\
	\hline
\multirow{5}{*}{Initial Mass Functions}	&	\citet{SalpeterLuminosityFunctionStellar1955}	&	\cellcolor{M05}	&	\cellcolor{BPASS}	&		&		&		\\
	&	\citet{Kroupavariationinitialmass2001}	&	\cellcolor{M05}	&	\cellcolor{BPASS}	&		&	\cellcolor{FSPS}	&	\cellcolor{EMILES}	\\
	&	\citet{ChabrierGalacticStellarSubstellar2003}	&		&	\cellcolor{BPASS}	&	\cellcolor{BC03}	&	\cellcolor{FSPS}	&	\cellcolor{EMILES}	\\
	&	\citet{vanDokkumEvidenceCosmicEvolution2008}	&		&		&		&	\cellcolor{FSPS}	&		\\
	&	Other	&		&	\cellcolor{BPASS}	&		&	\cellcolor{FSPS}	&	\cellcolor{EMILES}	\\
	\hline
    \end{tabular}
\end{table*}

\section{Redshift Sources} \label{app:redshifts}

\begin{figure}
    \centering
    \includegraphics[width = \linewidth]{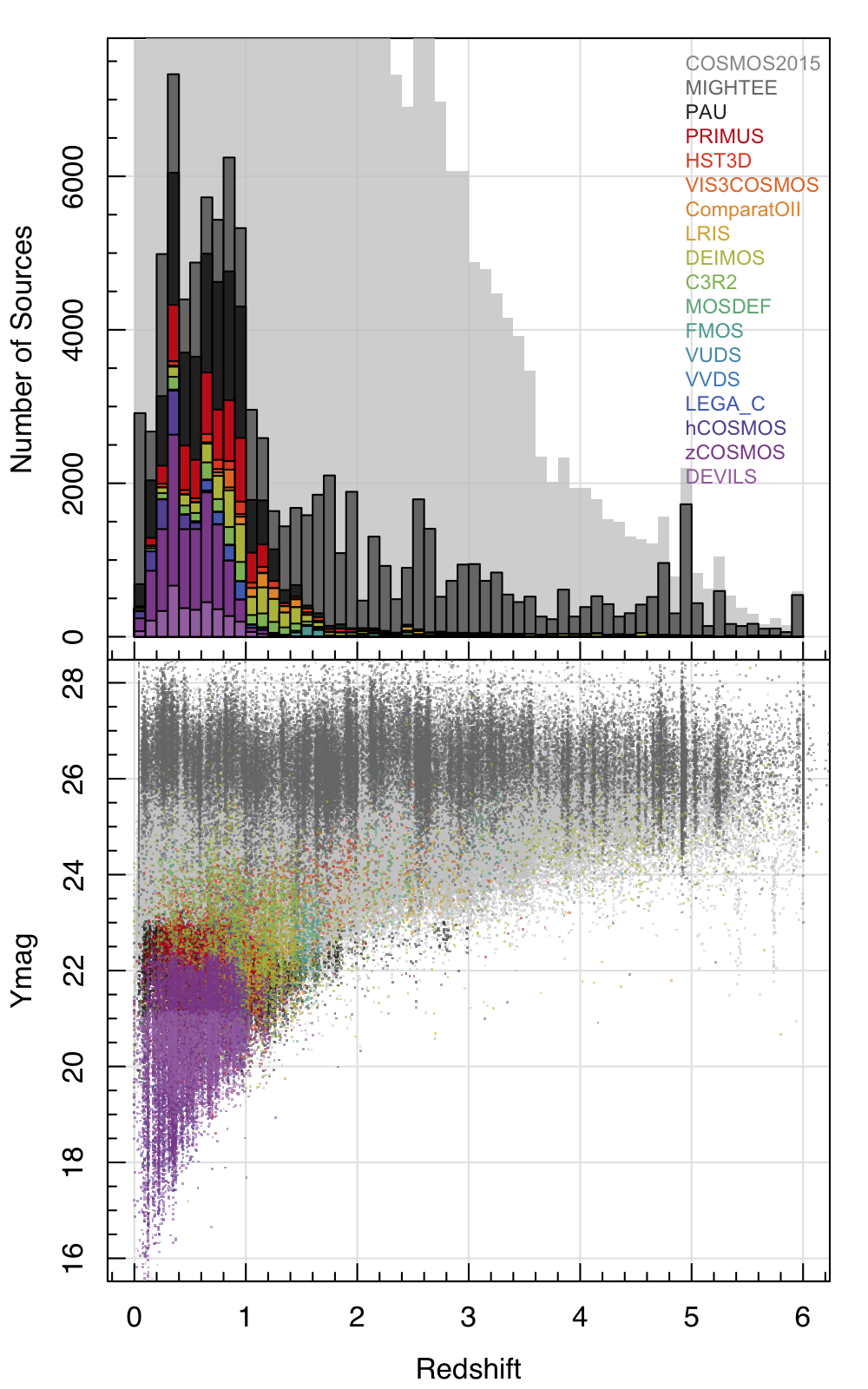}
    \caption{Stacked redshift distribution of the sample - (\textit{upper}) histogram of the redshifts used from $z=0$ to $z=6$. There are $\sim 500$ objects with $z>6$ of which 4 are spectroscopic (DEIMOS) and the rest photometric (MIGHTEE). The \textit{lower} panel shows the distribution of the sources as a function of Y band magnitude.}
    \label{fig:redshift_ymag}
\end{figure}

Here we present the references for each of the redshift catalogues that were compiled to make the redshift catalogue for this work. Table \ref{tab:redshifts} presents the references for each of the catalogues, the type of redshift, the number and distribution of redshifts used in this work. We also present the flag values from the original catalogues that were selected as good redshifts. In Figure \ref{fig:redshift_ymag} we show the distribution of these redshifts compared with the Y-band magnitude of the object from our photometry catalogue coloured by redshift source. 

\begin{table*}
    \centering
    \caption{Summary of the redshift sources used, the type of redshift measurement and reference, and the number and distribution of redshifts that make it into our final sample for fitting. The flag column shows the flags from the original authors' flag system that were deemed to be good spectroscopic redshift measurements. We show the photometric accuracy and outlier rate for each of the photometric redshift catalogues in the final two columns. The redshift sources are ranked by priority. }
    \label{tab:redshifts}
    \begin{tabular}{l l l l l l l l l }
    \hline
Redshift Source	&	Type	&	Reference	&	Nz	&	$z_\text{med}$	&	$z_\text{range}$	&	Flags	&	Accuracy  & Outlier Rate	\\
\hline
DEVILS	&	Spec	&	\cite{DaviesDeepExtragalacticVIsible2018}	&	3,394	&	0.509	&	[0.0002,1.240] 	&	Prob > 0.9	&	&	\\
zCOSMOS	&	Spec	&	\cite{LillyzCOSMOS10kBrightSpectroscopic2009}	&	9,774	&	0.494	&	[0.00, 4.447]	&	*  	&	&	\\
hCOSMOS	&	Spec	&	\cite{DamjanovhCOSMOSDenseSpectroscopic2018}	&	1,641	&	0.312	&	[0.00623,1.26471] 	&		&	&	\\
LEGA-C	&	Spec	&	\cite{StraatmanLargeEarlyGalaxy2018}	&	839	&	0.866	&	[0.359, 2.480]	&	0	&	&	\\
VVDS	&	Spec	&	\cite{LeFevreVIMOSVLTDeep2013}	&	0	&		&		&	3 or 4	&	&	\\
VUDS	&	Spec	&	\cite{LeFevreVIMOSUltraDeepSurvey2015}	&	126	&	2.510	&	[0.00, 4.908]	&	1.5,2,3,4,9	&	&	\\
FMOS	&	Spec	&	\cite{KashinoFMOSCOSMOSSurveyStarforming2019}	&	285	&	1.557	&	[0.895,2.486] 	&	3,4	&	&	\\
MOSDEF	&	Spec	&	\cite{KriekMOSFIREDeepEvolution2015}	&	318	&	2.280	&	[0.803, 3.712]	&	> 4	&	&	\\
C3R2	&	Spec	&	\cite{MastersCompleteCalibrationColorRedshift2019,MastersCompleteCalibrationColorRedshift2017}	&	2,242	&	0.890	&	[0.0625, 4.499]	&	3,3.5,4	&	&	\\
DEIMOS	&	Spec	&	\cite{HasingerDEIMOS10KSpectroscopic2018}	&	4,393	&	1.028	&	[0.00, 6.604]	&	1.5, 2	&	&	\\
LRIS	&	Spec	&	\cite{LeeFirstDataRelease2018}	&	217	&	2.530	&	[0.00, 3.029]	&	$ >= 3$	&	&	\\
ComparatOII	&	Spec	&	\cite{Comparat65evolutionbright2015}	&	883	&	1.172	&	[0.00, 4.816]	&		&	&	\\
VIS3COS	&	Spec	&	\cite{Paulino-AfonsoVIS3COSSurveyoverview2018}	&	348	&	0.839	&	[0.0248, 1.261]	&		&	&	\\
3D-HST	&	Grism	&	\cite{Momcheva3DHSTSurveyHubble2016}	&	1,369	&	0.962	&	[0.0529, 3.909]	&		&	&	\\
PRIMUS	&	Grism	&	\cite{CoolPRIsmMUltiobjectSurvey2013}	&	6,149	&	0.698	&	[0.0215, 3,485]	&		&	&	\\
PAU	&	Photo	&	\cite{AlarconPAUSurveyimproved2021}	&	15,563	&	0.692	&	[0.00, 2.990]	&		&	0.009	&  2\% \\
COSMOS2015	&	Photo	&	\cite{LaigleCOSMOS2015CATALOGEXPLORING2016}	&	411,472	&	1.295	&	[0.005,  5.995]	&		&	0.007/0.021$^\dagger$ &  0.5\%/13.2\%$^\dagger$	\\
MIGHTEE	&	Photo	&	\cite{AdamsrestframeUVluminosity2020} 	& 	45,744	&	1.944	& 	[0.04,  9] 	&		&	0.027& 3.9\% 	\\
\hline

    \end{tabular}
    
    * Z\_CC>2 \& Z\_CC<6, or Z\_CC>12 \& Z\_CC<16, or Z\_CC>22 \& Z\_CC<26
    
    $\dagger$ for comparison to $z < 1.2$ / $3 < z < 6$ respectively
\end{table*}

\begin{figure}
    \centering
    \includegraphics[width = \linewidth]{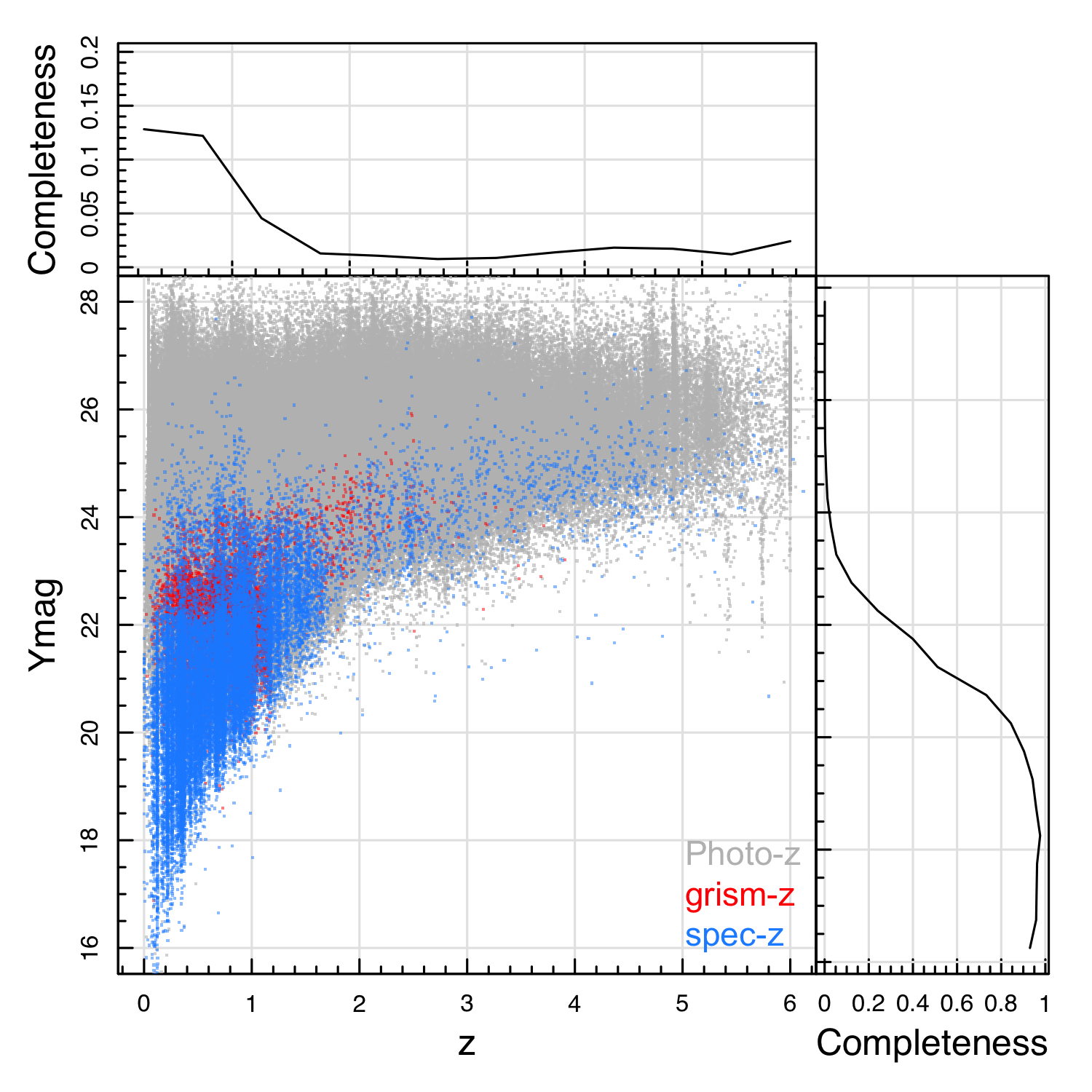}
    \caption{Distribution of sources of each redshift type in redshift and Y band magnitude. We show sources with photometric redshifts in grey, grism redshifts in red and spectroscopic redshifts in blue. We also show the fraction of sources that have spectroscopic or grism redshifts as a function of both redshift and Ymag.}
    \label{fig:ymag_source}
\end{figure}

\section{Fit Results and Regression Values}\label{App:Vals}

\begin{table*}
    \centering
    \caption{Parameters of the best fit model to the stellar mass function and stellar mass density out to $z=4.25$ as per Equation~\ref{eq:DoubleSchechter}. Full table available online.}
    \label{tab:SMFVals}
    \begin{tabular}{c c c c c c c}
    \hline
Redshift Range	&	$M^*$ 	&	$ \log_{10} (\phi_1^* /  M_\odot$ Mpc$^{-3}) $	&	$ \log_{10} (\phi_2^* /  M_\odot$ Mpc$^{-3}) $	&  $\alpha_1$	&	$\alpha_2$	& $\log_{10} ($ SMD / $M_\odot$ Mpc$^{-3}$) \\
\hline
0.02 < z <  0.08	&	10.4482	$\pm$	0.22	&	-2.6039	$\pm$	0.15	&	-3.618	$\pm$	0.19	&	-0.4864	$\pm$	0.096	&	-1.8497	$\pm$	0.045	&	8.5196	$\pm$	0.021	\\
0.08 < z <  0.14	&	10.7618	$\pm$	0.075	&	-2.4418	$\pm$	0.088	&	-3.0725	$\pm$	0.067	&	-0.521	$\pm$	0.095	&	-1.5937	$\pm$	0.023	&	8.4876	$\pm$	0.022	\\
0.14 < z <  0.20	&	10.8102	$\pm$	0.078	&	-2.8321	$\pm$	0.082	&	-3.5971	$\pm$	0.072	&	-0.4489	$\pm$	0.092	&	-1.7176	$\pm$	0.021	&	8.4578	$\pm$	0.023	\\
0.2 < z <  0.28	&	10.8063	$\pm$	0.045	&	-2.5776	$\pm$	0.052	&	-3.1587	$\pm$	0.05	&	-0.4473	$\pm$	0.087	&	-1.5434	$\pm$	0.018	&	8.4262	$\pm$	0.024	\\
0.28 < z <  0.36	&	10.7985	$\pm$	0.038	&	-2.5317	$\pm$	0.049	&	-2.9277	$\pm$	0.04	&	-0.391	$\pm$	0.088	&	-1.4541	$\pm$	0.016	&	8.3971	$\pm$	0.025	\\
0.36 < z <  0.45	&	10.7994	$\pm$	0.035	&	-2.6343	$\pm$	0.045	&	-3.0026	$\pm$	0.038	&	-0.3617	$\pm$	0.088	&	-1.4773	$\pm$	0.016	&	8.3701	$\pm$	0.026	\\
0.45 < z <  0.56	&	10.7717	$\pm$	0.03	&	-2.7099	$\pm$	0.033	&	-3.2329	$\pm$	0.045	&	-0.3412	$\pm$	0.083	&	-1.5399	$\pm$	0.02	&	8.3419	$\pm$	0.027	\\
0.56 < z <  0.68	&	10.7692	$\pm$	0.025	&	-2.644	$\pm$	0.024	&	-3.2558	$\pm$	0.046	&	-0.2845	$\pm$	0.076	&	-1.5621	$\pm$	0.023	&	8.311	$\pm$	0.028	\\
0.68 < z <  0.82	&	10.7289	$\pm$	0.021	&	-2.5994	$\pm$	0.018	&	-3.1678	$\pm$	0.042	&	-0.1251	$\pm$	0.075	&	-1.5336	$\pm$	0.023	&	8.2725	$\pm$	0.028	\\
0.82 < z <  1.00	&	10.7964	$\pm$	0.017	&	-2.49	$\pm$	0.016	&	-3.0578	$\pm$	0.052	&	-0.2	$\pm$	0.071	&	-1.4206	$\pm$	0.031	&	8.2144	$\pm$	0.029	\\
1.00 < z <  1.20	&	10.8076	$\pm$	0.02	&	-2.8084	$\pm$	0.021	&	-3.3484	$\pm$	0.056	&	-0.2212	$\pm$	0.08	&	-1.5264	$\pm$	0.037	&	8.1284	$\pm$	0.029	\\
1.20 < z <  1.45	&	10.8208	$\pm$	0.02	&	-2.8776	$\pm$	0.03	&	-3.347	$\pm$	0.078	&	-0.36	$\pm$	0.087	&	-1.4332	$\pm$	0.054	&	8.0039	$\pm$	0.028	\\
1.45 < z <  1.75	&	10.8309	$\pm$	0.019	&	-3.0228	$\pm$	0.083	&	-3.4117	$\pm$	0.2	&	-0.6499	$\pm$	0.11	&	-1.2988	$\pm$	0.1	&	7.8402	$\pm$	0.027	\\
1.75 < z <  2.20	&	10.8065	$\pm$	0.017	&	-3.1119	$\pm$	0.073	&	-3.8187	$\pm$	0.38	&	-0.8028	$\pm$	0.1	&	-1.3545	$\pm$	0.13	&	7.6388	$\pm$	0.026	\\
2.20 < z <  2.60	&	10.8055	$\pm$	0.04	&	-3.9046	$\pm$	0.35	&	-3.4468	$\pm$	0.076	&	-0.5408	$\pm$	0.11	&	-1.2697	$\pm$	0.072	&	7.494	$\pm$	0.024	\\
2.60 < z <  3.25	&	10.7558	$\pm$	0.034	&	-4.5695	$\pm$	0.31	&	-3.5757	$\pm$	0.044	&	-0.4979	$\pm$	0.1	&	-1.5065	$\pm$	0.058	&	7.3953	$\pm$	0.023	\\
3.25 < z <  3.75	&	10.5822	$\pm$	0.041	&	-4.677	$\pm$	0.37	&	-3.6266	$\pm$	0.054	&	-0.5004	$\pm$	0.1	&	-1.535	$\pm$	0.087	&	7.3452	$\pm$	0.024	\\
3.75 < z <  4.25	&	10.7851	$\pm$	0.064	&	-4.9476	$\pm$	0.36	&	-4.2674	$\pm$	0.07	&	-0.4985	$\pm$	0.1	&	-1.5213	$\pm$	0.091	&	7.3182	$\pm$	0.024	\\
	\hline
    \end{tabular}

\end{table*}

\begin{table*}
    \centering
    \caption{Regression functions displayed in Figure~\ref{fig:SMF_Evolving} for the two-component Schecheter function fits. Fits are linear in lookback time ($t_\text{lb}$), where the $A_i$ coefficient applies to the $i$th power of $t_\text{lb}$, except for the fits to the $\phi_1^*$ which are of the form $A_1\times t_\text{lb}^7 + A_0$.}
    \label{tab:my_label}
    \begin{tabular}{c c c}
    \hline
    Parameter & $A_1$ & $A_0$ \\
    \hline
$ M^* $	&	0.0006	$\pm$	0.001	&	10.7792	$\pm$	0.01	\\
$\log_{10} (\phi^*_1 )$	&	$-7\times10^{-8}$	$\pm$	0.001	&	-2.5825	$\pm$	0.022	\\
$ \log_{10} (\phi^*_2 )$	&	-0.0843	$\pm$	0.001	&	-2.6863	$\pm$	0.007	\\
$\alpha_1$	&	-0.0007	$\pm$	0.001	&	-0.3993	$\pm$	0.022	\\
$\alpha_2$	&	0.0013	$\pm$	0.001	&	-1.5138	$\pm$	0.008	\\
    \hline
    \end{tabular}
\end{table*}

\begin{table*}
    \centering
    \caption{Parameters of the best fit model to the star-formation main sequence from Equation~\ref{eq:LeeSFMS}. The full sample of galaxies is split into 20 redshift bins of $\sim 0.75$ Gyrs. Full table available online.  }
    \label{tab:SFMSVals}
    \begin{tabular}{c l l l l}
    \hline
    	Redshift Range	&	$S_0$ 	&	$\mathcal{M}_0$	&	$\alpha$	& $\beta$ \\
    \hline
0.02 < z <  0.08	&	0.064	$\pm$	0.0019	&	9.5971	$\pm$	0.011	&	0.9703	$\pm$	0.015	&	0.187	$\pm$	0.038	\\
0.08 < z <  0.14	&	0.139	$\pm$	0.00076	&	9.452	$\pm$	0.0047	&	1.1515	$\pm$	0.0074	&	0.1576	$\pm$	0.016	\\
0.14 < z <  0.20	&	0.3174	$\pm$	0.032	&	9.4452	$\pm$	0.056	&	1.1838	$\pm$	0.052	&	0.15	$\pm$	0.0017	\\
0.20 < z <  0.28	&	0.7597	$\pm$	0.02	&	10.1064	$\pm$	0.012	&	1.0728	$\pm$	0.013	&	0.15	$\pm$	0.017	\\
0.28 < z <  0.36	&	1.1795	$\pm$	0.019	&	10.6332	$\pm$	0.04	&	0.9618	$\pm$	0.016	&	0.1997	$\pm$	0.039	\\
0.36 < z <  0.45	&	0.8962	$\pm$	0.0087	&	10.2632	$\pm$	0.013	&	0.893	$\pm$	0.0065	&	0.2528	$\pm$	0.053	\\
0.45 < z <  0.56	&	0.5551	$\pm$	0.0045	&	9.4763	$\pm$	0.021	&	1.0229	$\pm$	0.039	&	0.15	$\pm$	0.015	\\
0.56 < z <  0.68	&	0.7267	$\pm$	0.0086	&	9.5132	$\pm$	0.02	&	1.0327	$\pm$	0.02	&	0.15	$\pm$	0.0046	\\
0.68 < z <  0.82	&	0.9698	$\pm$	0.016	&	9.7331	$\pm$	0.023	&	1.0036	$\pm$	0.027	&	0.1519	$\pm$	0.022	\\
0.82 < z <  1.00	&	1.1263	$\pm$	0.028	&	9.8598	$\pm$	0.045	&	0.972	$\pm$	0.027	&	0.1714	$\pm$	0.02	\\
1.00 < z <  1.20	&	1.3363	$\pm$	0.0053	&	10.0925	$\pm$	0.019	&	0.9298	$\pm$	0.02	&	0.1902	$\pm$	0.022	\\
1.20 < z <  1.45	&	1.4696	$\pm$	0.013	&	10.1153	$\pm$	0.021	&	0.9463	$\pm$	0.035	&	0.1784	$\pm$	0.036	\\
1.45 < z <  1.75	&	1.5302	$\pm$	0.0092	&	10.1547	$\pm$	0.018	&	0.9931	$\pm$	0.024	&	0.1662	$\pm$	0.02	\\
1.75 < z <  2.20	&	1.6857	$\pm$	0.0096	&	10.3276	$\pm$	0.021	&	0.9374	$\pm$	0.029	&	0.2101	$\pm$	0.047	\\
2.20 < z <  2.60	&	1.8791	$\pm$	0.007	&	10.4174	$\pm$	0.017	&	0.9589	$\pm$	0.016	&	0.1903	$\pm$	0.034	\\
2.6 < z <  3.25	&	1.8457	$\pm$	0.007	&	10.1967	$\pm$	0.018	&	0.8964	$\pm$	0.041	&	0.2021	$\pm$	0.042	\\
3.25 < z <  3.75	&	2.3238	$\pm$	0.0099	&	10.7865	$\pm$	0.013	&	0.9163	$\pm$	0.0098	&	0.2117	$\pm$	0.05	\\
3.75 < z <  4.25	&	2.4844	$\pm$	0.0043	&	10.9361	$\pm$	0.0078	&	0.9013	$\pm$	0.0045	&	0.2126	$\pm$	0.05	\\
4.25 < z <  5.00	&	2.6792	$\pm$	0.0057	&	11.1114	$\pm$	0.0035	&	0.9299	$\pm$	0.0029	&	0.2066	$\pm$	0.042	\\
5.00 < z <  9.00	&	2.7513	$\pm$	0.013	&	10.9998	$\pm$	0.0098	&	0.9861	$\pm$	0.011	&	0.2019	$\pm$	0.067	\\
\hline
    \end{tabular}

\end{table*}

\section{Impact of $\chi^2$ cuts on the derived stellar mass function}\label{App:chi2}

\begin{figure}
    \centering
    \includegraphics[width = \linewidth]{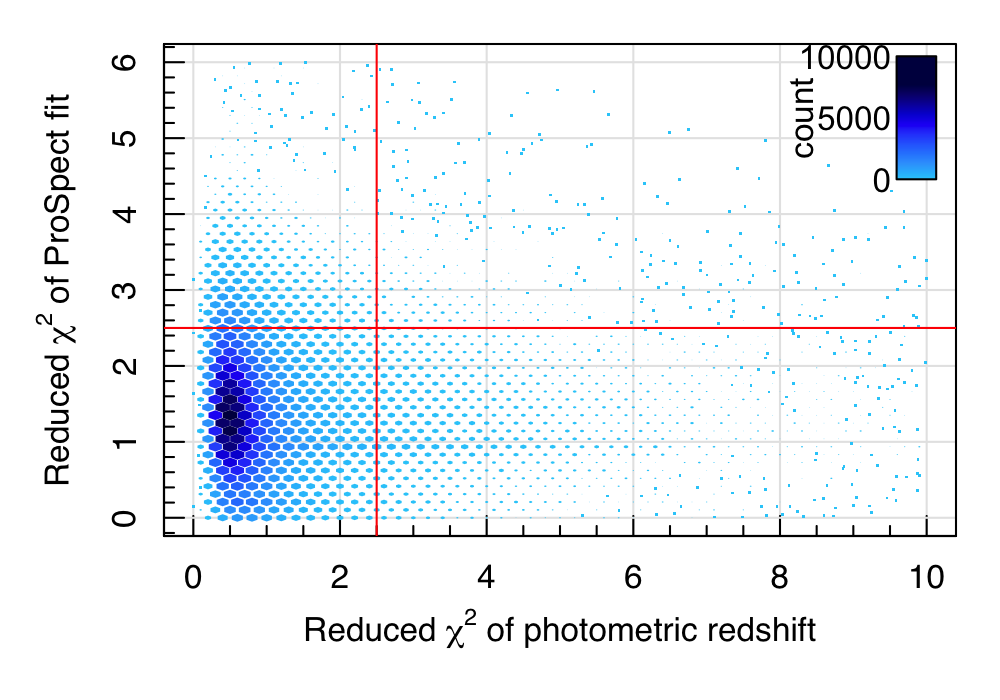}
    \caption{The reduced $\chi^2$ from the photometric redshift measurements compared to the converted reduced $\chi^2$ of the \textsc{ProSpect} fits. 
    The horizontal and vertical red lines show the $\chi^2$ cuts for the \textsc{ProSpect} and photometric redshift fits respectively as described in Appendix~\ref{App:chi2}.}
    \label{fig:Chi2dist}
\end{figure}

\begin{figure*}
    \centering
    \includegraphics[width = \linewidth]{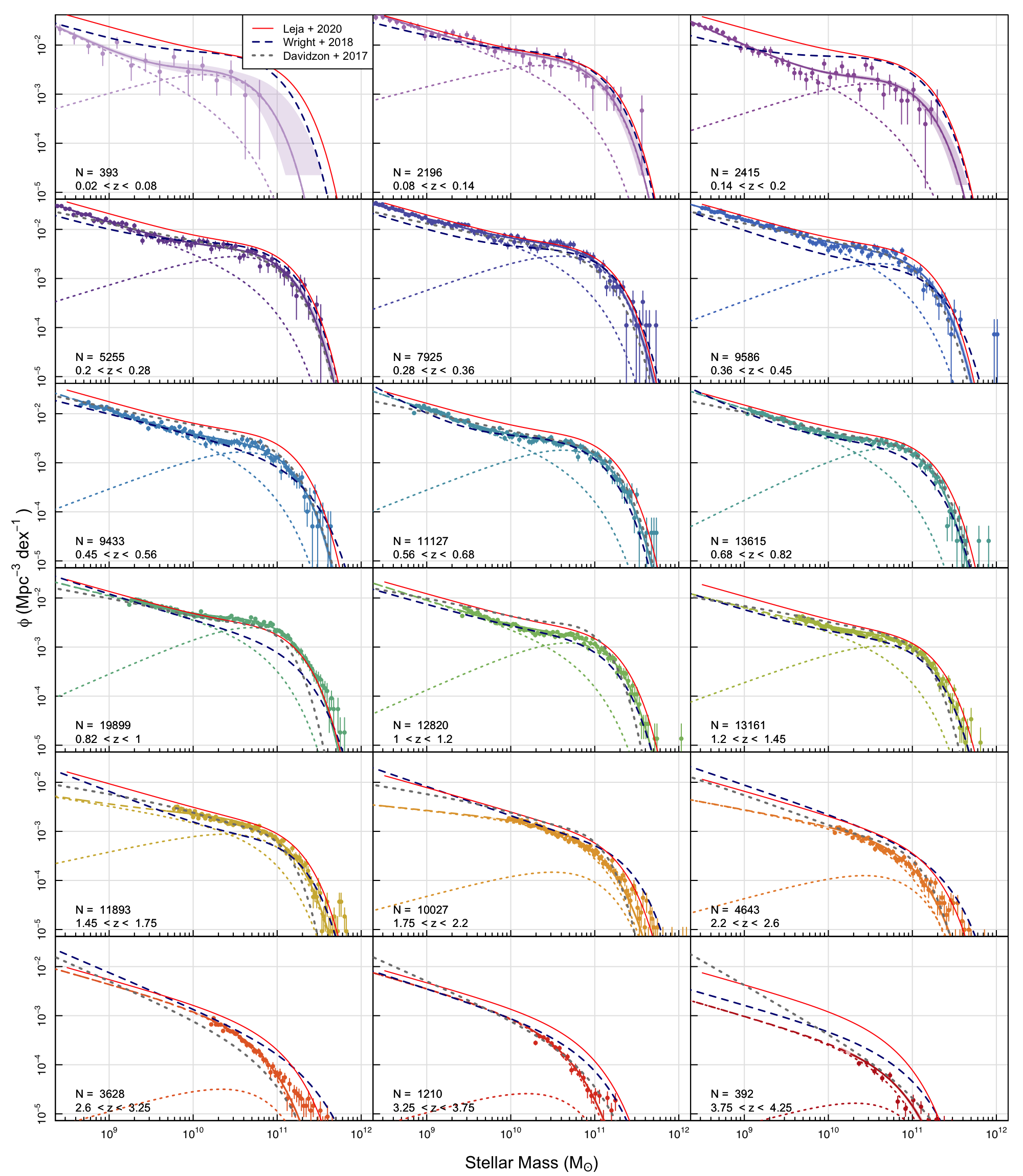}
    \caption{As per Figure~\ref{fig:SMF} but with objects removed if they have a reduced $\chi^2 > 2.5$ in either the photometric redshift fit or the \textsc{ProSpect} fit.} 
    \label{fig:SMFChi2}
\end{figure*}

To show the impact of removing galaxies with a `poor' fit in either the photometric redshift or \textsc{ProSpect} SED fit we
re-calculate the stellar mass function as per Section~\ref{sec:SMF} using the following cuts.
\citet{DavidzonCOSMOS2015galaxystellar2017} applies a $\chi^2 < 10$ cut on the photometric redshift fits which is found to remove 0.17 per cent of the data in our case. 
To understand the worst-case scenario of poor fits biasing our results we apply a more conservative cut at $\chi^2 = 2.5$ for both the photometric redshift fit and the \textsc{ProSpect} fit. 
A $\chi^2 = 2.5$ cut was selected as this is where the \textsc{ProSpect} fits begin to visually decline in quality.
The distribution of $\chi^2$ values is shown in Figure~\ref{fig:Chi2dist} with the two cuts shown as the red lines.
These cuts remove 37,854 galaxies, of which 17,632 are above our completeness cut (Equation~\ref{eq:MassComplete}). 
This represents a removal of 10 per cent of the objects above the completeness cut. 
We propagate this selection to our final stellar mass function fits and as seen in Figure~\ref{fig:SMFChi2}, it produces an entirely negligible impact on the results (fit parameters changing by much less than the estimated errors).

\begin{figure}
    \centering
    \includegraphics[width = \linewidth]{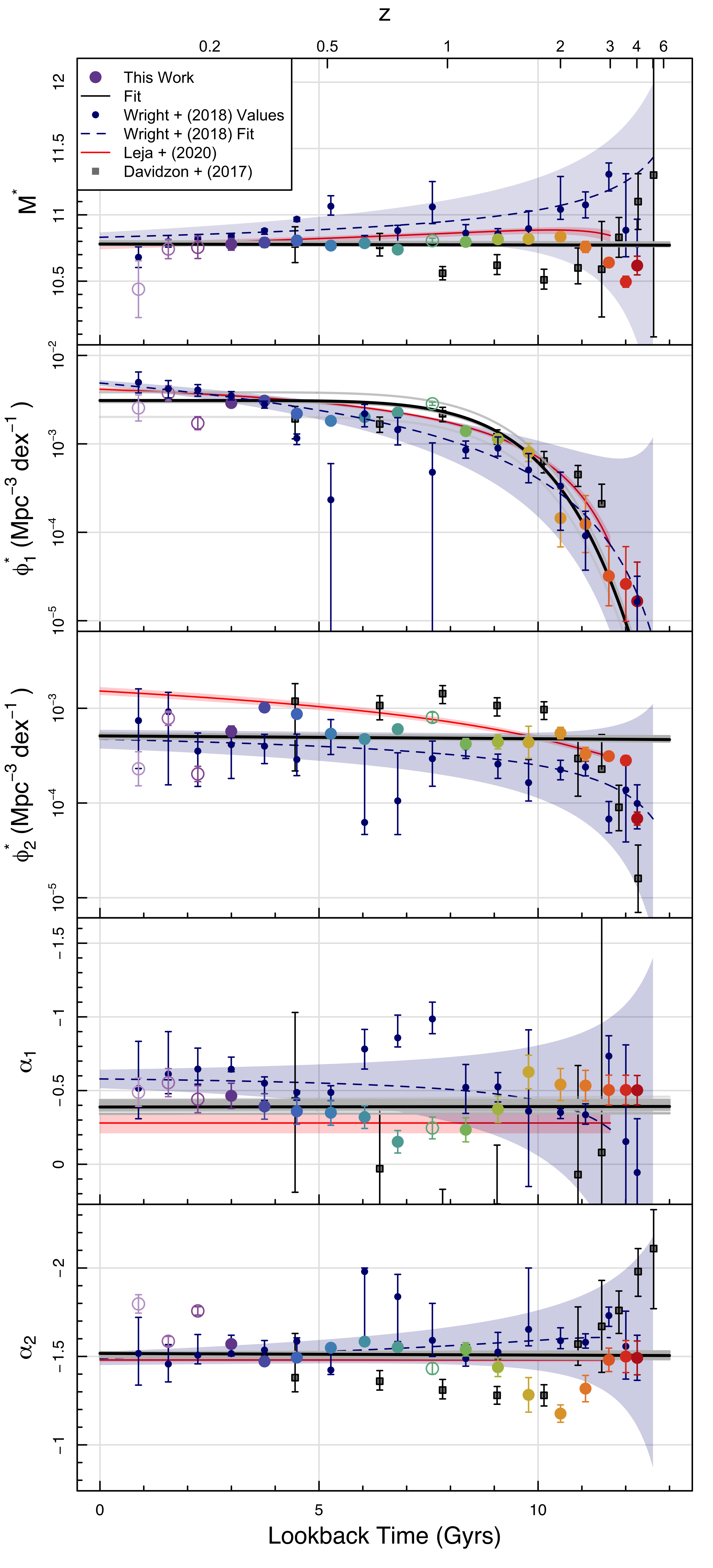}
    \caption{As per Figure~\ref{fig:SMF_Evolving} but with objects removed if they have a a reduced $\chi^2 > 2.5$ in either the photometric redshift fit or the \textsc{ProSpect} fit.
    We show the fit from the whole sample as the solid black line and the grey lines show samples from the posterior of the fit. }
    \label{fig:SMFEvoChi2}
\end{figure}


\bsp	
\label{lastpage}
\end{document}